\documentclass[traditabstract,longauth]{aa} 

\usepackage{graphicx}
\usepackage{amsmath,amsfonts,amssymb}
\usepackage[breaklinks,colorlinks,citecolor=blue,pdfa=true]{hyperref}
\usepackage{color}
\usepackage{fixltx2e}
\usepackage{natbib}
\usepackage{url}
\usepackage{multirow}

\usepackage{ifthen}

\usepackage[T1]{fontenc}
\usepackage{lmodern}
\usepackage{ifxetex,ifluatex}

\usepackage{dblfloatfix}
\usepackage{morefloats}

\usepackage[switch]{lineno}

\bibpunct{(}{)}{;}{a}{}{,}

\def\setsymbol#1#2{\expandafter\def\csname #1\endcsname{#2}}
\def\getsymbol#1{\csname #1\endcsname}

\def\Planck{\textit{Planck}}

\newbox\tablebox    \newdimen\tablewidth
\def\leaderfil{\leaders\hbox to 5pt{\hss.\hss}\hfil}
\def\endPlancktable{\tablewidth=\columnwidth 
    $$\hss\copy\tablebox\hss$$
    \vskip-\lastskip\vskip -2pt}
\def\endPlancktablewide{\tablewidth=\textwidth 
    $$\hss\copy\tablebox\hss$$
    \vskip-\lastskip\vskip -2pt}
\def\tablenote#1 #2\par{\begingroup \parindent=0.8em
    \abovedisplayshortskip=0pt\belowdisplayshortskip=0pt
    \noindent
    $$\hss\vbox{\hsize\tablewidth \hangindent=\parindent \hangafter=1 \noindent
    \hbox to \parindent{$^#1$\hss}\strut#2\strut\par}\hss$$
    \endgroup}
\def\doubleline{\vskip 3pt\hrule \vskip 1.5pt \hrule \vskip 5pt}

\def\L2{\ifmmode L_2\else $L_2$\fi}

\def\DeltaT{\ifmmode \Delta T\else $\Delta T$\fi}
\def\deltat{\ifmmode \Delta t\else $\Delta t$\fi}
\def\fknee{\ifmmode f_{\rm knee}\else $f_{\rm knee}$\fi}
\def\Fmax{\ifmmode F_{\rm max}\else $F_{\rm max}$\fi}
\def\solar{\ifmmode{\rm M}_{\mathord\odot}\else${\rm M}_{\mathord\odot}$\fi}
\def\Msolar{\ifmmode{\rm M}_{\mathord\odot}\else${\rm M}_{\mathord\odot}$\fi}
\def\Lsolar{\ifmmode{\rm L}_{\mathord\odot}\else${\rm L}_{\mathord\odot}$\fi}
\def\inv{\ifmmode^{-1}\else$^{-1}$\fi}
\def\mo{\ifmmode^{-1}\else$^{-1}$\fi}
\def\sup#1{\ifmmode ^{\rm #1}\else $^{\rm #1}$\fi}
\def\expo#1{\ifmmode \times 10^{#1}\else $\times 10^{#1}$\fi}
\def\,{\thinspace}
\def\lsim{\mathrel{\raise .4ex\hbox{\rlap{$<$}\lower 1.2ex\hbox{$\sim$}}}}
\def\gsim{\mathrel{\raise .4ex\hbox{\rlap{$>$}\lower 1.2ex\hbox{$\sim$}}}}

\def\simprop{\mathrel{\raise .4ex\hbox{\rlap{$\propto$}\lower 1.2ex\hbox{$\sim$}}}}
\def\deg{\ifmmode^\circ\else$^\circ$\fi}
\def\pdeg{\ifmmode $\setbox0=\hbox{$^{\circ}$}\rlap{\hskip.11\wd0 .}$^{\circ}
          \else \setbox0=\hbox{$^{\circ}$}\rlap{\hskip.11\wd0 .}$^{\circ}$\fi}
\def\arcs{\ifmmode {^{\scriptstyle\prime\prime}}
          \else $^{\scriptstyle\prime\prime}$\fi}
\def\arcm{\ifmmode {^{\scriptstyle\prime}}
          \else $^{\scriptstyle\prime}$\fi}
\newdimen\sa  \newdimen\sb
\def\parcs{\sa=.07em \sb=.03em
     \ifmmode \hbox{\rlap{.}}^{\scriptstyle\prime\kern -\sb\prime}\hbox{\kern -\sa}
     \else \rlap{.}$^{\scriptstyle\prime\kern -\sb\prime}$\kern -\sa\fi}
\def\parcm{\sa=.08em \sb=.03em
     \ifmmode \hbox{\rlap{.}\kern\sa}^{\scriptstyle\prime}\hbox{\kern-\sb}
     \else \rlap{.}\kern\sa$^{\scriptstyle\prime}$\kern-\sb\fi}
\def\ra[#1 #2 #3.#4]{#1\sup{h}#2\sup{m}#3\sup{s}\llap.#4}
\def\dec[#1 #2 #3.#4]{#1\deg#2\arcm#3\arcs\llap.#4}
\def\deco[#1 #2 #3]{#1\deg#2\arcm#3\arcs}
\def\rra[#1 #2]{#1\sup{h}#2\sup{m}}

\def\dots{\relax\ifmmode \ldots\else $\ldots$\fi}
\def\WHzsr{\ifmmode $W\,Hz\mo\,sr\mo$\else W\,Hz\mo\,sr\mo\fi}
\def\mHz{\ifmmode $\,mHz$\else \,mHz\fi}
\def\GHz{\ifmmode $\,GHz$\else \,GHz\fi}
\def\mKs{\ifmmode $\,mK\,s$^{1/2}\else \,mK\,s$^{1/2}$\fi}
\def\muKs{\ifmmode \,\mu$K\,s$^{1/2}\else \,$\mu$K\,s$^{1/2}$\fi}
\def\muKRJs{\ifmmode \,\mu$K$_{\rm RJ}$\,s$^{1/2}\else \,$\mu$K$_{\rm RJ}$\,s$^{1/2}$\fi}
\def\muKHz{\ifmmode \,\mu$K\,Hz$^{-1/2}\else \,$\mu$K\,Hz$^{-1/2}$\fi}
\def\MJysr{\ifmmode \,$MJy\,sr\mo$\else \,MJy\,sr\mo\fi}
\def\MJysrmK{\ifmmode \,$MJy\,sr\mo$\,mK$_{\rm CMB}\mo\else \,MJy\,sr\mo\,mK$_{\rm CMB}\mo$\fi}
\def\microns{\ifmmode \,\mu$m$\else \,$\mu$m\fi}

\def\muK{\ifmmode \,\mu$K$\else \,$\mu$\hbox{K}\fi}
\def\microK{\ifmmode \,\mu$K$\else \,$\mu$\hbox{K}\fi}
\def\muW{\ifmmode \,\mu$W$\else \,$\mu$\hbox{W}\fi}
\def\kms{\ifmmode $\,km\,s$^{-1}\else \,km\,s$^{-1}$\fi}
\def\kmsMpc{\ifmmode $\,\kms\,Mpc\mo$\else \,\kms\,Mpc\mo\fi}
\providecommand{\sorthelp}[1]{}

\newcommand\rmn{\textrm}
\newcommand{\Archeops}{{ARCHEOPS}}

\newcommand{\Iras}{{IRAS}}
\newcommand{\Wmap}{\wmap}
\newcommand{\wmap}{{WMAP}}
\newcommand{\hammurabi}{{\tt hammurabi}}
\newcommand{\planck}{\Planck}
\newcommand{\fermi}{\it Fermi\/}
\newcommand{\jf}{Jansson12}
\newcommand{\galprop}{\tt GALPROP}
\newcommand{\HEALPix}{\tt HEALPix}
\newcommand{\commander}{\tt Commander}

\newcommand{\added}[1]{#1}
\newcommand{\addedB}[1]{#1}
\newcommand{\removedB}[1]{} 
\newcommand{\addedC}[1]{{#1}}
\newcommand{\addedD}[1]{{#1}}

\renewcommand{\S}{Sect.}
\renewcommand{\ell}{l}

\newcommand{\cre}{CRL}  
\newcommand{\random}{random} 

\DeclareMathOperator{\sech}{sech}

\begin{document}
%
%
\title{\Planck\ intermediate results. XLII. Large-scale Galactic magnetic fields}
\titlerunning{Large-scale Galactic magnetic fields}
\authorrunning{Planck Collaboration}

%
%
%
\author{\small
Planck Collaboration: R.~Adam\inst{66}
\and
P.~A.~R.~Ade\inst{78}
\and
M.~I.~R.~Alves\inst{86, 8, 51}
\and
M.~Ashdown\inst{60, 4}
\and
J.~Aumont\inst{51}
\and
C.~Baccigalupi\inst{76}
\and
A.~J.~Banday\inst{86, 8}
\and
R.~B.~Barreiro\inst{56}
\and
N.~Bartolo\inst{25, 57}
\and
E.~Battaner\inst{88, 89}
% ...
}
\institute{\small
APC, AstroParticule et Cosmologie, Universit\'{e} Paris Diderot, CNRS/IN2P3, CEA/lrfu, Observatoire de Paris, Sorbonne Paris Cit\'{e}, 10, rue Alice Domon et L\'{e}onie Duquet, 75205 Paris Cedex 13, France\goodbreak
\and
African Institute for Mathematical Sciences, 6-8 Melrose Road, Muizenberg, Cape Town, South Africa\goodbreak
\and
Agenzia Spaziale Italiana Science Data Center, Via del Politecnico snc, 00133, Roma, Italy\goodbreak
\and
Astrophysics Group, Cavendish Laboratory, University of Cambridge, J J Thomson Avenue, Cambridge CB3 0HE, U.K.\goodbreak
\and
Astrophysics \& Cosmology Research Unit, School of Mathematics, Statistics \& Computer Science, University of KwaZulu-Natal, Westville Campus, Private Bag X54001, Durban 4000, South Africa\goodbreak
\and
CGEE, SCS Qd 9, Lote C, Torre C, 4$^{\circ}$ andar, Ed. Parque Cidade Corporate, CEP 70308-200, Bras\'{i}lia, DF, Brazil\goodbreak
\and
CITA, University of Toronto, 60 St. George St., Toronto, ON M5S 3H8, Canada\goodbreak
\and
CNRS, IRAP, 9 Av. colonel Roche, BP 44346, F-31028 Toulouse cedex 4, France\goodbreak
\and
California Institute of Technology, Pasadena, California, U.S.A.\goodbreak
\and
Centro de Estudios de F\'{i}sica del Cosmos de Arag\'{o}n (CEFCA), Plaza San Juan, 1, planta 2, E-44001, Teruel, Spain\goodbreak
\and
Computational Cosmology Center, Lawrence Berkeley National Laboratory, Berkeley, California, U.S.A.\goodbreak
\and
DSM/Irfu/SPP, CEA-Saclay, F-91191 Gif-sur-Yvette Cedex, France\goodbreak
\and
DTU Space, National Space Institute, Technical University of Denmark, Elektrovej 327, DK-2800 Kgs. Lyngby, Denmark\goodbreak
\and
D\'{e}partement de Physique Th\'{e}orique, Universit\'{e} de Gen\`{e}ve, 24, Quai E. Ansermet,1211 Gen\`{e}ve 4, Switzerland\goodbreak
\and
Departamento de Astrof\'{i}sica, Universidad de La Laguna (ULL), E-38206 La Laguna, Tenerife, Spain\goodbreak
\and
Departamento de F\'{\i}sica, Universidad de Oviedo, Avda. Calvo Sotelo s/n, Oviedo, Spain\goodbreak
\and
Department of Astrophysics/IMAPP, Radboud University Nijmegen, P.O. Box 9010, 6500 GL Nijmegen, The Netherlands\goodbreak
\and
Department of Physics \& Astronomy, University of British Columbia, 6224 Agricultural Road, Vancouver, British Columbia, Canada\goodbreak
\and
Department of Physics and Astronomy, Dana and David Dornsife College of Letter, Arts and Sciences, University of Southern California, Los Angeles, CA 90089, U.S.A.\goodbreak
\and
Department of Physics and Astronomy, University College London, London WC1E 6BT, U.K.\goodbreak
\and
Department of Physics, Gustaf H\"{a}llstr\"{o}min katu 2a, University of Helsinki, Helsinki, Finland\goodbreak
\and
Department of Physics, Princeton University, Princeton, New Jersey, U.S.A.\goodbreak
\and
Department of Physics, University of California, Santa Barbara, California, U.S.A.\goodbreak
\and
Department of Physics, University of Illinois at Urbana-Champaign, 1110 West Green Street, Urbana, Illinois, U.S.A.\goodbreak
\and
Dipartimento di Fisica e Astronomia G. Galilei, Universit\`{a} degli Studi di Padova, via Marzolo 8, 35131 Padova, Italy\goodbreak
\and
Dipartimento di Fisica e Scienze della Terra, Universit\`{a} di Ferrara, Via Saragat 1, 44122 Ferrara, Italy\goodbreak
\and
Dipartimento di Fisica, Universit\`{a} La Sapienza, P. le A. Moro 2, Roma, Italy\goodbreak
\and
Dipartimento di Fisica, Universit\`{a} degli Studi di Milano, Via Celoria, 16, Milano, Italy\goodbreak
\and
Dipartimento di Fisica, Universit\`{a} degli Studi di Trieste, via A. Valerio 2, Trieste, Italy\goodbreak
\and
Dipartimento di Matematica, Universit\`{a} di Roma Tor Vergata, Via della Ricerca Scientifica, 1, Roma, Italy\goodbreak
\and
Discovery Center, Niels Bohr Institute, Blegdamsvej 17, Copenhagen, Denmark\goodbreak
\and
Discovery Center, Niels Bohr Institute, Copenhagen University, Blegdamsvej 17, Copenhagen, Denmark\goodbreak
\and
European Space Agency, ESAC, Planck Science Office, Camino bajo del Castillo, s/n, Urbanizaci\'{o}n Villafranca del Castillo, Villanueva de la Ca\~{n}ada, Madrid, Spain\goodbreak
\and
European Space Agency, ESTEC, Keplerlaan 1, 2201 AZ Noordwijk, The Netherlands\goodbreak
\and
Gran Sasso Science Institute, INFN, viale F. Crispi 7, 67100 L'Aquila, Italy\goodbreak
\and
HGSFP and University of Heidelberg, Theoretical Physics Department, Philosophenweg 16, 69120, Heidelberg, Germany\goodbreak
\and
Helsinki Institute of Physics, Gustaf H\"{a}llstr\"{o}min katu 2, University of Helsinki, Helsinki, Finland\goodbreak
\and
INAF - Osservatorio Astronomico di Padova, Vicolo dell'Osservatorio 5, Padova, Italy\goodbreak
\and
INAF - Osservatorio Astronomico di Roma, via di Frascati 33, Monte Porzio Catone, Italy\goodbreak
\and
INAF - Osservatorio Astronomico di Trieste, Via G.B. Tiepolo 11, Trieste, Italy\goodbreak
\and
INAF/IASF Bologna, Via Gobetti 101, Bologna, Italy\goodbreak
\and
INAF/IASF Milano, Via E. Bassini 15, Milano, Italy\goodbreak
\and
INFN, Sezione di Bologna, viale Berti Pichat 6/2, 40127 Bologna, Italy\goodbreak
\and
INFN, Sezione di Ferrara, Via Saragat 1, 44122 Ferrara, Italy\goodbreak
\and
INFN, Sezione di Roma 1, Universit\`{a} di Roma Sapienza, Piazzale Aldo Moro 2, 00185, Roma, Italy\goodbreak
\and
INFN, Sezione di Roma 2, Universit\`{a} di Roma Tor Vergata, Via della Ricerca Scientifica, 1, Roma, Italy\goodbreak
\and
INFN/National Institute for Nuclear Physics, Via Valerio 2, I-34127 Trieste, Italy\goodbreak
\and
IPAG: Institut de Plan\'{e}tologie et d'Astrophysique de Grenoble, Universit\'{e} Grenoble Alpes, IPAG, F-38000 Grenoble, France, CNRS, IPAG, F-38000 Grenoble, France\goodbreak
\and
Imperial College London, Astrophysics group, Blackett Laboratory, Prince Consort Road, London, SW7 2AZ, U.K.\goodbreak
\and
Infrared Processing and Analysis Center, California Institute of Technology, Pasadena, CA 91125, U.S.A.\goodbreak
\and
Institut d'Astrophysique Spatiale, CNRS, Univ. Paris-Sud, Universit\'{e} Paris-Saclay, B\^{a}t. 121, 91405 Orsay cedex, France\goodbreak
\and
Institut d'Astrophysique de Paris, CNRS (UMR7095), 98 bis Boulevard Arago, F-75014, Paris, France\goodbreak
\and
Institute of Astronomy, University of Cambridge, Madingley Road, Cambridge CB3 0HA, U.K.\goodbreak
\and
Institute of Theoretical Astrophysics, University of Oslo, Blindern, Oslo, Norway\goodbreak
\and
Instituto de Astrof\'{\i}sica de Canarias, C/V\'{\i}a L\'{a}ctea s/n, La Laguna, Tenerife, Spain\goodbreak
\and
Instituto de F\'{\i}sica de Cantabria (CSIC-Universidad de Cantabria), Avda. de los Castros s/n, Santander, Spain\goodbreak
\and
Istituto Nazionale di Fisica Nucleare, Sezione di Padova, via Marzolo 8, I-35131 Padova, Italy\goodbreak
\and
Jet Propulsion Laboratory, California Institute of Technology, 4800 Oak Grove Drive, Pasadena, California, U.S.A.\goodbreak
\and
Jodrell Bank Centre for Astrophysics, Alan Turing Building, School of Physics and Astronomy, The University of Manchester, Oxford Road, Manchester, M13 9PL, U.K.\goodbreak
\and
Kavli Institute for Cosmology Cambridge, Madingley Road, Cambridge, CB3 0HA, U.K.\goodbreak
\and
Kazan Federal University, 18 Kremlyovskaya St., Kazan, 420008, Russia\goodbreak
\and
LAL, Universit\'{e} Paris-Sud, CNRS/IN2P3, Orsay, France\goodbreak
\and
LERMA, CNRS, Observatoire de Paris, 61 Avenue de l'Observatoire, Paris, France\goodbreak
\and
Laboratoire AIM, IRFU/Service d'Astrophysique - CEA/DSM - CNRS - Universit\'{e} Paris Diderot, B\^{a}t. 709, CEA-Saclay, F-91191 Gif-sur-Yvette Cedex, France\goodbreak
\and
Laboratoire Traitement et Communication de l'Information, CNRS (UMR 5141) and T\'{e}l\'{e}com ParisTech, 46 rue Barrault F-75634 Paris Cedex 13, France\goodbreak
\and
Laboratoire de Physique Subatomique et Cosmologie, Universit\'{e} Grenoble-Alpes, CNRS/IN2P3, 53, rue des Martyrs, 38026 Grenoble Cedex, France\goodbreak
\and
Laboratoire de Physique Th\'{e}orique, Universit\'{e} Paris-Sud 11 \& CNRS, B\^{a}timent 210, 91405 Orsay, France\goodbreak
\and
Lawrence Berkeley National Laboratory, Berkeley, California, U.S.A.\goodbreak
\and
Max-Planck-Institut f\"{u}r Astrophysik, Karl-Schwarzschild-Str. 1, 85741 Garching, Germany\goodbreak
\and
Max-Planck-Institut f\"{u}r Extraterrestrische Physik, Giessenbachstra{\ss}e, 85748 Garching, Germany\goodbreak
\and
National University of Ireland, Department of Experimental Physics, Maynooth, Co. Kildare, Ireland\goodbreak
\and
Nicolaus Copernicus Astronomical Center, Bartycka 18, 00-716 Warsaw, Poland\goodbreak
\and
Niels Bohr Institute, Blegdamsvej 17, Copenhagen, Denmark\goodbreak
\and
Niels Bohr Institute, Copenhagen University, Blegdamsvej 17, Copenhagen, Denmark\goodbreak
\and
Nordita (Nordic Institute for Theoretical Physics), Roslagstullsbacken 23, SE-106 91 Stockholm, Sweden\goodbreak
\and
SISSA, Astrophysics Sector, via Bonomea 265, 34136, Trieste, Italy\goodbreak
\and
SMARTEST Research Centre, Universit\`{a} degli Studi e-Campus, Via Isimbardi 10, Novedrate (CO), 22060, Italy\goodbreak
\and
School of Physics and Astronomy, Cardiff University, Queens Buildings, The Parade, Cardiff, CF24 3AA, U.K.\goodbreak
\and
Sorbonne Universit\'{e}-UPMC, UMR7095, Institut d'Astrophysique de Paris, 98 bis Boulevard Arago, F-75014, Paris, France\goodbreak
\and
Space Research Institute (IKI), Russian Academy of Sciences, Profsoyuznaya Str, 84/32, Moscow, 117997, Russia\goodbreak
\and
Space Sciences Laboratory, University of California, Berkeley, California, U.S.A.\goodbreak
\and
Special Astrophysical Observatory, Russian Academy of Sciences, Nizhnij Arkhyz, Zelenchukskiy region, Karachai-Cherkessian Republic, 369167, Russia\goodbreak
\and
Sub-Department of Astrophysics, University of Oxford, Keble Road, Oxford OX1 3RH, U.K.\goodbreak
\and
The Oskar Klein Centre for Cosmoparticle Physics, Department of Physics,Stockholm University, AlbaNova, SE-106 91 Stockholm, Sweden\goodbreak
\and
UPMC Univ Paris 06, UMR7095, 98 bis Boulevard Arago, F-75014, Paris, France\goodbreak
\and
Universit\'{e} de Toulouse, UPS-OMP, IRAP, F-31028 Toulouse cedex 4, France\goodbreak
\and
University Observatory, Ludwig Maximilian University of Munich, Scheinerstrasse 1, 81679 Munich, Germany\goodbreak
\and
University of Granada, Departamento de F\'{\i}sica Te\'{o}rica y del Cosmos, Facultad de Ciencias, Granada, Spain\goodbreak
\and
University of Granada, Instituto Carlos I de F\'{\i}sica Te\'{o}rica y Computacional, Granada, Spain\goodbreak
% ...
}


\abstract{

  Recent models for the large-scale Galactic magnetic fields in the
  literature have been largely constrained by synchrotron emission and
  Faraday rotation measures.  We use three different but
  representative models to compare their predicted polarized
  synchrotron and dust emission with that measured by the {\Planck}
  satellite.  We first update these models to match the {\planck}
  synchrotron products using a common model for the cosmic-ray
  leptons.  We discuss the impact on this analysis of the ongoing
  problems of component separation in the {\planck} microwave bands and of the
  uncertain cosmic-ray spectrum.  In particular, the \added{inferred} degree of
  ordering in the \added{magnetic} fields is sensitive to these systematic
  uncertainties, \addedC{and we further show the importance of
    considering the expected variations in the observables in addition
    to their mean morphology}.  We then compare the resulting \added{simulated
    emission} to the \added{observed} dust \addedC{polarization} and
  find that the dust predictions do not match the morphology in the
  {\planck} data \addedC{but underpredict the dust polarization away
    from the plane.}
  \addedC{We modify one of the models to roughly match both
    observables at high latitudes by increasing the field ordering in the thin
    disc near the observer.  Though this specific analysis is dependent on the component
    separation issues,} we present the improved model as a proof of concept \added{for how these studies can be
    advanced in
  future using complementary information from ongoing and planned
  observational projects.}

}


\maketitle

\section{Introduction}

The Galactic magnetic field is an important but ill-constrained
component of the interstellar medium (ISM) that plays a role in a
variety of astrophysical processes, such as molecular cloud collapse,
star formation, and cosmic-ray propagation.  Our knowledge of the
structure of the magnetic fields in our own Milky Way Galaxy is
limited by the difficulty interpreting indirect observational data and
by our position within the disc of the Galaxy.  We know that there are
both coherent and {\random} components of the magnetic fields and that
in external galaxies they tend to have a spiral structure similar to
that of the gas and stellar population (see \citealt{beck:2015vz} for a
review).  We do not, however, have an accurate view of the morphology
of these field components within either the disc or the halo
of our own Galaxy.  For a review, see \citet{haverkorn:2014}.  

There are many modelling analyses in the literature for the large-scale Galactic
magnetic fields, including work such as \cite{stanev:1997}, 
\citet{prouza:2003}, \citet{han:2006}, \citet{page:2007}, \citet[hereafter
Sun10]{Sun:2010}, \citet{ruizgranados:2010}, \citet[hereafter Fauvet12]{fauvet:2012}, \citet[hereafter
{\jf}]{jansson:2012c}, \citet[hereafter Jaffe13]{jaffe:2013}, and
\citet[hereafter Orlando13]{orlando:2013}. These studies have constrained properties of
the large-scale Galactic magnetic fields using complementary
observables that probe the magnetic fields in different ways.  Most of
the constraints so far have come from synchrotron emission, both total
and polarized, and Faraday rotation measures (RMs).  Thermal
dust emission is a useful complement for its different 
\added{dependence on the field strength and its different source
  particle distribution}. \citet{fauvet:2011} performed the first such
joint analysis making use of existing thermal dust polarization data
from the {\Archeops} balloon experiment.  Jaffe13 continued with an analysis
using Wilkinson Microwave Anisotropy Probe ({\wmap}) dust polarization
instead.  \addedD{Both of these data sets, however,} suffer
from low signal to noise or limited sky coverage.

The {\Planck}\footnote{\Planck\ (\url{http://www.esa.int/Planck}) is a project of the European Space Agency (ESA) with instruments provided by two scientific consortia funded by ESA member states and led by Principal Investigators from France and Italy, telescope reflectors provided through a collaboration between ESA and a scientific consortium led and funded by Denmark, and additional contributions from NASA (USA). } data provide a
new opportunity to constrain the magnetic fields using the most
sensitive full-sky maps to date of both total and polarized dust
emission in the sub-mm bands as well as an alternative synchrotron probe in
the low frequency bands.  Our aim is therefore to add the information
from the {\planck} full-sky polarized dust emission maps to our
magnetic field modelling and to use their complementary geometry to
better constrain the properties of the magnetic fields.  

The preliminary work by Jaffe13 in the Galactic plane suggests that
constructing a single global model of the Galactic magnetic fields
that reproduces both the polarized synchrotron and dust emission over
the full sky will be difficult.  In this work, we make a first attempt
by taking several models in the literature that have been constrained
largely by the synchrotron emission and RMs and comparing the corresponding
dust prediction to the {\planck} data.  Such simple comparisons of the
morphology of the resulting polarization sky maps will give insight
into how the models can be improved.

Using the comparison of the data to the model predictions for both synchrotron and
dust emission, we will perform simple updates to the models where the
morphologies do not match and where we can study the physical parameters such as the scale
heights and scale radii of the different ISM components.  Although
constructing new analytic forms is beyond the scope of this work, our
analysis will point the way to how we can improve the large-scale
field modelling and progress towards a global model that can reproduce
all observables.

We will also discuss the difficulties with these analyses,
particularly the problem of component separation and the uncertainty
in the synchrotron spectral variations over the sky.  The models we
present here are based on {\planck} component separation products, and
we discuss the limitations of these products and therefore of 
the resulting models.   We will also discuss information from other observables
and how the situation will improve in the future based on
ongoing and next-generation surveys.  

In \S\,\ref{sec:methods} we review the data and methods used, referring
to appendices for discussion of the {\planck} polarization systematics
and component separation issues.  In
\S\,\ref{sec:synch_modeling}, we describe the synchrotron modelling
that, along with RM studies, has led to the development
of the magnetic field models we use from the literature.  We discuss
how they were constructed, on what cosmic-ray lepton ({\cre}) model
they depend, how they compare to each other, and how they need to be
updated.  In \S\,\ref{sec:dust}, we present the comparison of the
updated models with the {\planck} data for dust polarization and
discuss the implications.  Lastly, in \S\,\ref{sec:discussion}, we
discuss how we expect this work to be improved in the future.

\section{Data and methods}
\label{sec:methods}

\subsection{Observations}
\label{sec:data}

\subsubsection{ {\planck} data}
\label{sec:planck_data}

The results presented in this paper are based on the 2015 data release\footnote{\url{http://pla.esac.esa.int/}}
described in \citet{planck2014-a01}, \citet{planck2014-a03}, and \citet{planck2014-a09} including both intensity and
polarization results.   Because of the presence of numerous
astrophysical components at each frequency, we use the {\commander}
component separation estimates of the synchrotron and dust total
intensities described in \citet{planck2014-a12}.  In the case of
dust, the purpose is to remove confusion from \addedB{an intensity} offset or cosmic
infrared background (CIB).  For synchrotron, however, this is more
complicated, as the low-frequency total intensity includes several
different components in addition to synchrotron emission.  The
importance of this choice is further discussed in
\S\,\ref{sec:data_comps}.  The products are given
as maps in the {\HEALPix}\footnote{\url{http://healpix.sourceforge.net}} \citep{healpix} pixelization scheme in units of
$\rmn{mK}_\rmn{RJ}$ (i.e., brightness temperature), and we
downgrade\footnote{We simply average the high-resolution pixels in each
  lower-resolution pixel.  This is done for each Stokes parameter,
  which does not take into account the rotation of the polarization
  reference frame (see, e.g., \citealt{planck2014-XIX}).  
  This effect is only significant at the highest
  latitudes, however, and has no impact on our results.  }  
  these to a low resolution of
$N_\rmn{side}=16$ for comparison with the models.  

These products consist of spatial information at a reference frequency
and a prescription for the spectral model, which we combine to generate
the correct prediction for synchrotron emission at 30\,GHz and for
dust emission at 353\,GHz as described in Table\,4 of \cite{planck2014-a12}.

We also compare to the full-mission maps of the Low Frequency
Instrument (LFI, \citealt{planck2014-a07}) at a frequency of
30\,{\GHz} and the 
High Frequency Instrument (HFI, \citealt{planck2014-a09}) at 353\,{\GHz}.  Those maps are
given in $\rmn{K}_\rmn{CMB}$ units\footnote{Temperature units referring to
  the cosmic microwave background (CMB) blackbody spectrum.} and
require a \added{unit conversion in addition to colour} \addedB{and leakage corrections} \added{based on the
  instrument bandpasses, as described in \citet{planck2014-a03} and \citet{planck2014-a08}. }   We also make use of the other HFI
polarization channels as well as several different methods to correct
for systematics, as described in Appendix\,\ref{sec:appendix_systematics}.

We note that \Planck\ products and the results in this paper are
expressed in Stokes $I$, $Q$, and $U$ using the same convention
followed by HEALPix for the polarization angle (or equivalently, the sign of $U$) rather
than the IAU convention.

\subsubsection{Ancillary data}
\label{sec:ancillary}

We compare the {\Planck} synchrotron solution to those from the
{\wmap} analysis by \citet{gold:2011}.  They used two component
separation methods, with several versions each, and we will compare
with their basic Markov chain Monte Carlo (MCMC) solution.

We also compare to the 408\,MHz map of \citet{haslam:1982} reprocessed
as described by \citet{remazeilles:2015}.  For
comparison with previous work such as Jaffe13, we subtract
the offset determined by \cite{lawson:1987} to account for the
extragalactic components.  There is an uncertainty in the calibration
zero-level of about $3$\,K for this survey, which will be further
discussed in Appendix\,\ref{sec:appendix_compsep}.  We also
subtract the {\planck} {\commander} free-free estimate from the 408\,MHz map, which
is still significant along the Galactic plane.  \added{The
result is then almost identical to the {\commander} synchrotron solution
except for a $1\,\sigma$ shift in the zero level.  }

These ancillary data are available on the LAMBDA\footnote{\url{http://lambda.gsfc.nasa.gov/
}} website.  \added{For comparison to the {\planck} component separation
products, see Appendix\,\ref{sec:appendix_compsep}.}

\subsubsection{Data caveats}
\label{sec:data_comps}

Ideally, studies of the Galactic magnetic fields using synchrotron
emission would compare the total and polarized emission at the same
frequency in order to measure the degree of ordering in the fields.  
In order to probe the full structure of the Galactic disc, \addedD{however,} we need to
study the emission in the mid-plane where there are two complications.
At radio frequencies (below roughly 3\,GHz), the synchrotron emission
is depolarized by Faraday effects.  These impose a so-called
polarization horizon\footnote{The distance is dependent on the
  frequency and the telescope beam but typically of order a few kpc
  for radio surveys; see, e.g., \citet{uyaniker:2003}.} beyond which
all diffuse polarization information is effectively lost due to the
Faraday screen of the magnetized and turbulent ISM.  In the
microwave bands \addedC{(tens of GHz)}, 
where Faraday effects are negligible, the total intensity is
dominated along the Galactic plane by free-free and anomalous
microwave emission (AME);  \addedB{see, e.g., \citet{planck2014-XXIII}}.  These components have steep spectra in the
microwave bands, which makes them difficult to separate from the
synchrotron emission.  

Therefore, there are two options for this sort
of study:   
\begin{itemize}
\item use the radio frequency for total intensity and microwave
  frequency for polarization, which subjects the analysis to the
  uncertainty of assuming a spectral  behaviour over a large frequency
  range that magnifies even a small uncertainty in the spectrum into a
  large uncertainty in the amplitude and morphology (e.g., Sun10,
  Orlando13, and Jaffe13);
\item or use the microwave frequencies for both, which subjects the
  analysis to the significant uncertainty of the component separation
  in the Galactic plane (e.g., {\jf}). 
\end{itemize}
These issues are also discussed in \citet{planck2014-a31}.

We choose to use the {\planck} {\commander} component separation results,
and though this sounds like the second option, it is effectively the
first.  The {\commander} analysis fits a model for the synchrotron total
intensity based on the 408\,MHz map as an emission template and
assumes a constant synchrotron spectrum across the sky.  That spectrum
(see \S\,\ref{sec:cres}) is in turn the result of a model for the
large-scale Galactic magnetic field as well as the {\cre}
distribution, and we use the same model for the latter to be as
consistent as possible while studying the former.  It must be noted, \addedD{however,} that
there is an inconsistency in the analysis.  Ideally, the component
separation should be a part of the astrophysical modelling, but this is
not feasible.  An iterative approach would be the next best option,
and our analysis here can be considered the first iteration.

It is important to recognize that the various models for the large-scale
Galactic magnetic fields in the literature have been developed based
on different approaches to these issues. In order to compare these
models, the different choices made must be considered.  We discuss
this further and compare the data sets explicitly in
Appendix\,\ref{sec:appendix_compsep}.  

\added{It is also unclear what effect small but nearby (and therefore
  large angular-scale) structures have on such analyses.  Clearly,
  models of the large-scale fields will not reproduce individual
  features such as supernova remnants, but these features may bias our
  model fitting.  We discuss some of these features in
  \S\,\ref{sec:loops}, but we cannot reliably quantify how
  large an effect they may be having without a better understanding of
  what these features are.  Only when looking through the full
  Galactic disc in the midplane, where 
  \addedC{such features are small compared to the integrated emission}, can we be sure that the resulting models are largely
  unaffected.  We also exclude known regions of
  localized emission or average over large areas of the sky in order to minimize their impact.}

\subsection {The {\hammurabi} code}
\label{sec:hammurabi}

The
{\hammurabi}\footnote{\url{http://sourceforge.net/projects/hammurabicode/}}
\citep{waelkens:2009} code simulates synchrotron
and dust emission in full Stokes parameters as well as associated
observables such as Faraday RM, emission measure, and dispersion measure.  It includes analytic forms for the
components of the magnetized ISM (magnetic fields, thermal electrons,
{\cre}s, etc.) or can be given an external file that specifies those
components over a spatial grid.

The Sun10 and Fauvet12 magnetic field models are implemented in the publicly
available version of {\hammurabi}, while the Jaffe13 and {\jf}
models \addedC{will be included in the next release.}  

We model the {\random} field component using a Gaussian random field
(GRF) simulation characterized by a power-law power spectrum and an
outer scale of turbulence.  In order to compute this component with
the highest possible resolution, we split the integration into two
steps: firstly from the observer out to a \addedB{heliocentric} distance of
$R<2$\,kpc, and then for $R>2$\,kpc.  For the latter, we simulate the
full Galaxy in a 40\,kpc by 40\,kpc by 8\,kpc grid of 1024 by 1024 by 256 bins,
i.e., with a resolution of roughly $40$\,pc.  For the $R<2$\,kpc case, we
compute the GRF in a cube 4\,kpc and 1024 bins on a side, giving a
resolution of 4\,pc.  We have in both cases run tests with a
resolution a factor of two higher in each dimension (requiring several
tens of GB of memory) and found the result to be qualitatively
unaffected by the resolution.  \added{The high-resolution, local part of the
simulation has a Kolmogorov-like power spectrum, 
$P(k)\propto k^{-5/3}$, 
and in both
cases we use an outer scale of turbulence of $100$\,pc (see
\citealt{haverkorn:2013} and reference therein).  While the nearby
simulation samples different scales, the resolution of the full-Galaxy
simulation is too low to be more than effectively single-scale.  The
ensemble average emission maps are not sensitive to these parameters
of the turbulence (though the predicted uncertainty can be, as
discussed in \S\,\ref{sec:galactic_variance}).  In both regimes, the GRF
is normalized to have the same total rms variation (configurable as shown in 
Appendix\,\ref{sec:appendix_params}).  This GRF is then
rescaled as a function of position in the Galaxy depending on the model
(e.g., with an exponential profile in Galacto-centric $r$ or $z$).}

The {\HEALPix}-based integration grid is done at an observed resolution of
$N_\rmn{side}=64$, i.e., roughly $1\degr$ pixels.  As described by 
\citet{waelkens:2009}, the integration grid is refined 
successively \added{along the line of sight (LOS) to maintain a roughly constant integration
bin size.  We set the integration resolution parameters 
to match the resolution of the Cartesian grid for the GRF.}\footnote{\added{
  {\hammurabi} uses a configurable number of shells defined by
  $N_\rmn{shells}$.  For each shell, its {\HEALPix} $N_\rmn{side}$ defines a constant angular
  width for each bin, while an
  independent variable controls the bin length along the LOS in $\Delta
  R$.  The length is constant along the entire LOS, but the width then
  varies within each shell (see Fig.\,A.1 of
  \citealt{waelkens:2009}.)  For the $R<2$\,kpc integration, we use
  $\Delta R=2$\,pc and $N_\rmn{shells}=4$, so that at the last
  shell the $N_\rmn{side}=512$ pixels range from $2$ to $4$\,pc wide
  from the front to the back of the shell.
  For the $R>2$\,kpc integration, we use $\Delta R=32$\,pc and
  $N_\rmn{shells}=5$, so that at the last shell the
  $N_\rmn{side}=1024$ pixels range from $16$ to $32$\,pc wide.}}

In the case of synchrotron emission, we have explicitly compared the
results of a set of GRF simulations with the results from the analytic
method used in {\jf} and verified that the ensemble average is the
same.  For the dust, we have no analytic expression for the expected
emission, so we use the numerical method of GRF realizations for all
of the main results of this paper.  We compute 10 independent realizations of each
model and compare the mean in each pixel to the data in that pixel.
We use the variation among the realizations in each pixel as the
uncertainty due to the galactic variance, as discussed in
\S\,\ref{sec:galactic_variance}.

\subsection{Parameter exploration}

Though ideally we would perform a complete search over the full
parameter space to determine the best values of all parameters, this is 
computationally not feasible.  Such searches have been performed in the past by, e.g.,
\citet{jaffe:2010} and \citet{jansson:2012b}.  In the first case, the
number of parameters was limited and the analysis restricted to the
plane.  The full 3D optimization is far more difficult, even excluding
the dust emission.  In the second case, the fit was performed by using
an analytic expression for the synchrotron emission from the
{\random} field components, which allows for a very fast computation but
does not correctly take into account the variations produced by the
modelled {\random} fields.  Furthermore, this analytic approach is not
possible for dust emission, as there is no correspondingly simple
closed-form expression for the ensemble average.  (The
synchrotron case requires assuming the dependence of the emissivity on
$B^2$, or equivalently that the {\cre} spectrum is a power law,
$N(E)\propto E^{-p}$, with index $p=3$.)  

Therefore, the updated models discussed in this paper are only
approximations arrived at by visual comparison, focusing on the
longitude profiles along the plane and the latitude profile in the
inner two quadrants, where the data represent the integration through
most of the Galaxy.  We accept that in the outer Galaxy away from the
plane, the models may not match observations well, but this region represents much
less of the Galaxy.  We vary key parameters \added{such as the degree of
ordering in the fields, the relative strengths of disc and halo
components, the scale heights and scale radii of these components, and
individual arm amplitudes that affect the emission on large scales.
The changes are} motivated by the data, but this is 
subjective rather than quantitative.  A complete parameter
optimization remains a significant computation challenge for future
work.

\section{Synchrotron modelling}
\label{sec:synch_modeling}

From an observational point of view, the magnetic field can be
considered as having three components that contribute to observables
differently depending on the sensitivity to orientation and/or
direction.  \addedB{(See fig.\,1 in \citealt{jaffe:2010}.)}  The coherent component (e.g., an axisymmetric spiral)
contributes to all observables, since by definition, it always adds
coherently.  The isotropic {\random} component contributes only to the
average total intensity, which co-adds without dependence on orientation;  to polarization
and RM, it \added{does not contribute to the ensemble average but only
  to the ensemble variance}.  A third component, \added{which we call an ``ordered'' random component following
\citet{jaffe:2010} but which was called ``striated'' by
\cite{jansson:2012b}}, contributes to polarization, which is sensitive
to orientation, but not to RM, which is additionally sensitive to direction.
This third component represents the anisotropy in the {\random} fields
thought to arise due to differential rotation and/or compression of
the turbulence in the spiral arms of the Galaxy.  \addedC{(See, e.g.,
\citealt{Brown:2001jt} or \citealt{beck:2015vz}.)} These components can
only be separated unambiguously using a combination of complementary
observables.\footnote{The literature often refers to ``random'' and
  ``regular'' fields, which means that the third component is
  ``random'' in the case of RM observations but is ``regular'' in the
  case of polarized emission.  We prefer to avoid the ambiguous use of
the word ``regular''.}

The large-scale magnetic field models in the literature are most
commonly constrained for the coherent field component.  The {\random} field component is not treated
specifically in some of these models, nor is the anisotropy in this
component often considered, though it has been shown by, e.g.,
\citet{jaffe:2010}, \citet{jansson:2012b}, and Orlando13 to be
comparable in strength.  Depending on the scientific aims, the
{\random} components may either be treated as noise or must be modelled
for unambiguous interpretation of the field strengths.  
 
In addition to the model for the magnetic field, we also require
a model for the {\cre}s that produce the synchrotron emission.  The following sub-sections describe these
elements of the modelling and how changes in each affect the results.


\subsection{Cosmic-ray leptons}  
\label{sec:cres}

For the synchrotron computation, we require both the spatial and
spectral distribution of {\cre}s\footnote{These are mostly electrons but include a
  non-negligible contribution 
from positrons.  We therefore refer to leptons rather than, as is
common, simply electrons.}.
The topic of cosmic ray (CR) acceleration and propagation is a complicated one
(see, e.g., \citealt{grenier:2015}), and
there are degeneracies in the space of CR injection and propagation
parameters, which can approximately reproduce the synchrotron or
$\gamma$-ray data that are the primary probes of the particle
distributions.  It is not the purpose of this work to constrain the
particle distributions, so we choose a representative model for the
{\cre}s and discuss how some of our conclusions are subject to
the uncertainties in this input.  For a further discussion of this
topic and in particular the impact on the observed synchrotron
emission, see Orlando13. 

For this work, we use \added{ a model of the {\cre} distribution as published
in Orlando13 and generated using the
{\galprop}}\footnote{\url{http://galprop.stanford.edu}}$^,$\footnote{\url{http://sourceforge.net/projects/galprop}} \added{CR
propagation code.  This takes as input the spatial and spectral
distributions of the injected primary particles 
and the magnetic field.  It then models the
propagation of CRs accounting for energy losses, reacceleration processes,
and generation of  secondary particles, including positrons.  
In addition to the primary electrons, our {\galprop} model also
includes protons and helium in the propagation in order to properly
account for the production of secondary leptons.}\footnote{\added{There is
now firm observational evidence for the existence of primary
positrons. For references to the observations and demonstration of
their primary nature see, e.g., \cite{gaggero:2014} or \cite{Boudaud:2015ep}. The lepton spectrum
used in the present work reproduces the total electron plus positron
measurements. However, positrons do not become significant compared to
the electrons until energies well above 20\,GeV, which corresponds to
synchrotron frequencies much higher than those we consider here, so
the question is not directly relevant to this work.}}

\subsubsection{{\cre} spatial distributions}

\begin{figure*}
\begin{center}
\includegraphics[]{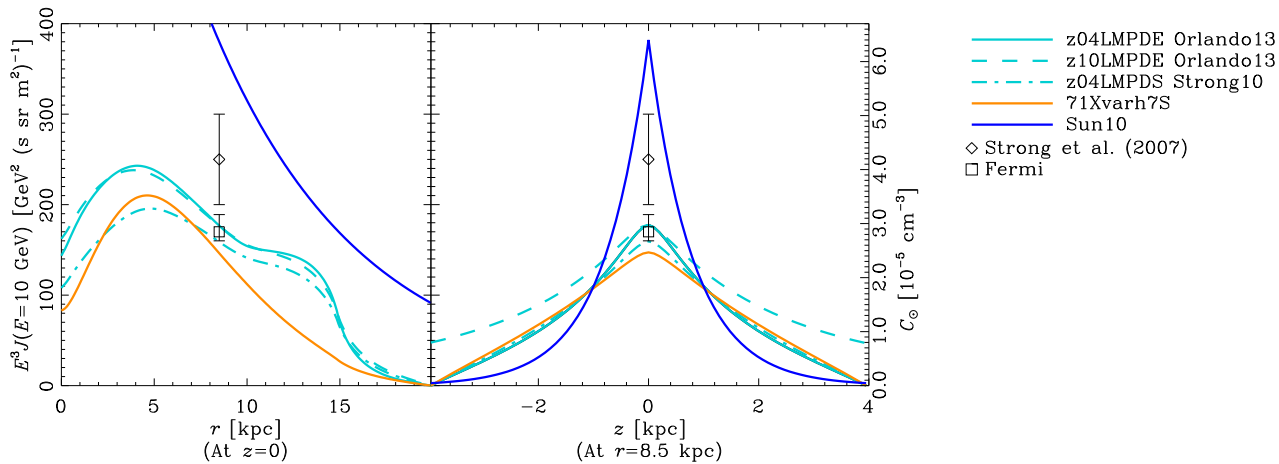} 
\end{center}
\caption{Comparison of {\cre} distributions.  The profile \added{of
    the {\cre} density at a reference energy of 10\,GeV is shown on the 
    left as a function of Galacto-centric radius for $z=0$ and on
  the right as a function of height 
  at the Solar radius ($8.5$\,kpc).}  The Sun10 curve does not include the local enhancement
  described in \S\,\ref{sec:sun10_model}.  The conversion between the
  units on the vertical scales at left and right are explained in
  \citet{jaffe:2010}.  The {\it diamond} and {\it square} symbols show
  the directly-measured {\cre} fluxes from \citet{strong:2007} and from
  \cite{fermi:2010}, respectively.  (The former point and its error
  bar are estimated by eye from their Fig.\,4.)  The
  ``z04LMPDE\,Orlando13'' and ``z10LMPDE\,Orlando13'' models are the more recent
  versions from Orlando13 for different {\cre} scale heights
  of 4 and 10\,kpc, respectively, while the ``z04LMPDS Strong10'' is the
  older version from \cite{strong:2010}.  (The z04LMPD model
  extends to $|z|=10$\,kpc, though the plot is cut off at
  $z=\pm4$\,kpc.)
\label{fig:cre_comp}
}
\end{figure*}

The Sun10 analysis used a simple exponential disc distribution.  
{\jf} used (A. Strong, D. Khurana, private communications) the
spatial distribution of {\cre}s from a slightly modified
%
%
version of the ``71Xvarh7S'' {\galprop} model discussed in, e.g., \cite{Abdo:2010}.
The Jaffe13 model was based on the ``z04LMPDS'' {\galprop} 
model of \cite{strong:2010}.  For reference, the spatial distributions of these
different {\cre} models \added{(computed with a common magnetic field model,
the {\jf})} are compared in Fig.\,\ref{fig:cre_comp}.

%
%

\subsubsection{{\cre} energy spectrum}

Previous works such as Sun10 and \citet{jaffe:2010} used power-law
spectra of a fixed index ($N(E)\propto E^{-p}$, where $p=3$).  This
\addedB{is a reasonable approximation above frequencies of a few GeV
  and} was
arguably sufficient for early studies of the field morphologies at the
largest scales, but it is now insufficiently accurate for the
increasing amounts of data available, as demonstrated by
\cite{jaffe:2011} and \citet{strong:2011}.

\added{For updating the magnetic field models to match the {\planck}
  {\commander} synchrotron maps, we use the ``z10LMPD\_SUNfE''
  {\galprop} {\cre} distribution, as derived in Orlando13.  This
  distribution is the latest result of a long-running project
  including \cite{strong:2010}, \cite{strong:2011}, and Orlando13 to
  develop a model for the spatial and spectral distribution of
  {\cre}s.  In particular, Orlando13 used synchrotron observations and
  updated not only the {\cre} scale height but also the turbulent
  and coherent magnetic field parameters.  Various existing magnetic
  field models were investigated with synchrotron observations, in
  both temperature and polarization, in the context of CR source and
  propagation models.  The lepton spectrum was adjusted by \citet{strong:2011} to fit the
  {\fermi} electron and positron direct measurements \citep{fermi:2010}, while the
  spatial distribution was the one found to better reproduce the
  Galactic latitude and longitude profiles of synchrotron emission,
  after fitting the intensities of the random 
  (isotropic and ordered) and coherent magnetic
  field components (based on Sun10 for their best fit) to {\wmap} synchrotron and
  408\,MHz maps.  (We note that this analysis remains subject to
  degeneracies in the parameter space, in particular at the
  $E\lesssim10$\,GeV region of the {\cre} spectrum where the direct measurements of {\cre}s
  are affected by solar modulation.)  Because the original name
  reflects the {\em resulting} {\cre}s using the Sun10 magnetic field model, while we of
  course explore different field models, we will refer to the model for the injected CRs simply as
  ``z10LMPDE''.}

\begin{figure}
\includegraphics[]{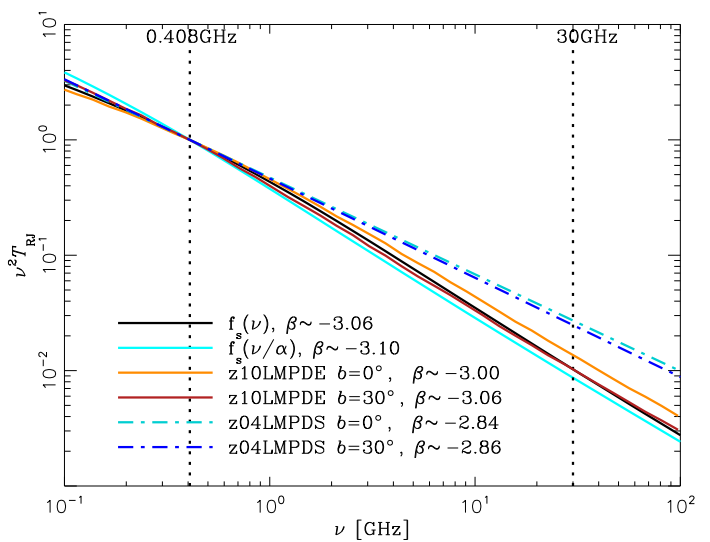}
\caption{Comparison of synchrotron spectra for different
  {\cre} models, all normalized to one at 408\,MHz.  The black solid
  curve shows the original spectral template used in the {\commander}
  analysis, while the cyan solid curve shows the shifted template as
  described in \citet{planck2014-a12}.   In orange is the
  resulting SED on the plane for synchrotron emission computed using
  the z10LMPDE {\cre} model on which the {\commander}
  template  is based.  This curve shows the average curve for the Galactic plane, while the
    brick red is the average for the pixels on a ring at $b=30\degr$.  For
  these spectra, the effective  spectral index $\beta=\log\left(A_{30}/A_{0.408}\right)/\log\left(30/0.408\right)$ is
  computed.  Lastly, the light and dark blue dot-dashed lines
  show the power law with the effective indices for the z04LMPDS
  model averaged at the two latitudes.  
\label{fig:cre_spec_comp}
}
\end{figure}

\added{This is the base model for the synchrotron spectral template used in the {\planck} {\commander} analysis,
which assumed a constant spectrum derived from this
{\cre} model and fitted only a shift in frequency space
\citep{planck2014-a12}.  }
We choose this {\cre} model to be as consistent as possible with the
component separation, but it is not exactly consistent, since the
component separation analysis chose a single synchrotron spectrum to be
representative of the sky away from the Galactic plane (see
Fig.\,\ref{fig:cre_spec_comp}) and allowed it to shift in frequency
space in a full-sky analysis.\footnote{The logarithmic shift
  in frequency space by a factor of $\alpha=0.26$ was not given an associated
uncertainty in \citet{planck2014-a12}.  As noted in that paper, this shift was
highly dependent on other parameters \added{ and was barely detected as
different from unity}.}  Here, we use the full spatial and
spectral distribution produced by {\galprop}, \added{varying the
magnetic field but keeping the same CR injection model.}  This is discussed further in
\S\,\ref{sec:synch_spec}. 

Figure\,\ref{fig:cre_spec_comp} compares several synchrotron spectra.
The spectral template used in the {\commander} analysis is shown with and without the spectral shift.
This is compared to results computed with {\hammurabi} from the
same CR source distribution.  Since this varies on the sky, an
average synchrotron spectrum is computed along the Galactic plane and
also for the pixels at $b=30\degr$ for comparison.  From these, we
compute effective power-law spectral indices, $\beta$, from 408\,MHz
to 30\,GHz.  \added{The resulting {\cre} distribution and
  therefore the synchrotron spectrum depend not only on
the injected CRs but also on the magnetic field model
assumed through synchrotron losses.  The comparison
of our results using the z10LMPDE injection model with the $f(\nu)$ curves is thus approximate.  (We compared
several of the field models with these injection parameters and found
the resulting synchrotron spectra to vary around
$\beta=-3\pm0.05$.  The plotted
curve is based on the {\jf} model.)}

We see that the single spectral template used in the {\commander}
method has a steep spectrum, with a net $\beta=-3.1$.  The spectral
template without the shift \added{is fractionally harder},
$\beta=-3.06$.  \added{The z10LMPDE spectrum (with the same CR
  injection parameters as underlies the $f(\nu)$ template but now including the spatial variations)
  predicts a steeper spectrum at $b=30\degr$ than on the plane.  See
  also Fig.\,14 in \citet{planck2014-a31}.}
%
%
Also shown for comparison is the {\cre} model used in the Jaffe13
model, the older z04LMPDS model, which has an effective index that
is hardest at $\beta\approx-2.84$ (which is due to the harder
intermediate-energy CR injection spectrum;  see Table\,\ref{tab:params_cres}).

When comparing the effective spectral indices, $\beta$, in
Fig.\,\ref{fig:cre_spec_comp}, note that a difference in the effective
spectrum of $\Delta\beta=0.04$ (e.g., from the shift in the {\commander}
template) corresponds to a difference in the synchrotron intensity
extrapolated from 408\,MHz to 30\,GHz of roughly $20\,\%$.  A difference
of $\Delta\beta=0.1$ corresponds to an intensity difference of roughly
$50\,\%$.  These numbers illustrate the uncertainty in the resulting
analysis in the Galactic plane based on the uncertainty in the
{\cre} spectrum in the plane, which is closely related to the issue of
component separation.  

In what follows, we will consider the possible extremes and see what
statements about the magnetic fields are robust despite this uncertainty.

%
%


\subsection{Magnetic field models from the literature}
\label{sec:models}

We choose three models of the large-scale Galactic magnetic field in
the literature to be compared with the {\Planck} data (LFI and HFI):
\added{the Sun10, {\jf}, and Jaffe13 models.  This is not meant to be a
  comprehensive review of the literature.  A variety of models have
  been published, though most tend to be morphologically similar to
  one of these three.  The models used in \citet{page:2007} and
  \citet{fauvet:2012}, for example, are axisymmetric spirals like
  the Sun10 model, only without the reversal (because they do not make
  use of RM information).  The models of \cite{stanev:1997} and
  \citet{prouza:2003} include spiral arms, either axi- or bisymmetric,
  and can be considered special cases similar to the Jaffe13 model.
  The {\jf} model is a more generic parametrization that can reproduce the
  largest-scale features of most of these models.}

We review these models here, but we do not compare the
precise original models with the {\planck} data but rather update them
as described in the next section.\footnote{In testing the original
  models, we found it difficult to reproduce precisely the synchrotron
  intensity normalization according to the respective papers in the
  case of the Sun10 and {\jf} models.  This is likely related to the
  different {\cre} models used and is degenerate with the uncertain
  {\cre} normalization.  It does not affect the results of the current
  work, since we use a more recent {\cre} model and are interested in
  these models for their morphology.}

\subsubsection{Sun10}
\label{sec:sun10_model}

The ``ASS+RING'' model of \cite{sun:2008} and Sun10 is a simple
axisymmetric spiral field that is reversed in a Galacto-centric
ring and in the inner 5\,kpc in order to model the RMs in addition to polarized synchrotron
emission.  The spatial distributions of both {\cre} density and
coherent field strength are modelled with exponential discs.  The
{\cre} spectrum is assumed to be a power law with $p=3$.  The {\cre}
density model also includes a local enhancement near the Sun's
position to increase the high-latitude emission.  
This field model also includes a homogeneous and
isotropic {\random} component.  The model was adjusted by visual
comparison with RM data, 408\,MHz total synchrotron intensity, and
{\wmap} polarized synchrotron intensity.

The assumed {\cre} density normalization in the Sun10 analysis is significantly
higher than usually assumed.  Figure\,\ref{fig:cre_comp} compares the
{\cre} models, where the normalization at the Galacto-centric radius
of the Sun was set to $C_{\sun}=6.4\times 10^{-5}\,\mathrm{cm}^{-3}$ (at
10\,GeV) for the
Sun10 model.  It is unclear to what degree
local values can be considered
typical of the Galactic average, but \added{{\fermi}'s direct
measurements from Earth orbit} near 10\,GeV are roughly $3\times
10^{-5}\,\mathrm{cm}^{-3}$ (\citealt{fermi:2010}).  Furthermore, the
Sun10 model requires an additional enhancement in the form of a
$250\,\%$ relative increase in a sphere of radius $1$\,kpc near the Sun's
position (shifted $560$\,pc towards longitude $45\degr$).  This makes
their assumed {\cre} density even more incompatible with \added{{\fermi}'s direct
measurements in Earth orbit.}

\subsubsection{\jf}

The \cite{jansson:2012b,jansson:2012c} model consists of independently
fitted spiral segments in a thin disc \addedB{(each of which runs from
  the inner molecular ring region to the outer Galaxy)}, a toroidal thick disc (or
``halo''), and an x-shaped poloidal halo component.  This model was
optimized in an MCMC analysis in comparison to the RM data as well as
synchrotron total and polarized emission from {\wmap}.   Their analysis includes an
analytic treatment of the anisotropic turbulent fields.  The average
emission from this component (which they call ``striated'' and we call
``ordered random'') is
computed by scaling up the contribution from the coherent field
component appropriately.  For total intensity, the expected average
emission from the isotropic {\random} component is computed
straightforwardly from the assumption of isotropy.  This field was
developed using a modified version of a {\cre} prediction from
{\galprop} discussed in \S\,\ref{sec:cres}.

One interesting comment made by these authors is to note the
importance of what \citet{jaffe:2010} call the ``galactic variance'' (GV),
i.e., the expected variation of the synchrotron emission due to the
{\random} magnetic field components.  They use the data to estimate
this variation and use that estimate as the uncertainty in their
fitting, but because they use an analytic treatment of these
components, they cannot directly model this variance.  This will be
further discussed in \S\,\ref{sec:galactic_variance}.


We note that this model was fitted excluding the plane region from the
synchrotron polarization analysis.  The disc components of the
coherent  and ordered random  
fields were then determined by the RM data, while the synchrotron data
constrained only the local and halo components.  For the synchrotron total
intensity analysis, the plane was included in order to fit the random
field components in the disc.

\subsubsection{Jaffe13}

The \cite{jaffe:2010,jaffe:2013} model, fitted only in the Galactic
plane, consists of four independent spiral arms and a ring
component and was optimized with an MCMC analysis.  It includes a
numerical treatment of the isotropic and ordered {\random} fields,
which are constrained by the combination of RM with total and
polarized synchrotron emission.  The scale heights had not been constrained before.

This model includes the enhancement of the field strength of all
components (coherent, isotropic {\random}, and ordered {\random}) in
the spiral arms.  The arms lie roughly coincident with those of the
thermal electron density model of \cite{cordes:2002}.  \addedB{(See
  \citealt{jaffe:2010}, Figs.\,2 and 4 for how these features appear
  when viewed along the Galactic plane and for the definitions of 
  the spiral arms, respectively.)}  The ordered {\random} field, representing the
anisotropy in the turbulence, is generated by adding a component
\added{with the same amplitude as the isotropic {\random} component}
but with an orientation aligned with the coherent field.

This model was developed using the ``z04LMPDS'' {\cre} prediction from
{\galprop} described in \S\,\ref{sec:cres}.  It was fitted in the
Galactic plane to the RM data, the 408\,MHz synchrotron
intensity (corrected for free-free emission as described in
\citealt{jaffe:2011}), and the 23\,GHz polarization data.  This
modelling was self-consistent in the sense that the magnetic field
model was first used in the {\cre} propagation, and then the resulting
spatial and spectral distribution of {\cre}s was used in the
synchrotron modelling at the two observed frequencies.

\citet{jaffe:2010} find evidence for the need to include the
ordered {\random} component in order to fit the three complementary
observables, while \citet{sun:2008} do not. This is largely due to
different assumptions about the {\cre} density.  The Sun10 model
assumes a {\cre} density in the disc more than twice as high as Jaffe13,
which means that the coherent field consistent with the RMs then
contributes enough synchrotron emissivity to reproduce all of the
polarization signal;  this is not the case if one assumes the 
level of {\cre}s from the {\galprop}-based model that matches the
{\fermi} CR data.


\subsection{Updated magnetic field models}
\label{sec:models_updated}

\begin{figure*}$\begin{array}{ccc}
%
   \includegraphics[width=60mm]{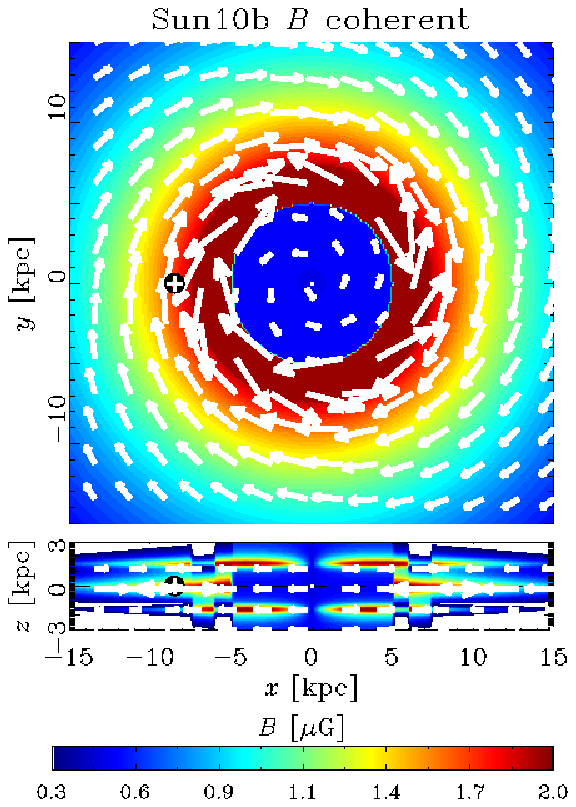} & 
   \includegraphics[width=60mm]{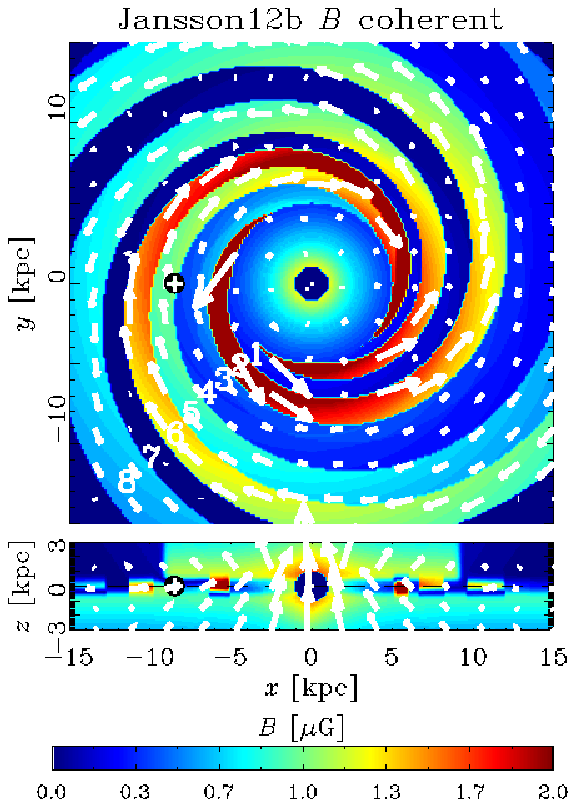} & 
   \includegraphics[width=60mm]{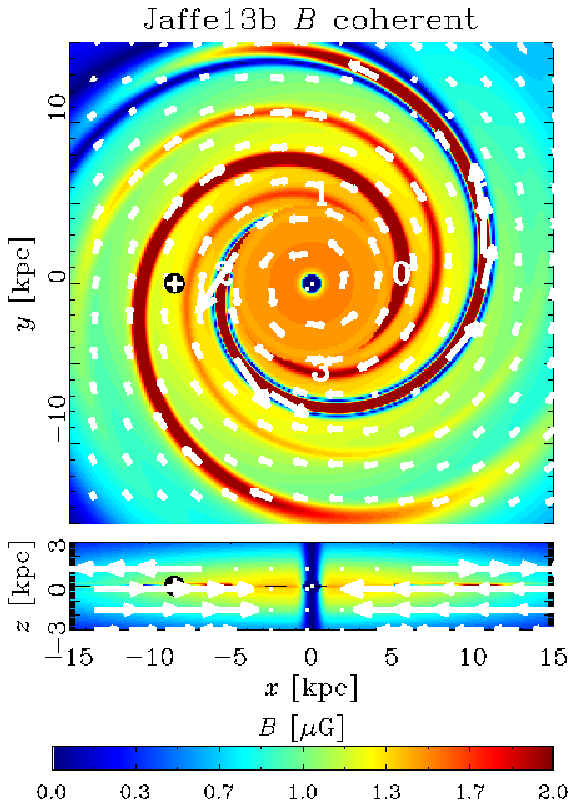} \\
   \includegraphics[width=60mm]{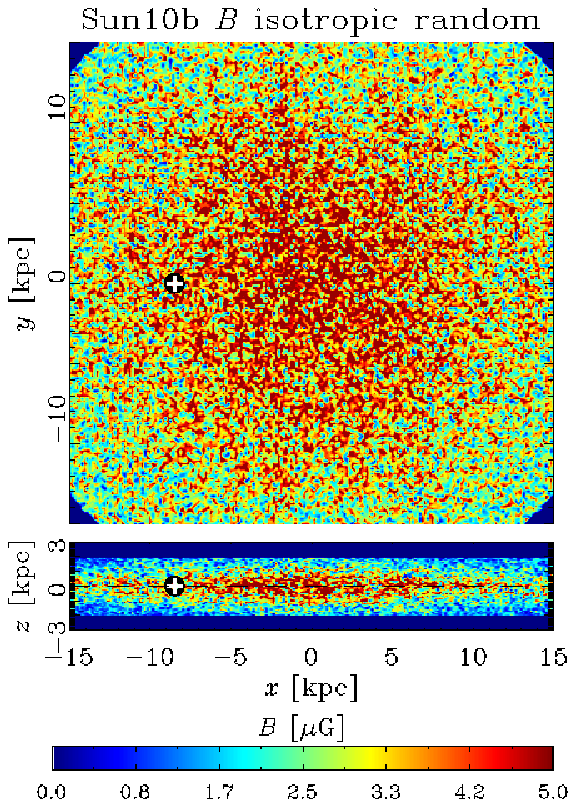} & 
   \includegraphics[width=60mm]{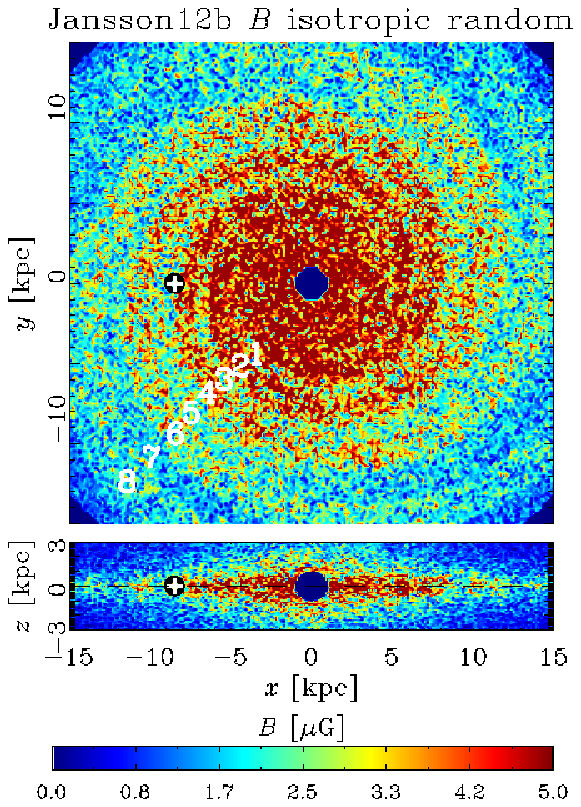} & 
   \includegraphics[width=60mm]{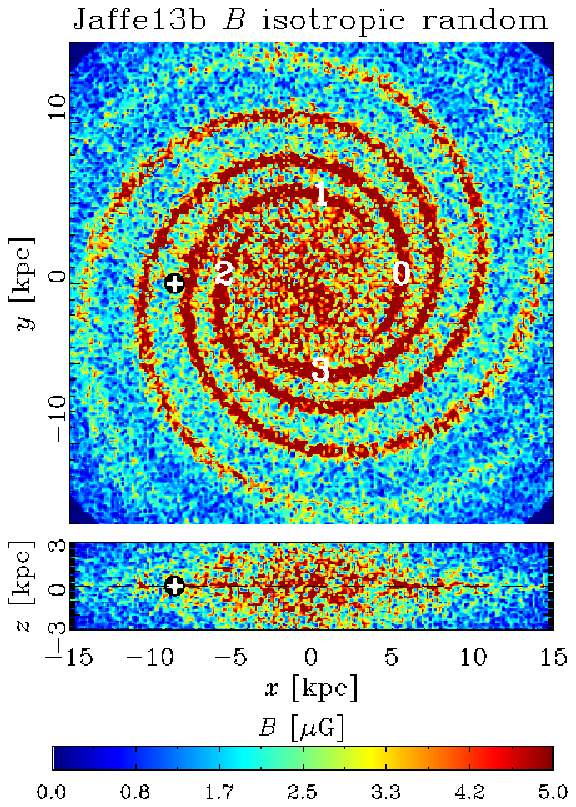} 

\end{array}$
\caption{ \added{Comparison 
  of the \addedB{updated} magnetic field models} described in \S\,\ref{sec:models_updated}.  Each {\it column} shows
  one of the models.  The top row shows both the coherent field
  amplitude in colour \added{(on a common scale)} and the projected direction shown by the arrows.  The top
  portion of each panel shows the $x$-$y$ plane at $z=0$, while the
  bottom portion shows the $x$-$z$ plane at $y=0$.  The bottom row
  shows the amplitude of a single realization of the isotropic random
  field component.  \added{The white cross in a black circle shows the
    position of the observer.}
\label{fig:models}
}
\end{figure*}

The updated models will be referred to as ``Sun10b'', ``Jansson12b'',
and ``Jaffe13b''.  They are shown in Fig.\,\ref{fig:models},
\addedC{which makes clear the morphological differences among the
  three magnetic field models that cannot yet be distinguished using
  observables integrated through the entire LOS.  We note that such a plot of the
  original models would look visually quite similar.}

Here we detail the changes made to each of the models described above
in order to match the {\planck} synchrotron solution at 30\,GHz in
conjunction with the z10LMPDE {\cre} model.  We focus on the
longitude profiles along the plane and the latitude profiles \added{averaged over} the
inner Galaxy ($-90\degr<\ell<90\degr$).  These changes are summarized
in Table\,\ref{tab:param_updates}.  The specific values of all changes
were simply chosen to approximately match by eye the profiles in
Figs.\,\ref{fig:lat_prof_comp_synch} and \ref{fig:lon_prof_comp_synch}
and as such have no associated uncertainties, nor do they necessarily
represent the unique or best solution.

For all models, the first change is to the degree of field ordering in
order to match the different synchrotron total intensity estimates in
the microwave bands.  This firstly requires a global change in the
average amounts of random versus ordered fields but also requires
morphological changes, since the different field components each
combine differently in total and polarized intensities.  We attempt to
change the smallest number of parameters that still capture the global
morphology approximately, such as scale radii and heights, or which
project onto a large part of the sky (e.g., the Perseus arm dominates
the outer Galaxy).  
We leave unchanged most of the parameters that 
affect the coherent field, since those were optimized compared
to the Faraday RM data that remain the best tracer of this component.
\added{In some cases, however, changes were needed, but we have checked that
the RM morphology remains roughly the same.}

\subsubsection{Sun10b}  

\added{This model was previously updated in Orlando13 to be consistent with
synchrotron polarization from {\wmap} and total intensity from the
408\,MHz data.  Our update here is quite similar to this but not
identical.  In particular, we use a different morphological form for
the random field component in the disc.  }

The original Sun10 model used a uniform distribution of the {\random} field
component over the simulation box and a {\cre} model sharply peaked at
$z=0$.  The {\galprop} {\cre} model is not as sharply peaked (see
Fig.\,\ref{fig:cre_comp}), so the synchrotron distribution within
$|b|\lesssim 10\degr$ requires a modified {\random} field model.  We
try an exponential disc proportional to $\exp{(-r/r_0)}
\sech^2(z/z_0)$ consisting of two components:
%
a narrow disc with $z_0=1$\,kpc and a
thicker disc with $z_0=3$\,kpc.  (The height of the thick
disc is somewhat but not entirely degenerate with a linear offset in
the total intensity.)  We find that we can fit well the latitude
profile, as shown in Fig.\,\ref{fig:lat_prof_comp_synch}, using the
two-disc model.  The amplitude is \addedC{slightly higher than} that in Orlando13.  

We add an ordered \added{random component following
  \citet{jaffe:2010}, \citet{jansson:2012b}, and Orlando13, each of whom found that this additional component is needed to
reproduce the polarized emission with a realistic {\cre} model.  As in
Orlando13, we add a component simply proportional to the coherent component
using the same approach as in the {\jf} model.  We find a slightly
\addedC{higher} amplitude than Orlando13 for this component.
\addedC{These differences reflect} the
different data sets used and are likely related to the additional spectral
shift in the {\commander} component separation solution.}
%

\subsubsection{Jansson12b}
\label{sec:jansson_updates}

As with all models, first the random component amplitude has to change to
correct the degree of ordering in the field near the plane.  This also
matches much better the galactic variance discussed in
\S\,\ref{sec:galactic_variance}.
With only this change, the morphology no longer
matches well, not only because of the different {\cre} distribution,
but also because the coherent and {\random} fields have different
distributions, and the change in their relative strengths changes the
morphology of the sum.  

With the new {\cre} distribution, the high-latitude
synchrotron polarization is too high.  We therefore lower the
amplitude of the x-shaped field component.  (This is degenerate with
other parameters such as the amplitudes of the toroidal halo
components.)  Along the plane, the polarization is also too strong in
the outer Galaxy, so we drop the coherent field amplitude of the
Perseus arm (segment number six in Fig.\,\ref{fig:models} and Table\,\ref{tab:param_updates}).  (This is degenerate with the
increased {\cre} density in the outer Galaxy.)

The results of the parameter optimization in {\jf}
include a set of spiral segments for the {\random} field
components that are dominated by a single arm.  One arm is more than
twice as strong as the next strongest, and in terms of synchrotron
emissivity, which goes roughly as $B^2$, this is then a factor of 4 higher.
In other words, the synchrotron total intensity is dominated by a
single spiral arm segment in the {\jf} model.  The quoted
uncertainties on their fit parameters do not take into account the
systematic uncertainties in \added{ the component-separated map that they
used for synchrotron total intensity}, and we consider these parameters to be unreliable in detail.
Because of this and the physically unlikely result of one dominant
turbulent arm, we further modify this model to distribute the random
component more evenly through \addedB{alternating} spiral arm segments.  As discussed
by Jaffe13, the distribution of the synchrotron emission in
latitude and longitude is not very sensitive to precisely where the
disordered fields lie in the disc.  (The {\jf} fit that resulted in
one dominant arm segment was likely driven by individual features that
may or may not be reliable tracers of large-scale morphology.)  The
precise {\em relative} distributions of ordered and disordered fields 
make a larger difference for the dust, however, and this will be
discussed further in \S\,\ref{sec:dust_field_models}.

As discussed above, we replace the analytic estimate for
the total synchrotron emission from the isotropic random component
with numerical simulations of a GRF.  We retain, however, the simple
generation of their ordered random component by simply scaling up the
coherent component.  This means that we are missing the ordered random 
field's contribution to the galactic variance.

\subsubsection{Jaffe13b}

For the Jaffe13 model, the different components' scale heights need to
be adjusted, since these had not been constrained by the previous
analysis confined to the Galactic plane.  We now use the values listed
in Table\,\ref{tab:param_updates}.  To match the synchrotron latitude
profiles in total and polarized intensities, we now use two
exponential discs, as for Sun10b, one a thin disc and one thick, or ``halo'', 
component for each of the coherent and random field components.  We
also flip the sign of the axisymmetric components (disc and halo, but
not the arms) above the plane to match the RM asymmetry as discussed
in \citet{sun:2008}.  

The combination of a different method for estimating the total
synchrotron intensity and the updated {\cre} model require
a corresponding change in the degree of field ordering.  We therefore decrease
the amplitude of the random component.  

Because of the difference in the {\cre} distribution between
the inner and outer Galaxy, we adjust slightly some of the arm
amplitudes.  \added{The field amplitude in the Scutum arm drops, as it does 
in the molecular ring.  }

Lastly, we do not include the shift \added{in the spiral arm pattern
  introduced in Jaffe13} between the arm ridges of the
isotropic random field component and
the rest of the components.  \added{This shift was introduced to
increase the dust polarization and was motivated by observations of
external galaxies}.  As we will see in \S\,\ref{sec:dust_results},
the updated models produce more strongly polarized dust emission
without this \added{additional complexity}.

%
\begin{figure*}
\includegraphics[width=\linewidth,clip,trim=0 0 0 0.2cm]{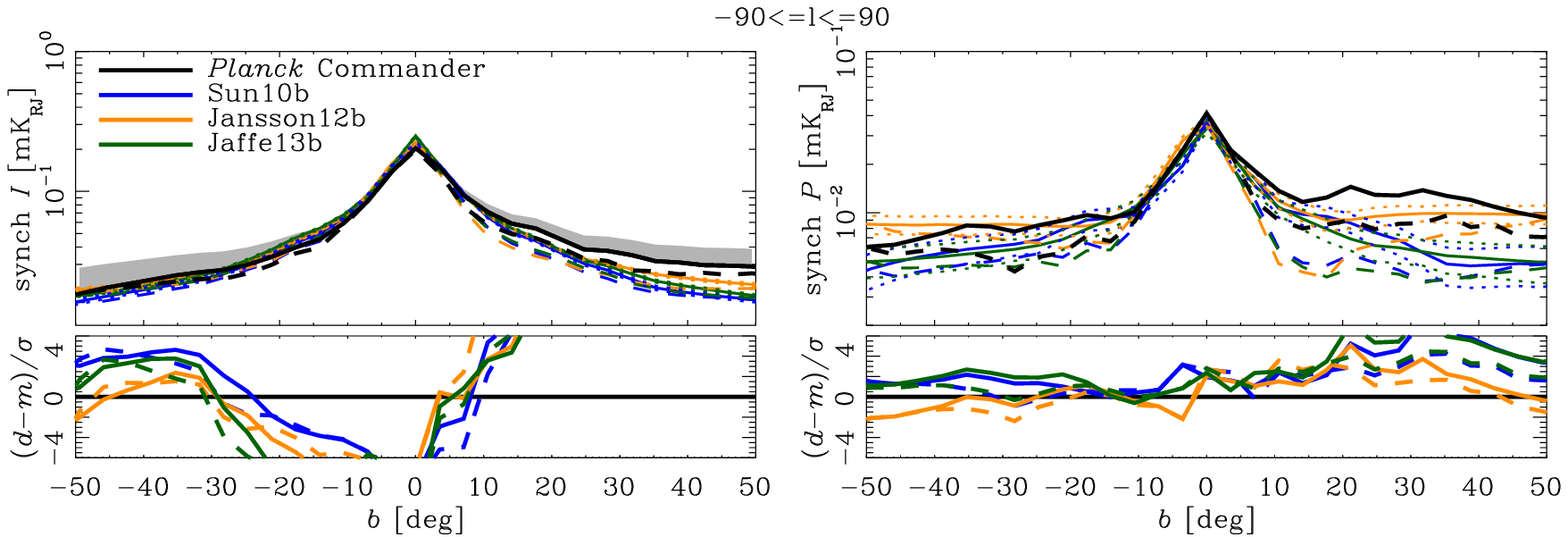}
\includegraphics[width=\linewidth,clip,trim=0 0 0 0.2cm]{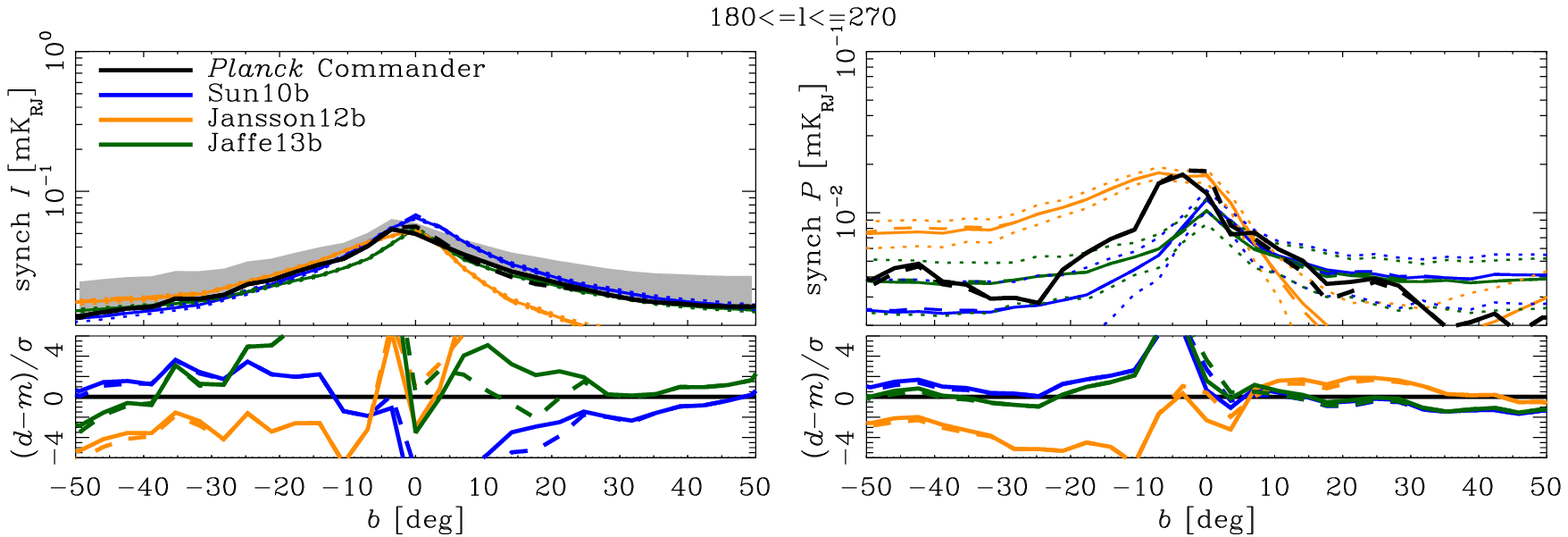}
\caption{ \added{ Synchrotron latitude profiles for the
  data and the updated models' ensemble average.  The observables are
  averaged over a range of longitudes for a given latitude bin, and on
  the left is total intensity and on the right polarized intensity.  The top shows the inner Galaxy (i.e.,
  $-90\degr<\ell<90\degr$), while the bottom shows the third
  quadrant ($180\degr<\ell<270\degr$, i.e., the outer Galaxy excluding
  the Fan region).}  The dotted coloured lines show \added{the model mean
plus or minus} the expected variation predicted by the models
\addedB{(though these are often too close to the solid lines to be visible)}.  This variation is also the $\sigma$ used to
  compute the significance of the residuals in the bottom panel of
  each row.   \added{The dashed curves show the profiles excluding the
   loops and spurs discussed in \S\,\ref{sec:loops}.  The grey band shows the $\pm3$\,K zero-level
  uncertainty of the data at 408\,MHz extrapolated with $\beta=-3.1$.}
\label{fig:lat_prof_comp_synch}
}
\end{figure*}

\begin{figure*}
\includegraphics[]{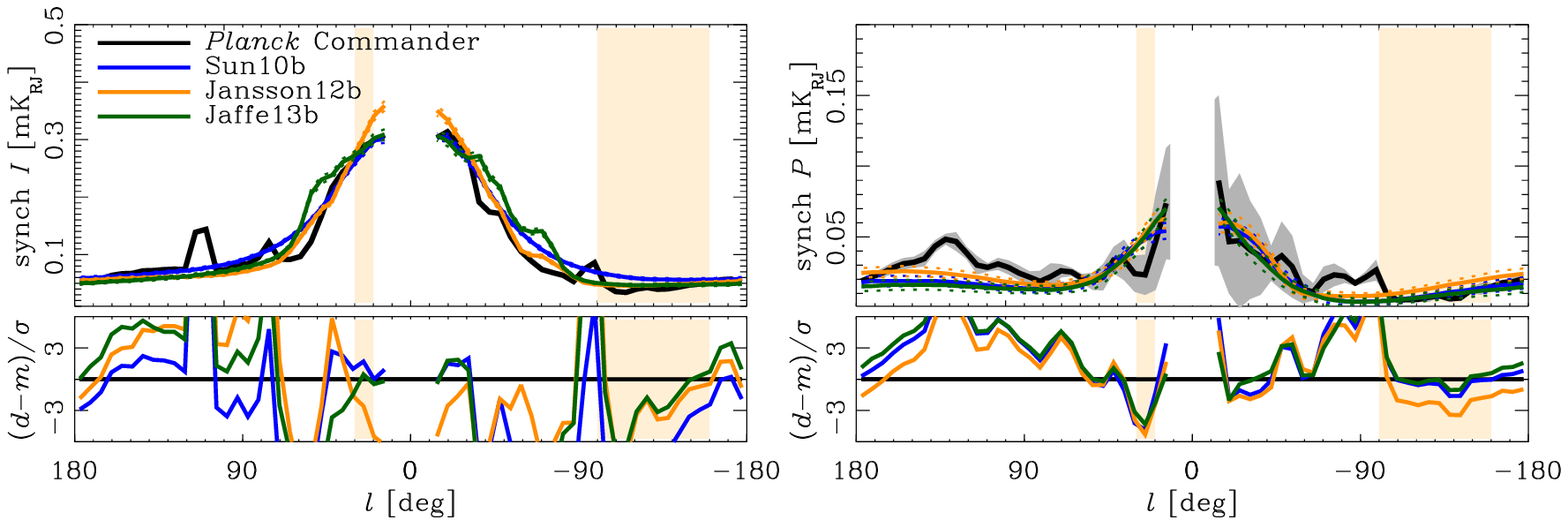} \\
\caption{ Longitude profiles for synchrotron for the updated models as
  described in \S\,\ref{sec:models_updated}.  See
  Fig.\,\ref{fig:lat_prof_comp_synch}.   The grey band is an estimate
  of the uncertainty primarily due to bandpass leakage discussed in
  Appendix\,\ref{sec:appendix_systematics_lfi}.  \added{The pale orange
  vertical bands highlight longitude ranges where the plane crosses
  any of the loops discussed in \S\,\ref{sec:loops}.}
\label{fig:lon_prof_comp_synch}
}
\end{figure*}

 \subsection{Synchrotron results}  
\label{sec:synch_results}

\begin{figure*}
$\begin{array}{cccc} 
%
%
& { I} & { Q} & { U} \\

& \multicolumn{3}{c}{      \rmn{Data, }d} \\

& 
\includegraphics[trim=0.2cm 0 0.3cm 0, height=0.28\linewidth,angle=90,clip]{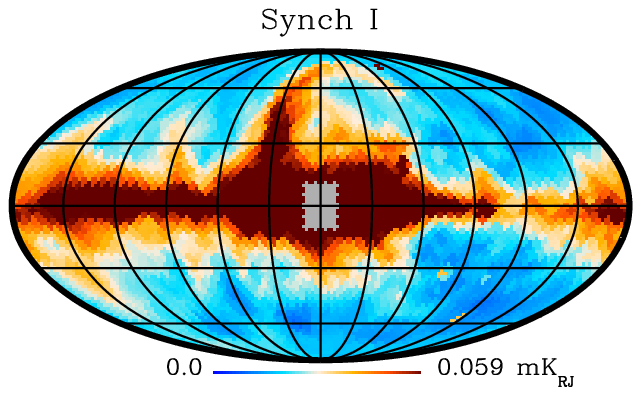} &
\includegraphics[trim=0.2cm 0 0.3cm 0, height=0.28\linewidth,angle=90,clip]{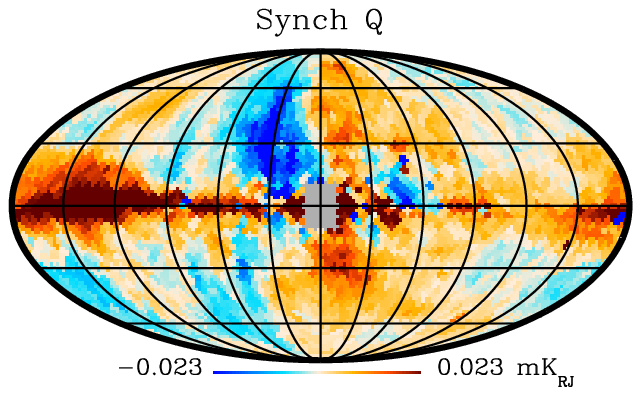} &
\includegraphics[trim=0.2cm 0 0.3cm 0, height=0.28\linewidth,angle=90,clip]{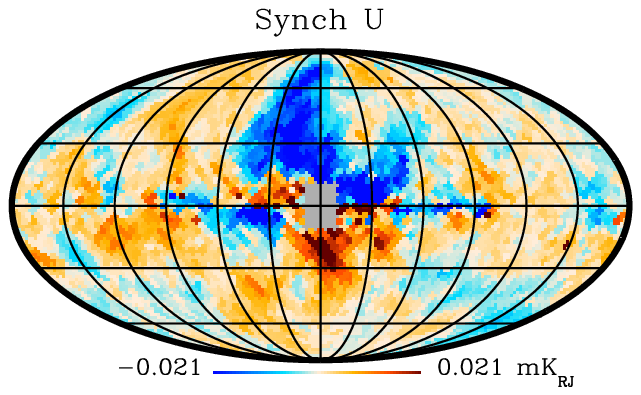} \\ 

& \multicolumn{3}{c}{      \rmn{Models, }m} \\

\rotatebox{90}{{\hspace{1cm} Sun10b\hspace{0.5cm}}} & 
\includegraphics[trim=0.2cm 0 0.3cm 0, height=0.28\linewidth,angle=90,clip]{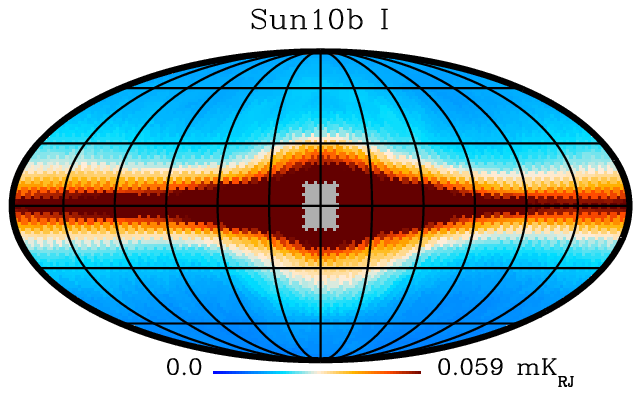} &
\includegraphics[trim=0.2cm 0 0.3cm 0, height=0.28\linewidth,angle=90,clip]{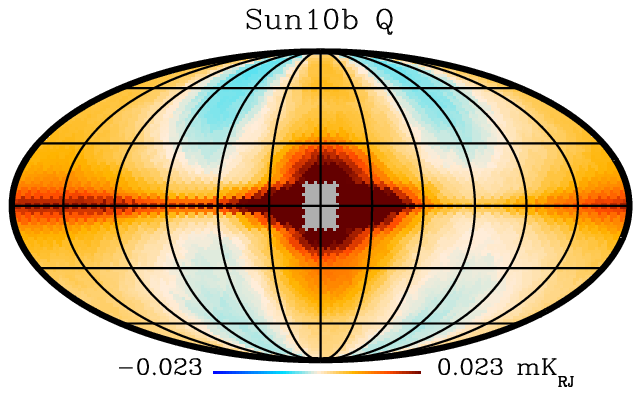} &
\includegraphics[trim=0.2cm 0 0.3cm 0, height=0.28\linewidth,angle=90,clip]{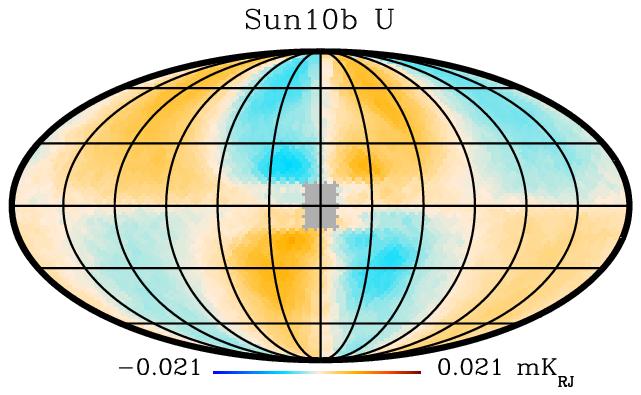}  \\

\rotatebox{90}{{\hspace{0.5cm} {\jf}b\hspace{0.5cm}}} & 
\includegraphics[trim=0.2cm 0 0.3cm 0, height=0.28\linewidth,angle=90,clip]{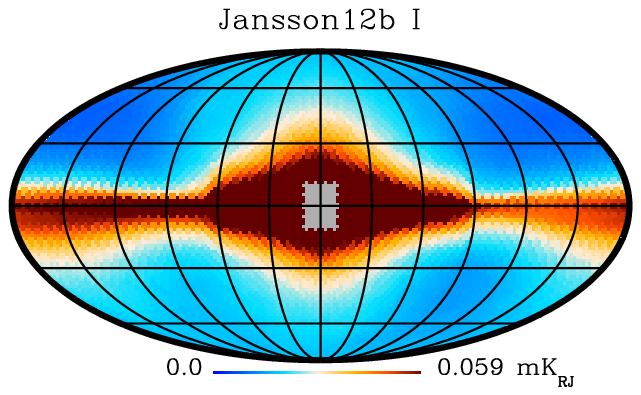} &
\includegraphics[trim=0.2cm 0 0.3cm 0, height=0.28\linewidth,angle=90,clip]{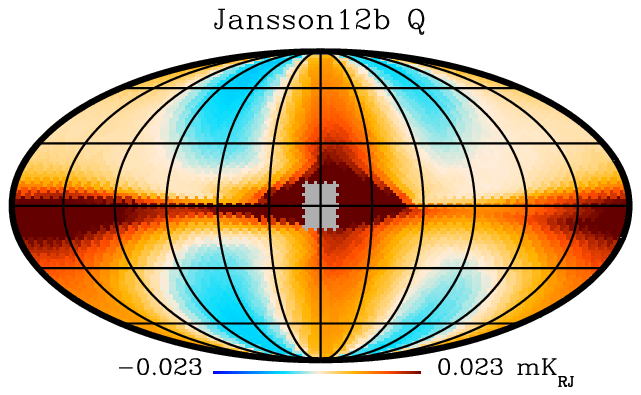} &
\includegraphics[trim=0.2cm 0 0.3cm 0, height=0.28\linewidth,angle=90,clip]{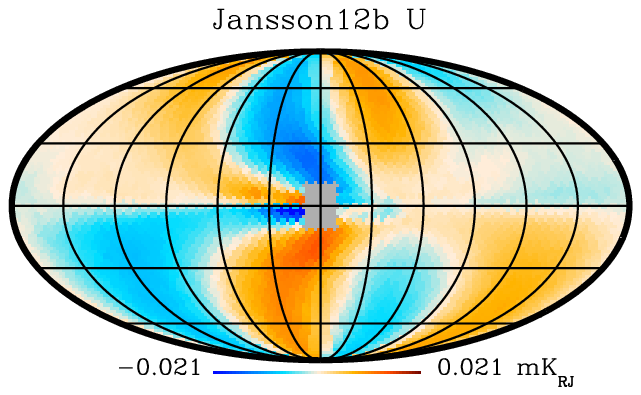} \\

\rotatebox{90}{{\hspace{1cm}  Jaffe13b\hspace{0.5cm}}} & 
\includegraphics[trim=0.2cm 0 0.3cm 0, height=0.28\linewidth,angle=90,clip]{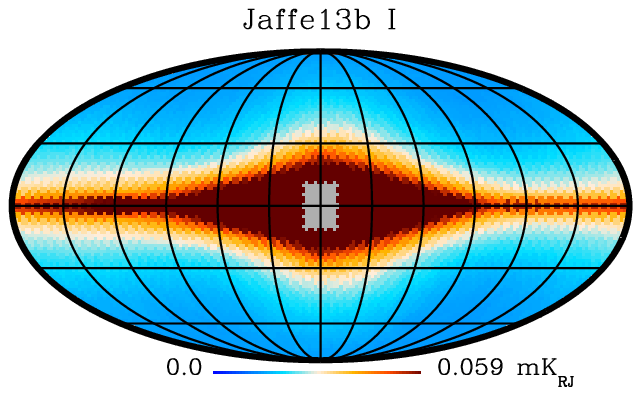} &
\includegraphics[trim=0.2cm 0 0.3cm 0, height=0.28\linewidth,angle=90,clip]{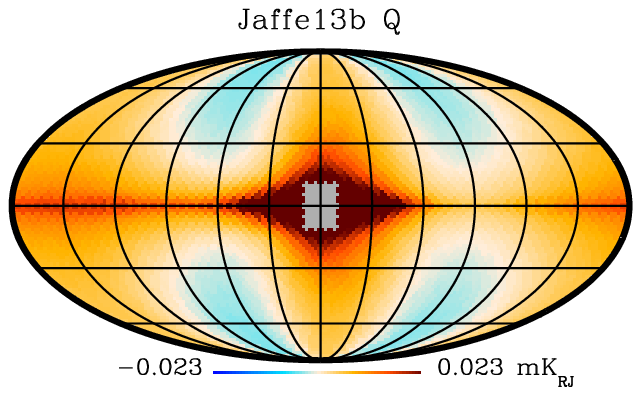} &
\includegraphics[trim=0.2cm 0 0.3cm 0, height=0.28\linewidth,angle=90,clip]{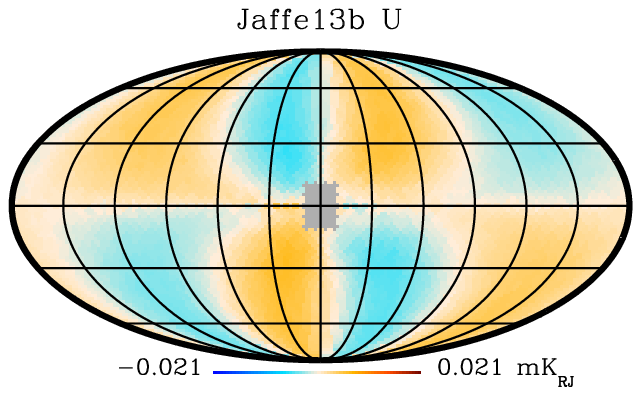}\\

&  \multicolumn{3}{c}{ \rmn{Residuals, } (d-m)/\sigma } \\

\rotatebox{90}{\hspace{1cm} Sun10b\hspace{0.5cm}} & 
\includegraphics[trim=0.2cm 0 0.3cm 0, height=0.28\linewidth,angle=90,clip]{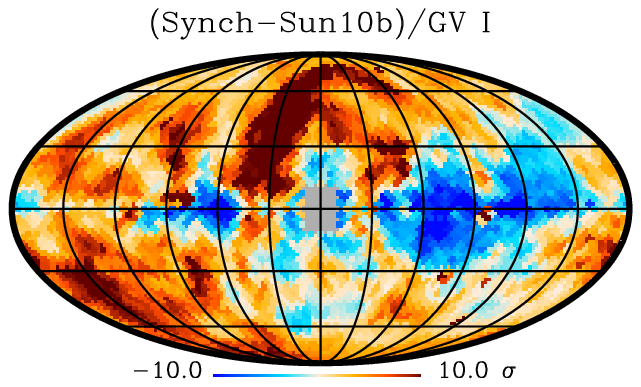} &
\includegraphics[trim=0.2cm 0 0.3cm 0, height=0.28\linewidth,angle=90,clip]{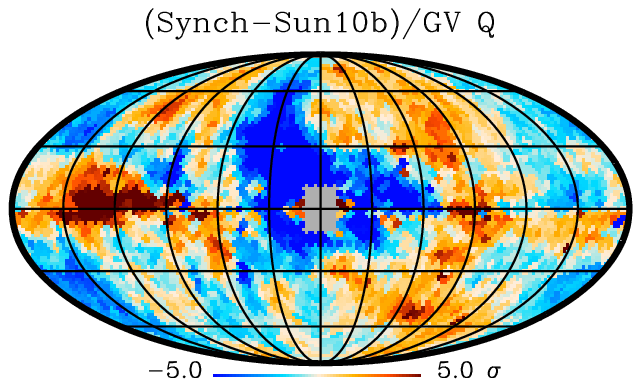} &
\includegraphics[trim=0.2cm 0 0.3cm 0, height=0.28\linewidth,angle=90,clip]{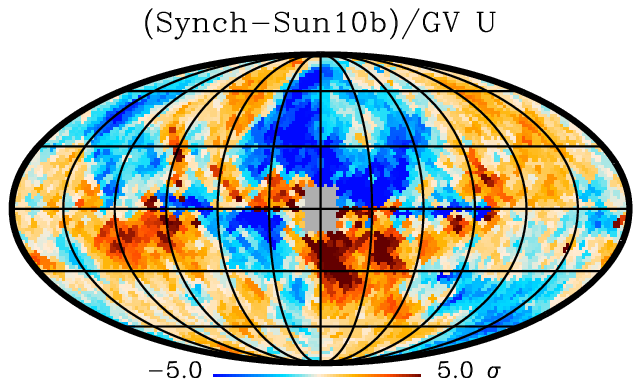}  \\

\rotatebox{90}{\hspace{0.5cm} {\jf}b\hspace{0.5cm}} & 
\includegraphics[trim=0.2cm 0 0.3cm 0, height=0.28\linewidth,angle=90,clip]{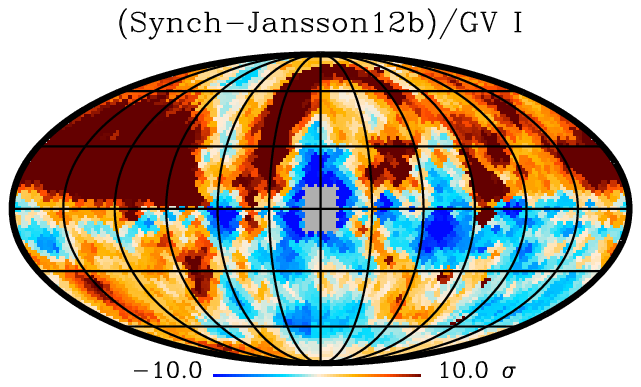} &
\includegraphics[trim=0.2cm 0 0.3cm 0, height=0.28\linewidth,angle=90,clip]{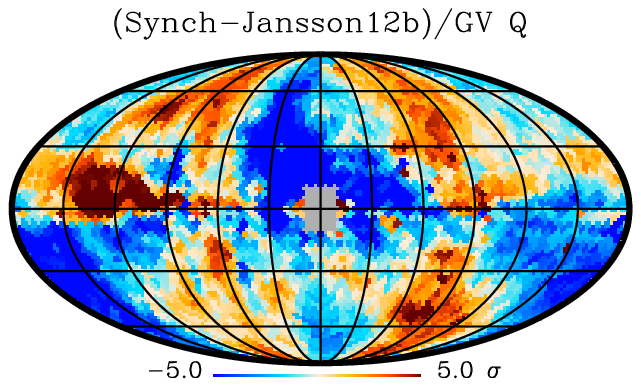} &
\includegraphics[trim=0.2cm 0 0.3cm 0, height=0.28\linewidth,angle=90,clip]{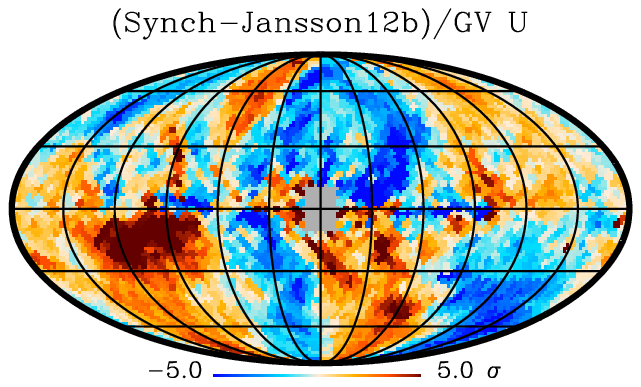}\\

\rotatebox{90}{\hspace{0.8cm} Jaffe13b} & 
\includegraphics[trim=0.2cm 0 0.3cm 0, height=0.28\linewidth,angle=90,clip]{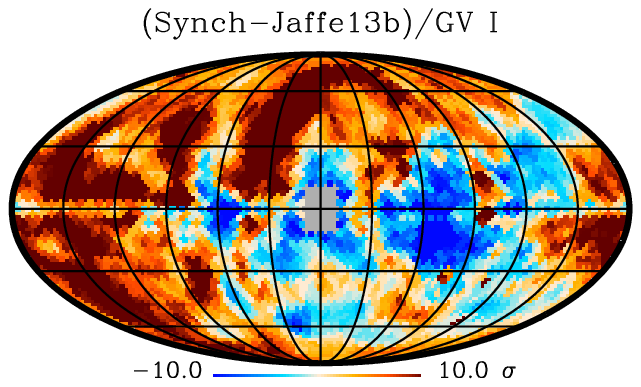} &
\includegraphics[trim=0.2cm 0 0.3cm 0, height=0.28\linewidth,angle=90,clip]{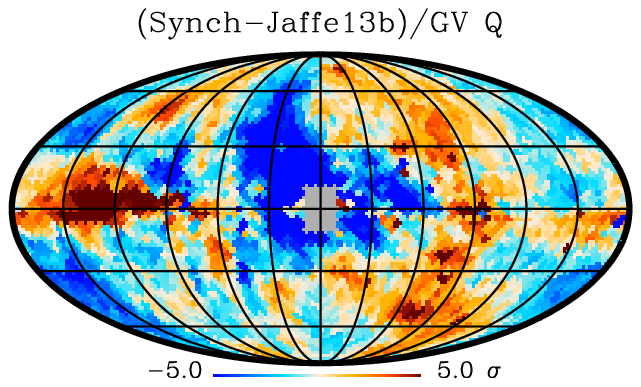} &
\includegraphics[trim=0.2cm 0 0.3cm 0, height=0.28\linewidth,angle=90,clip]{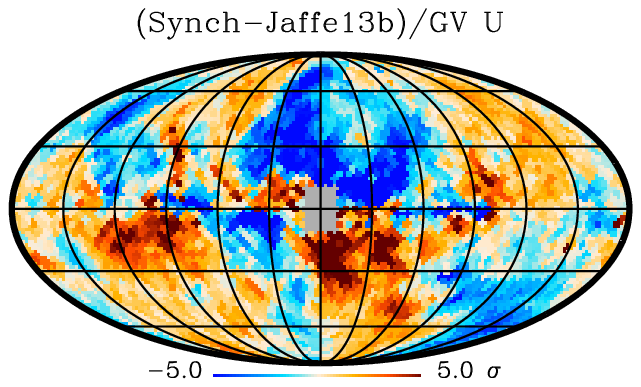}\\
\\

\end{array}$
\caption{ Comparison of the 
 model predictions for synchrotron emission and the {\Planck}
 synchrotron maps.  The columns from left
  to right are for Stokes $I$, $Q$, and $U$, while the rows are the
  data followed by the prediction for each model, and lastly the
  difference between model and data divided by model uncertainty
  (galactic variance).   
\label{fig:chisq_maps_synch} 
}
\end{figure*}

Figures \ref{fig:lat_prof_comp_synch} and
\ref{fig:lon_prof_comp_synch} \added{compare the data with the 
  ensemble average models in profiles in longitude and
  latitude.}  They demonstrate that each of the three Galactic
magnetic field models can be configured to reproduce roughly the right
amount of emission in total and polarized intensity \added{towards the inner
Galaxy (which covers most of the Galactic disc)}, despite the
significant morphological differences in the field models \added{shown
  in Fig.\,\ref{fig:models}}.  They do,
of course, differ in detail, including polarization angles not
visible in those plots, \added{and they do not fit well in the outer Galaxy}.  Figure\,\ref{fig:chisq_maps_synch} shows maps
of the data and models in Stokes $I$, $Q$, and $U$ as well
as the differences.  

\added{ In both profiles and maps, we also plot the residuals as
  differences divided by the expected galactic variations computed
  from the models.  These variations are model-dependent, since the
  amplitude of the random field component impacts not only the mean
  total intensity of synchrotron emission but also the expected
  variation of our single Galaxy realization from the mean.  This
  means that the significance of the residuals is model-dependent and
  should be treated carefully.  The question of this galactic variance
  is discussed as an observable in itself in
  \S\,\ref{sec:galactic_variance}.  }

%
%

These residuals show clearly the North Polar Spur (NPS) and exclude the Galactic
centre region, neither of which is treated explicitly in the
modelling.  \added{We can see an excess of total intensity emission at
high latitudes in the data compared to all models, which may be 
due to a missing isotropic component in the models} or to the uncertain offset level of
the 408\,MHz map used in the {\commander} synchrotron total intensity
solution.  In polarization, we see strong residuals in all models in
the so-called Fan region in the second quadrant near the plane.
\addedC{In the following
sections, we discuss the most important aspects of the synchrotron
modelling:  the information in the galactic variance and the impact of
a varying synchrotron spectrum.}  We emphasize the dependence of 
these results on our choice to base the modelling on the {\planck}
{\commander} component-separation products.


\subsubsection{Galactic variance}
\label{sec:galactic_variance}

\begin{figure*}
$
\begin{array}{cccc}
%

&  I & Q\, \rmn{[{\wmap}]} & Q\, \rmn{[}Planck\rmn{]} \\

\rotatebox{90}{{\hspace{0.5cm} Data rms}} & 
\includegraphics[trim=0 0 0.3cm 0,  clip , height=0.28\linewidth,angle=90]{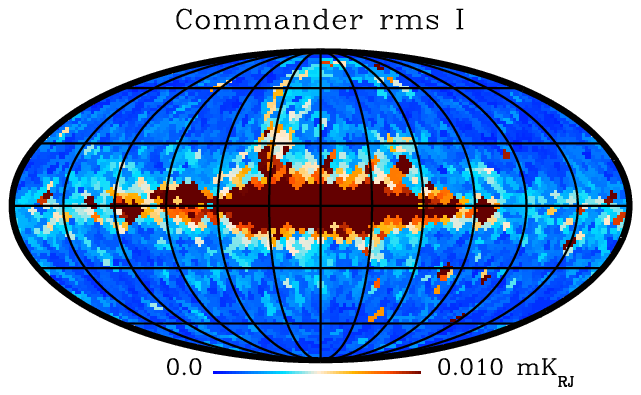} & 
\includegraphics[trim=0 0 0.3cm 0,  clip , height=0.28\linewidth,angle=90]{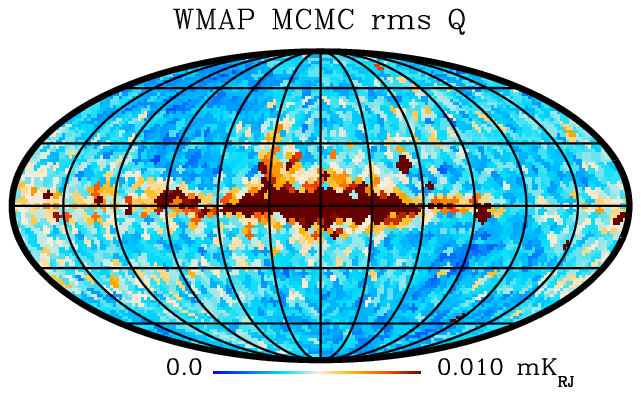} & 
\includegraphics[trim=0 0 0.3cm 0,  clip , height=0.28\linewidth,angle=90]{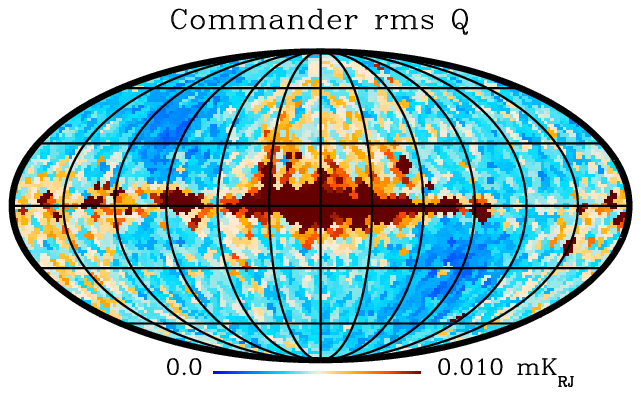} \\

\rotatebox{90}{{\hspace{0.5cm} Sun10b rms}} & 
\includegraphics[trim=0 0 0.3cm 0, clip , height=0.28\linewidth,angle=90]{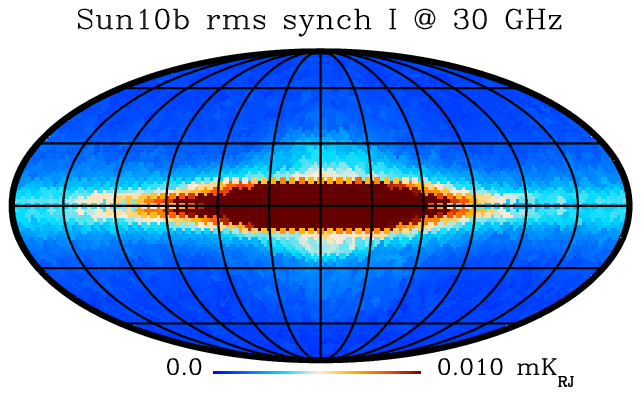} & 
&
\includegraphics[trim=0 0 0.3cm 0, clip , height=0.28\linewidth,angle=90]{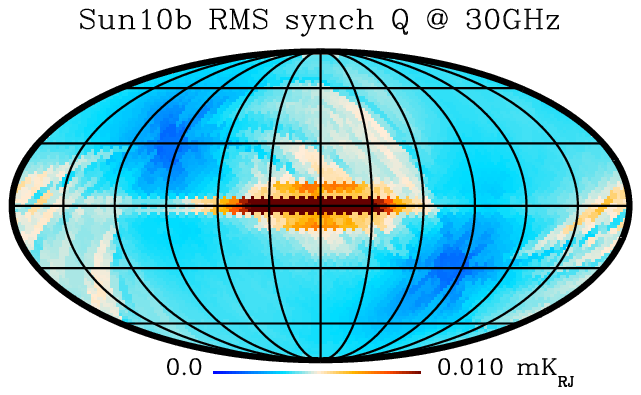} \\ 

\rotatebox{90}{{\hspace{0.5cm} Jaffe13b rms}} & 
\includegraphics[trim=0 0 0.3cm 0, clip , height=0.28\linewidth,angle=90]{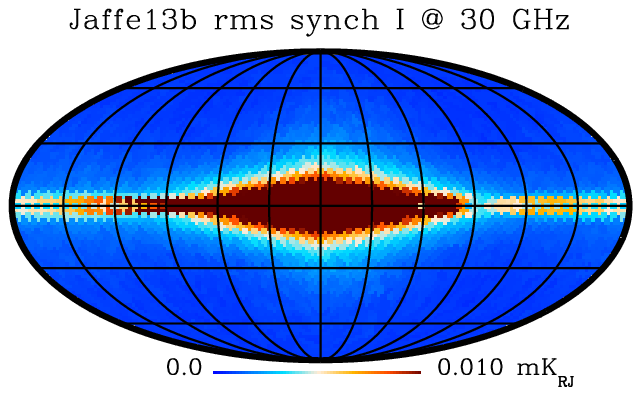} & 
&
\includegraphics[trim=0 0 0.3cm 0, clip , height=0.28\linewidth,angle=90]{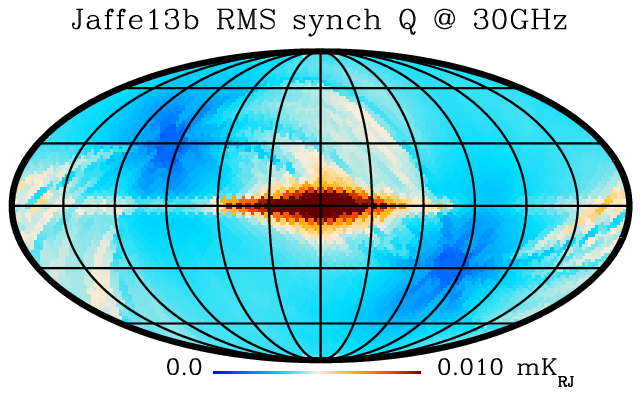} \\ 

\rotatebox{90}{{\hspace{0.5cm} Jansson12(b) rms}} & 
\includegraphics[trim=0 0 0.3cm 0, clip , height=0.28\linewidth,angle=90]{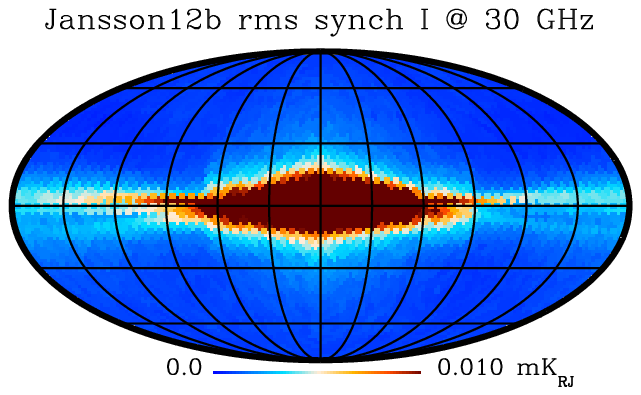} & 
\includegraphics[trim=0 0 0.3cm 0, clip , height=0.28\linewidth,angle=90]{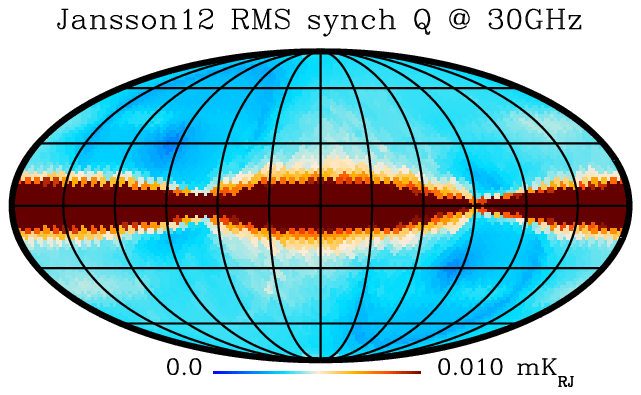} & 
\includegraphics[trim=0 0 0.3cm 0, clip , height=0.28\linewidth,angle=90]{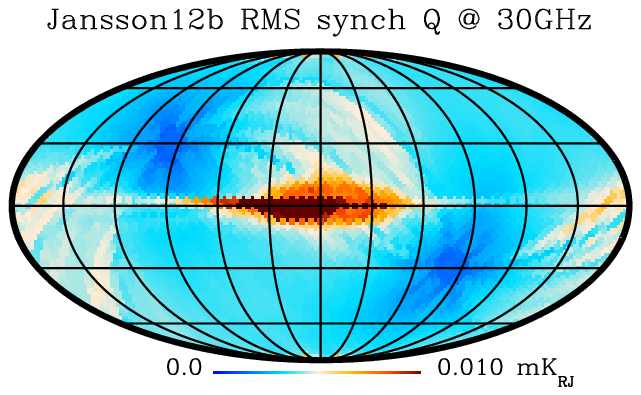} \\

\rotatebox{90}{{\hspace{0.5cm} Jansson12(b) GV}} &
\includegraphics[trim=0 0 0.3cm 0, clip , height=0.28\linewidth,angle=90]{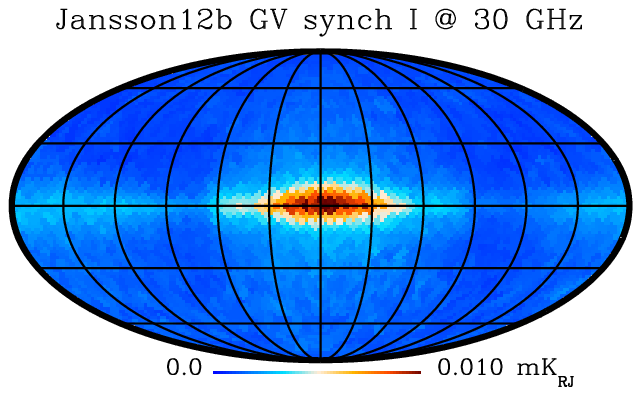} & 
\includegraphics[trim=0 0 0.3cm 0, clip , height=0.28\linewidth,angle=90]{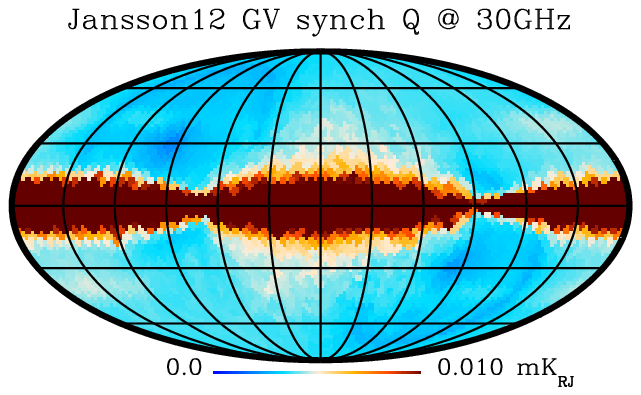} & 
\includegraphics[trim=0 0 0.3cm 0, clip , height=0.28\linewidth,angle=90]{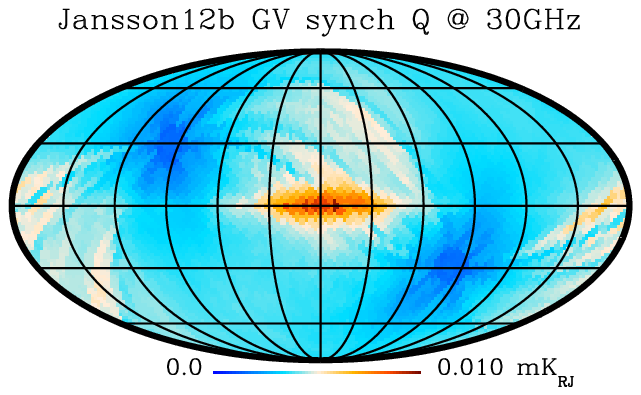} \\

\end{array}$
\caption{Comparison of estimates for galactic variance in data and models.  The top row shows estimates from the data, while the following rows show
  the model predictions.   Excepting the
  last row, these estimates are based on the rms variations in each low-resolution
  ($N_\mathrm{side}=16$) pixel.   From left to right, the top
    row shows the estimates from the synchrotron total 
  intensity from {\commander}, the synchrotron $Q$ map from {\wmap} MCMC
  (extrapolated to 30\,GHz), and
  the synchrotron $Q$ map from {\commander}.  (The Stokes
  $U$ maps, not shown, look very similar to those for $Q$.) 
The updated Sun10b model is on the second row, the updated Jaffe13b
model 
on the third row, and two versions of the {\jf} model are in each of the 
    fourth and bottom rows.  To each model prediction of the ensemble
  variance is added simulated noise.  
  In the case of
  Sun10b and Jaffe13b, we only show the updated model but for {\jf}, we compare the
  original model ({\it middle column}) optimized with {\wmap} MCMC $I$ and
  $Q$ and the updated {\jf}b model ({\it left and right}) optimized with the {\commander} synchrotron solution. 
  For comparison, the
  last row shows the full galactic variance in each pixel for
  the {\jf} models, as described in the text.  
\label{fig:model_data_rms_comp}
}
\end{figure*}

In comparing model predictions with the observables in the presence of
a {\random} field component, we must take into account the fact that the
observables do not represent the ensemble average galaxy.  Instead,
they represent one turbulent realization, i.e., our Galaxy, and therefore we do not
expect the models to match precisely.  The models do, however, predict
the degree of variation due to the {\random} magnetic fields.  We refer
to this as galactic variance.  These predictions are not only
necessary for estimating the significance of residuals but also an
additional observable in and of themselves.

\cite{jansson:2012b} computed their model entirely analytically and
therefore obtained no prediction for the galactic variance.  They
recognize its importance as an observable,
but they only estimate this variance from their
data to use in their likelihood analysis.  For each low-resolution
$N_\rmn{side}=16$ pixel, they compute the rms variation of the data at
its nominal resolution ($1\degr$ in the case of {\wmap} foreground
products).  We test this approach by comparing with the results of the identical
operation on a set of simulated galaxy realizations.  Specifically,
for each realization, we compute the rms in each large pixel, and then
take the average among the realizations.   The Sun10 
analysis did compute a random component but did not look at this issue.  The
analyses on which the Jaffe13 model is based did compute such
realizations and the resulting variance, and it was used in the model
comparison plots and the likelihood computation \added{but not
  examined as an observable in itself}.  


\cite{jansson:2012b} do not show whether their model for the isotropic
{\random} component of the magnetic field in {\jf} results in a variance
similar to what they measure with their $N_\rmn{side}=16$ pixel-based
variance estimate.  Figure\,\ref{fig:model_data_rms_comp} shows this
comparison explicitly.

The top row of Fig.\,\ref{fig:model_data_rms_comp} gives the data rms
\addedC{variation} using the method from \cite{jansson:2012b,jansson:2012c}.
From left to right, we show this rms for synchrotron total $I$ from
{\commander}, for Stokes $Q$ from {\wmap} MCMC (extrapolated to 30\,GHz
assuming $\beta=-3$), and Stokes $Q$ at 30\,GHz from LFI.  We do not
show Stokes $U$, which is very similar to $Q$.

The second through fourth rows of Fig.\,\ref{fig:model_data_rms_comp}
show the $N_\rmn{side}=16$ rms variations predicted by the three
models for comparison with the data.  Because the variations in the
data include both GV and noise variations, \added{we add to the models
  in quadrature the expected noise level of LFI computed from the
  diagonal elements of the published covariance matrix.  
For comparison with {\wmap} in the middle column, we add an estimate of the noise level 
  computed from the published $\sigma_0$ and the $N_\rmn{obs}$
for the K-band at 23\,GHz.  For both surveys, the noise has a
  quite distinct morphology from the GV in the models, with noise
  minima near the ecliptic poles due to the scan pattern visible
  similarly in both observed and simulated maps.  We include the noise
  for comparison, but it is the GV that is of interest.}

These comparisons \addedC{of models and data} show some significant differences.  
The original {\jf} model significantly overpredicts the variation in
the synchrotron polarized emission, while the updated
model somewhat underpredicts the variation.  This implies in each case an 
incorrect degree of ordering in the fields.  

We note, however, that the method using the sky rms has little sensitivity to 
fluctuations significantly larger than the $N_\rmn{side}=16$ pixels, which are roughly
$4\degr$ wide.  If the outer scale of turbulence is roughly $100$\,pc, then fluctuations
on these scales are not fully accounted for when nearer than about $1.5$\,kpc.  In other words, the sky rms method is not representative
of the emission variations due to local structures within this
distance, but this applies equally to the data and to the models.  

\added{The high
level of random field in the original {\jf} model
was likely caused by the contamination
of the microwave-band total intensity synchrotron observables by
anomalous dust emission.}  The updated model may underpredict the emission because
it is too far the other way due to the \added{steep spectral index assumed
for the synchrotron
spectrum} in the {\planck} {\commander} solution.  (In the case of the
Sun10b and {\jf} models, the simulations do not include the variation
due to the ordered random component.  They are therefore missing some of
the expected physical variations.  The Jaffe13b model, however, does
include this in the simulations and shows a similar degree of variation.)
 
The {\jf} analysis uses these estimates of the uncertainty in the $\chi^2$
computation in their likelihood exploration of the parameter space.
Given that the variations are overpredicted in polarization, this
could easily allow incorrect models to fit with unrealistically low
values of $\chi^2$ giving the appearance of a good fit.  

We also show for comparison on the bottom row of
Fig.\,\ref{fig:model_data_rms_comp} the actual galactic variance
computed for the models by using a set of realizations of each.  In
this case, we compute the variation {\em among} the different
realizations for each full-resolution pixel and then downgrade (i.e.,
average) the result to $N_\rmn{side}=16$.  This shows the true
galactic variance in the simulations, including the largest angular scales, which is a
quantity we cannot compute for reality but which is interesting to
compare.  It is the uncertainty we use in this work for comparing
models to the data, since it expresses how much we expect the real sky
to deviate from the model average.  Unlike the rms method, the GV
method does include local structures as long as they are
\addedB{resolved by} the simulation (see \S\,\ref{sec:hammurabi}).   On the other hand, the
rms measured on the real sky includes variations down to
arbitrarily small scales (as long as they are \added{far enough away to be
sampled within the size of the $N_\rmn{side}=16$ pixel).}
For the models, this is limited by the resolution of the simulation,
and so the model estimates using both rms and GV will always be
missing some variations at small spatial scales.

The rms
method does, however, predict more variation in the Galactic plane
than the GV method does (compare the last two rows on the left of
Fig.\,\ref{fig:model_data_rms_comp}).  This is not a simulation
resolution issue, since both are based on the same simulations.
Instead, this is due to the fact that the Galactic emission components
all have steep gradients at low latitudes, and this contributes
variance within the large pixel in the rms method that is not due to
the turbulent field component.  It is therefore impossible to directly
compare the different methods.  (This \addedC{gradient contributes}
to the rms of both simulations and sky, however, so the comparison of those two
remains valid.)

%
We have also tested the effect of the simulation resolution on
these estimates of the GV.  We dropped the simulation resolution by a
factor of 2 as well as increasing it by a factor of 2. 
The lowest resolution does significantly affect the analysis, but our
chosen resolution is within a few percent of the highest resolution estimate
over almost all of the sky, differing by up to $20\,\%$ in the inner
Galactic plane only.  



The variance discussed here is related to the strength of the
\addedB{isotropic} {\random} magnetic field component \addedB{relative to the coherent
  and ordered components}, which, as discussed in
\S\,\ref{sec:models_updated}, is related to the estimate for the
synchrotron total intensity in the microwave bands.  This in turn is a
function of the {\cre} spectrum assumed, which is highly uncertain and
varies on the sky.  The original {\jf} model (in the middle column of
the fourth row of Fig.\,\ref{fig:model_data_rms_comp}) shows the hardest
spectrum considered, since it is based on the {\wmap} MCMC solution
that effectively assumes $\beta=-2.6$ in the Galactic plane and therefore has
the highest level of {\random} fields and variance.  The updated models
shown on the right of that figure are tuned to match the {\planck}
{\commander} synchrotron solution that assumes a spectrum with an
effective index of $\beta=-3.1$.  This is at the steep end of
reasonable for the sky as a whole and may be too steep
for the Galactic plane region.  And indeed, the original model
overpredicts and the updated model underpredicts the variance in the
polarization. 

\added{The fact that the updated models appear to underestimate
  somewhat the variations implies that the residuals computed as
  $(d-m)/\sigma$ will appear more significant than they perhaps are.
  It is important, \addedD{however,} to keep in mind when looking at the residuals
  how the uncertainties themselves are model-dependent, and therefore
  so is the significance of any residual.  For example, one could make
  polarization residuals appear less significant by increasing the random
  component.  An explicit likelihood-space exploration should take this into account in the parameter
  estimation (see, e.g., Eq.\,14 of \citealt{jaffe:2010} where
  this was done), but our approximate fitting here does not.  (Nor
  was this done in the {\jf} analysis.)  This will have
  to be dealt with correctly in any future analysis with a correct
  parameter estimation once the component-separation problem has been
  solved.}

Lastly, we note that the modelled GV is also a function of
other properties of the {\random} field component such as the outer
scale of turbulence and the power-law index.  We have tested the
effects of varying these parameters as much as possible given the
dynamic range in our simulations.  A larger turbulence scale results
in a larger GV, since the GV is partly a function of
the number of turbulent cells in each observed pixel.  Likewise, a
steeper turbulence spectrum (i.e., more dominated by the largest scales)
causes an increase in the GV, though this effect is fairly weak.  
These effects should be kept in mind when looking at the predicted
amount of GV for each model, but the chosen parameters are
well motivated by observations of the ISM, as discussed by 
\citet{haverkorn:2013}.

%

\subsubsection{The synchrotron spectrum}
\label{sec:synch_spec}

We have adopted in our analysis the {\planck} {\commander} synchrotron
solution, which assumes a constant spectrum on the sky that can shift
in frequency space (which effectively steepens or \addedC{hardens} it).  The
component separation is only sensitive to the effective spectral index
between the microwave regime and the 408\,MHz total intensity
template.  \added{This assumed synchrotron spectrum was originally based on an
analysis of radio data at intermediate latitudes in
\citet{strong:2011}.  (The Orlando13 follow-up studied the influence of the
magnetic field models but did not change the spectral parameters of
the injected electrons.)  }

The resulting {\commander} synchrotron spectrum is quite steep, with an
effective $\beta=-3.1$ from 408\,MHz to the microwave bands.
This fit is likely driven by the intermediate- and high-latitude sky; 
\added{near the plane other components (free-free and AME) are
strongly correlated, while the higher latitudes are dominated by
synchrotron, particularly strongly emitting regions like the NPS.}
This result should therefore not be taken as evidence for such a steep
spectrum of synchrotron emission in the Galactic plane.  \added{On the
  contrary, the use of this spectrum ignores
the fact that there is evidence for a global hardening of the spectrum
in the Galactic plane and also for a larger curvature in the spectrum at low
frequencies. }

There are a variety of studies that find evidence for the steepening
of the spectrum with Galactic latitude, such as \citet{fuskeland:2014}
and references therein.  Their results imply that the spectrum {\em
  within} the {\wmap} bands themselves hardens by $\Delta\beta =
0.14$ in the plane compared to the rest of the sky.  More recently, a
similar steepening of about $0.2$ in the microwave bands off the plane was
found by the QUIET project \citep{quiet:2015}.  Both find a steeper
index of $\beta\approx-3.1$ off the plane, consistent with what {\commander}
finds, while the index in the plane can be $\beta\approx-2.98$ 
\citep{fuskeland:2014} or as hard as $\beta\approx-2.9$ in the QUIET data.  \added{The
steepening seen by \citet{fuskeland:2014} is measured above $|b|\approx 15\degr$, as
they analysed the whole sky with a set of large regions.  The QUIET
analysis, however, found the steepening as close as $|b|>2.5\degr$
from the plane.  It is therefore unclear in how narrow a region the
microwave spectrum hardens.} 

Furthermore, evidence for the hardening of the synchrotron spectrum at
low frequencies in the plane comes from \citet{planck2014-XXIII},
which found that the synchrotron emission on the plane has a spectral
index in the radio regime of $\beta=-2.7$ between 408\,MHz and
2.3\,GHz.  This is a separate question to that of the difference
between the plane and the higher-latitude sky in the microwave bands.
\added{This paper also identifies two distinct synchrotron-emitting regions:
a narrow $|b|\approx1\degr$ component and a wider
$2\degr\le|b|\le4\degr$, which they interpret as having different
origins.  Their \addedC{hard} spectrum applies again to this very narrow region
along the plane. } 

If these two results are correct, i.e., that in the radio regime the
spectrum in the plane is $\beta=-2.7$ and in the microwave regime the
spectrum in the plane is $\beta=-2.9$, then the total effective
spectrum from 408\,MHz to 30\,GHz is $\beta\approx-2.8$ depending on
exactly where the turnover occurs.
%

This would imply that the {\planck} {\commander} synchrotron solution
underpredicts the synchrotron total intensity in the plane by nearly a factor of
4.  
%
%

The $\beta=-2.6$ hardening in the plane in the
{\wmap} solution, \addedD{however,} may partly be due to contamination by AME and free-free emission.  The former is not explicitly
included in the {\wmap} MCMC component separation used by {\jf}.  

\begin{figure}
\includegraphics[clip,trim=0 0.25cm 0 0]{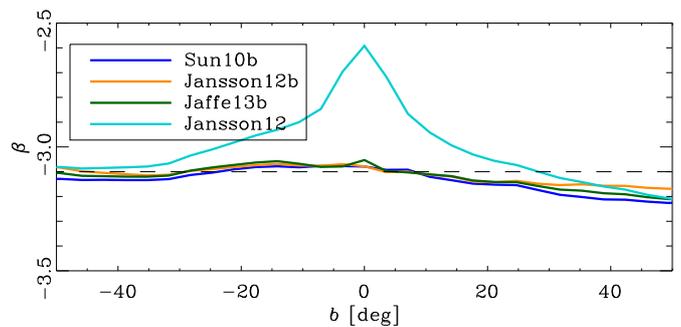}
%
\caption{ Effective synchrotron spectral index, $\beta$, between
  models at 30\,GHz and the data at 408\,MHz as averaged over latitude
  bins in the inner Galaxy ($-90\degr\leq\ell\leq90\degr$). 
\label{fig:lat_prof_beta_synch}
}
\end{figure}

Figure\,\ref{fig:lat_prof_beta_synch} shows the effective synchrotron
spectral index from the models at 30\,GHz to the data at 408\,MHz.  In
other words, we compute $\beta$ from the maps by averaging over the
inner Galaxy for each latitude bin and computing
$$
\beta=\frac{ \ln{(m_{30}/d_{0.408}}) }{ \ln{(30/0.408)} } .  
$$
The updated models are all around $\beta\approx-3.1$
with variations where the models do not quite match the morphology of
the data, particularly in the north around the NPS.\footnote{\added{Because we use the full {\galprop} {\cre} spatial distribution, not
  the single spectral template used in the {\commander} analysis, the 
  models do include a variation of the synchrotron spectral index on
  the sky of $\Delta\beta\lesssim0.05$.  This does not enter into our
  analysis, which is confined to a single synchrotron frequency, but if one took
  our resulting model and generated the prediction at 408\,MHz, it would differ
  from the 408\,MHz data due to these variations.}}  
The original {\jf} model (developed to fit
the {\wmap} MCMC synchrotron) implies a much harder index at low
latitudes.

Reality is therefore likely to be somewhere in between the steep
spectrum of the {\planck} {\commander} solution and the hard spectrum of
the {\wmap} MCMC solution.  The Galactic magnetic field models,
similarly, may be considered as bracketing reality.  The original
{\jf} model had too much {\random} magnetic field, while the updates
here based on {\planck} {\commander} results likely have too little.  If
this is indeed the case, it might explain why our residuals are much
larger than the model variance, i.e., the $(d-m)/\sigma$ plots in
Figs.\,\ref{fig:lat_prof_comp_synch} through \ref{fig:chisq_maps_synch}
have a large range; the models may well underestimate the expected
variance.

\citet{fuskeland:2014} also find that the synchrotron 
spectrum is hardest when looking tangentially to the local spiral arm
($\ell\approx\pm90\degr$) of the Galaxy and is steepest towards both the
Galactic centre and anti-centre.  Such a large variation \added{is not
reproduced by the {\galprop} model, implying something incorrect in the
spatial modelling of the CR injection or propagation.}  The difference is of order $\Delta\beta\approx 0.2$,
and this variation also affects the determination of the average
spectrum in the Galactic plane.  Taking it into account, the average spectrum
could then be as \addedC{hard} as $-2.85$ with a corresponding impact on the
implied synchrotron intensity in the plane.

\subsubsection{Radio loops and spurs}
\label{sec:loops}

\begin{figure}
\includegraphics[trim=0 0 1.2cm 0, clip,height=\linewidth,angle=90]{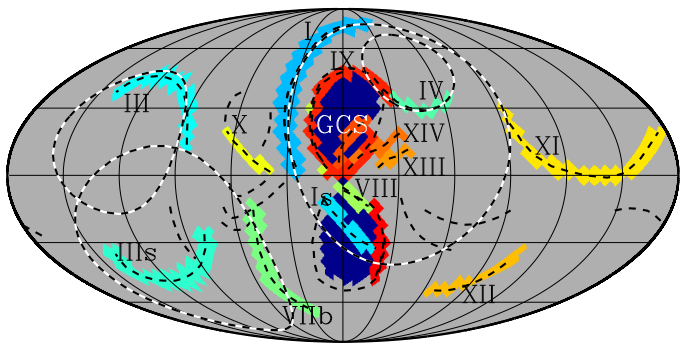} 
\caption{The regions masked for each of the loops and
  spurs defined by \citet{planck2014-a31} and \citet{vidal:2015} as well as for the region
  on the edge of (red) or inside of  (dark blue) the {\fermi} bubbles.  The dashed lines delineate the
  emission ridges (black) and the four  largest radio loops at their original locations (black and
    white).   
\label{fig:spurs_mask} 
}
\end{figure}

As pointed out in \citet{planck2014-a31}, the inner regions of the
Galactic plane show a thickened disc in total intensity that does not
have a counterpart in polarization.  The latter instead shows only a
thin disc and a set of loops and spurs that cross the plane.  \added{These
loops and spurs are indicated in Fig.\,\ref{fig:spurs_mask}, along
with the outline of the {\fermi} bubbles.  }

\begin{figure}
$\begin{array}{l}
\includegraphics[clip,trim=0 5mm 0 5mm]{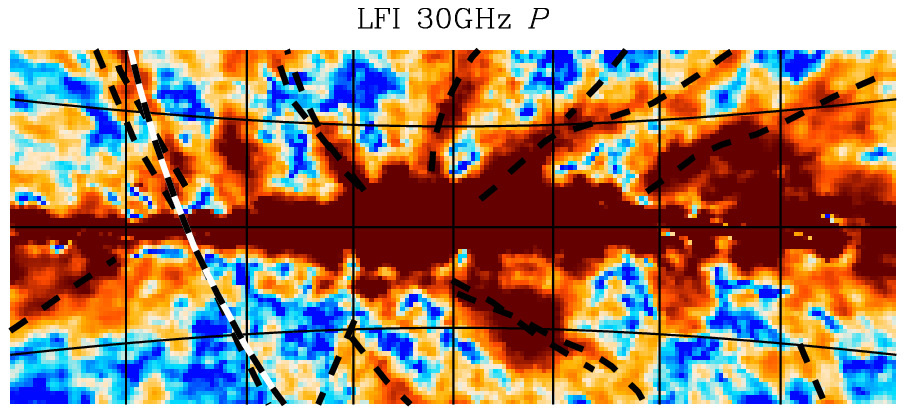} \\
\includegraphics[clip,trim=0 1cm 0 22cm,width=\linewidth]{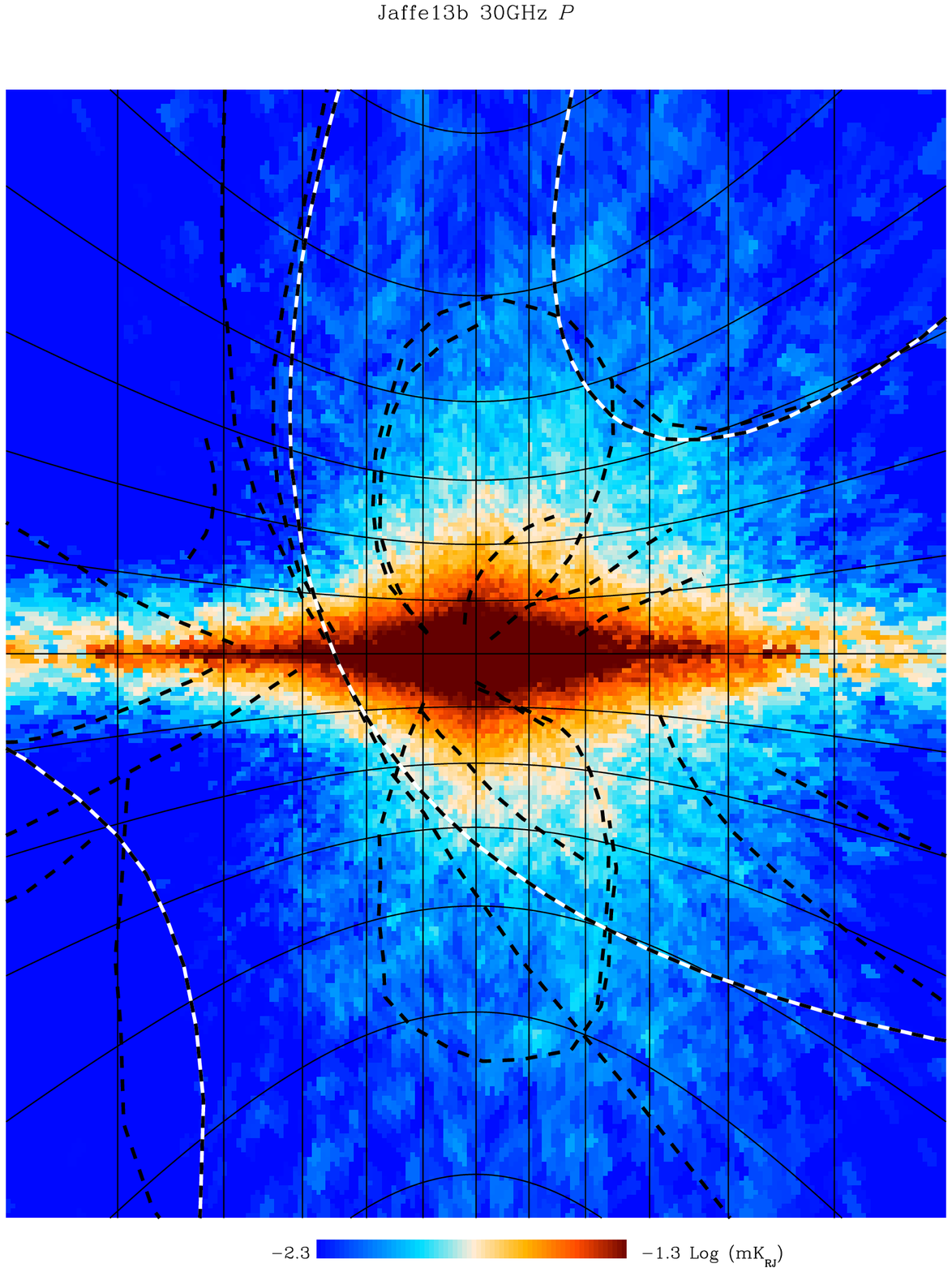} \\
\textrm{(a) LFI 30 GHz}\ \textit{P} \\ \\
\includegraphics[clip,trim=0 5mm 0 5mm]{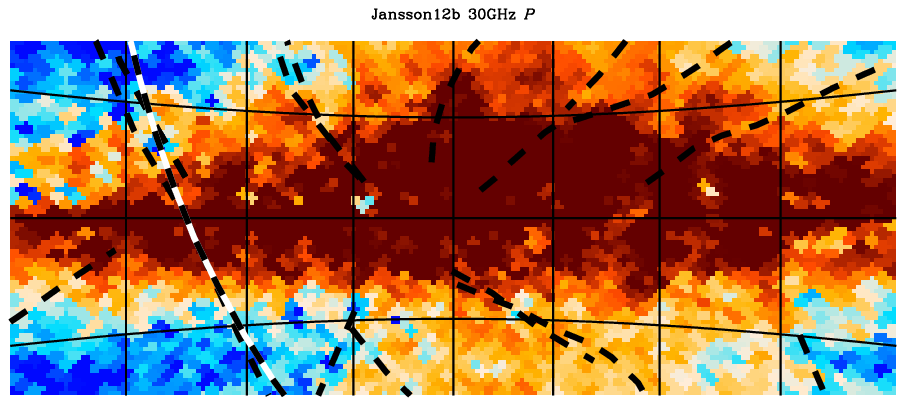} \\
\includegraphics[clip,trim=0 1cm 0 22cm, width=\linewidth]{figs/vidal_zoom_Jaffe13b_one2.ps} \\
\textrm{(b) Jansson12b 30 GHz}\ \textit{P} \\ \\
\includegraphics[clip,trim=0 5mm 0 5mm]{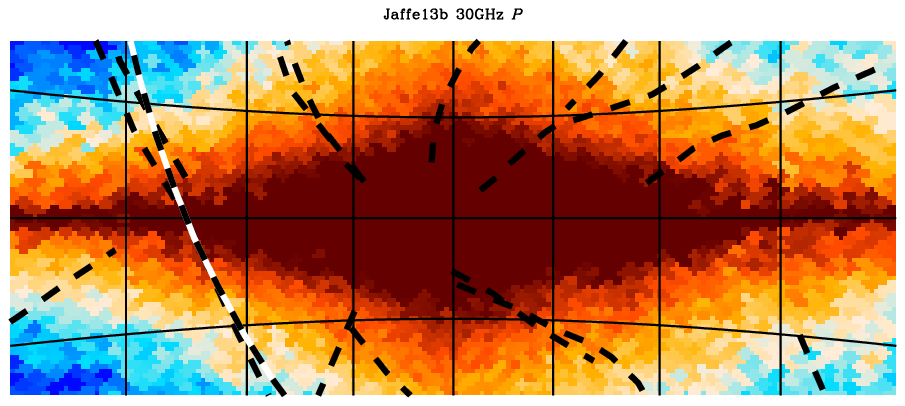} \\
\includegraphics[clip,trim=0 1cm 0 22cm, width=\linewidth]{figs/vidal_zoom_Jaffe13b_one2.ps} \\
\textrm{(c) Jaffe13b 30 GHz}\ \textit{P} \\
\end{array}$

\caption{ Zooms centred on the inner Galactic region (in a Gnomonic
  projection with a $10\degr$ grid) for the $P$ of the data
  (top) and two of the models (one realization of each).  The dashed lines are the spurs
  and loops as in Fig.\,\ref{fig:spurs_mask}.  (Recall that in the
  profile plots, the region where $|\ell|<10\degr$ and $|b|<10\degr$
  is masked.) 
\label{fig:spurs_zooms}  
}
\end{figure}

\added{Figure\,\ref{fig:spurs_zooms} shows a zoom of the inner Galaxy in
synchrotron polarized intensity for the data at 30\,GHz and for
two of the models.  The ridges of the spurs and loops as defined in
\citet{planck2014-a31} are
over-plotted.   (See also \citealt{vidal:2015}.)  The thickness of the disc visible in polarized
emission between the spurs is clearly narrower \addedC{in the data}
than in the models.  The
latitude profiles in Fig.\,\ref{fig:lat_prof_comp_synch} that show a
rough match for the data when averaged over a broad range in longitude
are therefore somewhat misleading, as they average over
these structures as well.  The ordered fields may
be distributed in a narrower disc than the current models.}

We test the effect of removing the brightest parts of these features by
applying the mask shown in Fig.\,\ref{fig:spurs_mask}.  \added{This is a
downgraded version of the mask shown in figure~2 of \citet{vidal:2015}
and includes a mask for the edges of the {\fermi} bubbles.}   \added{We show the
profiles in latitude when excluding these regions as the
dashed lines in Fig.\,\ref{fig:lat_prof_comp_synch}.  For the
longitude profiles, the masking would exclude the regions denoted by
the two vertical bands in Fig.\,\ref{fig:lon_prof_comp_synch}.} 

\added{ Two regions in the Galactic plane are removed by this mask:
  the region near $\ell\approx 30\degr$ where the NPS intersects the
  plane and another from $-160\degr\lesssim\ell\lesssim-110\degr$.  In
  the first region, the models overpredict the signal significantly compared to
  the data.  This region may be depolarized due to the fact that the
  orientation of the polarization in the spur is perpendicular to that
  of the diffuse emission in the plane, and there is a cancellation
  along the LOS.  In the second region at negative longitudes,
  the loop is roughly parallel to the plane where it
  intersects, and so a similar structure should co-add rather than
  cancel with the diffuse emission, but little effect is seen.  The
  latitude profiles show how much the emission both to the north and
  south is reduced by the exclusion of the brightest ridges.  The
  significance of the residuals drops, which is unsurprising when we
  mask out bright and clearly localized regions not reproduced by the
  models, but the residuals remain higher in the north than the south.}

These comparisons show how the presence of the loops and spurs can
affect the \added{large-scale modelling by either cancelling} or
adding polarized emission to the diffuse component.  \added{Though we
  can mask the brightest of these features, the interiors also contain
  emission that is visibly related to the loop.  In the case of the
  NPS, the interior likely covers a significant fraction of the sky
  and may have a substantial effect on attempts to fit large-scale magnetic
  field models.
}

%

\section{Dust modelling} 
\label{sec:dust}

We now take the models whose synchrotron emission we have examined
above and look at the predicted dust emission and compare it to the
{\planck} observations in total and polarized intensity at 353\,GHz.  

\added{Fauvet12 performed the first fitting of the large-scale
field to dust emission.}  They 
fitted a magnetic field model to both polarized synchrotron emission
and polarized thermal dust emission using the {\Wmap} and {\Archeops}
data, respectively.  One important aspect of their analysis is the use
of intensity templates to account for localized variations.  Fauvet12
used the simple exponential distributions of the magnetized ISM
components and {\hammurabi} to create maps of Stokes parameters, but
these were not directly compared to observations.  Instead, for
synchrotron and dust emission, they multiplied an observed total
intensity template by the simulated polarization fraction in order to
simulate the polarization data.  This means that the assumed particle
distributions did not have to be very accurate, and yet the resulting
simulation of polarization could be made to match well.  In order
to unambiguously constrain the large-scale properties of the field, \addedD{however,} we
prefer to directly compare the morphologies
of models and data.

\subsection{Dust distribution and polarization properties}
\label{sec:dust_model}

For the computation of the thermal dust emission, we require a model
for the spatial distribution of the dust.  We start from the
dust distribution model of Jaffe13, which is similar to
that of \citet{drimmel:2001} in its parametrization.  The parameters
required updating, particularly the scale heights, which had not been
constrained in Jaffe13, since that analysis was confined to the
plane.  

We find that using that parametrization with two scale heights, one
for a thicker axisymmetric component and one for a narrower spiral arm
component, \added{fits the low-latitude emission but 
overpredicts the emission for $|b|\gtrsim 30\degr$}.  This is likely due to the
absence of dust emission in the Solar neighbourhood studied by, e.g.,
\cite{lallement:2014} and references therein.  We therefore add a
feature to our dust distribution:  a cylinder centred on the
Sun's position with a configurable radius, height, and dust density
damping factor.  \added{We set the dust density to zero within a region of
150\,pc in radius and 200\,pc in height.  This cylinder is a crude
approximation to the ``local bubble'', a tilted low-density region studied in
\citet{lallement:2014}.}  The amount of dust left in this region
relative to the large-scale model is largely degenerate with its
extent, and we do not attempt a detailed modelling here.  \added{This
  is the only such small-scale structure in the model, but it is
  apparently necessary because of the large effect such a local 
  structure can have on the high-latitude sky.  Elsewhere, the model is 
  effectively an average over a Galaxy full of such small-scale
  structures that the analysis is not sensitive to.}  

We also adjust several parameters to fit the longitude profile \addedB{of the
dust intensity} along
the plane because the morphology is not quite the same at 353\,GHz
\addedB{(used here)} as at
94\,GHz \addedB{(used in Jaffe13)}.  In particular, we damp the two outer arms relative to the two
inner arms, and we reduce the scale radii for both the smooth and spiral
arm components.  

This leads to the model that approximately matches the data, as shown in Figs.\,\ref{fig:lat_prof_comp_dust}
through \ref{fig:chisq_maps_dust}.  \added{As with the magnetic field models, a complete
  exploration of the parameter space is not
  performed here.  This would, for example, improve the locations and amplitudes of
  the shoulder-features in the longitude profiles seen in
  Fig.\,\ref{fig:lon_prof_comp_dust}.  This distribution would be interesting to
  compare to the original \citet{drimmel:2001} model, since the older
  model was based on {\Iras} data at higher frequencies, and as we see
  in M31 \citep{planck2014-XXV}, the apparent profile of the dust
  emission depends on frequency.  The analysis here, \addedD{however,} is not
  sensitive to the details of this distribution, since the 
  uncertainties in the magnetic field modelling are larger than the
  uncertainties in the dust distribution.  } 
This approximate model is sufficient to study the degree of
polarization and the implications for the magnetic fields towards the
inner Galaxy.  The outer Galaxy latitude profiles show that the model
is too narrow, but we focus on the inner Galaxy, where we are looking
through most of the disc.  

We give an updated table of
our dust model parameters in Table\,\ref{tab:params_dust}.  

The polarization is modelled as described by Fauvet12, where
the degree of polarization drops with the angle to the LOS, $\alpha$,
as $\sin^2\alpha$.  In this case, we omit the additional factor of
$\sin{\alpha}$ that was used there as an approximation for grain
misalignment, but it does not make a significant
difference to the results.   The intrinsic (i.e., sub-grid) polarization fraction is set to
$20\,\%$ following the results of \citet{planck2014-XIX}.  \added{This
  is the lower limit for the {\em maximum} polarization fraction observed in the diffuse
  emission by {\planck}.  In our modelling, this parameter folds in all sub-grid
  effects, i.e., any variations in the polarization properties on
  small scales and any correlations between the polarization
  properties and the dust density or emissivity.  This parameter is a
  large systematic uncertainty in the analysis and can effectively
  scale the polarization independently of all other parameters.}

\subsection{Dust predictions from synchrotron-based models}
\label{sec:dust_results}

\begin{figure*}
\includegraphics[clip,trim=0 0 0 0.25cm]{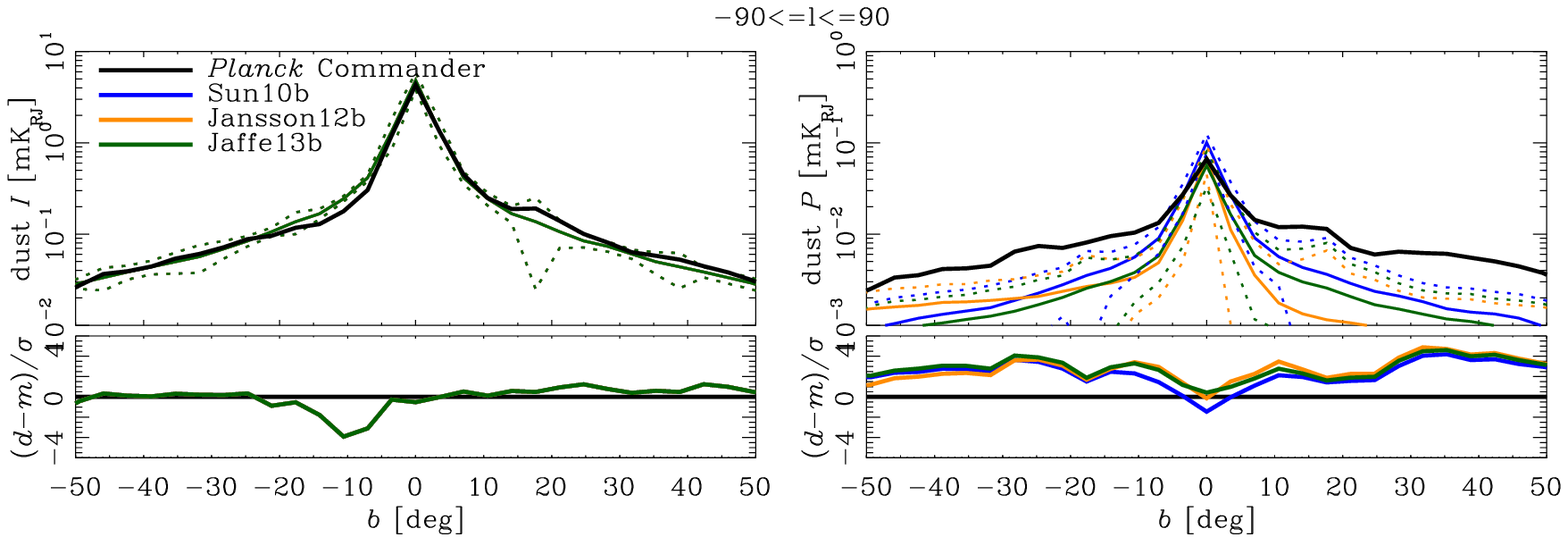}
\includegraphics[clip,trim=0 0 0 0.25cm]{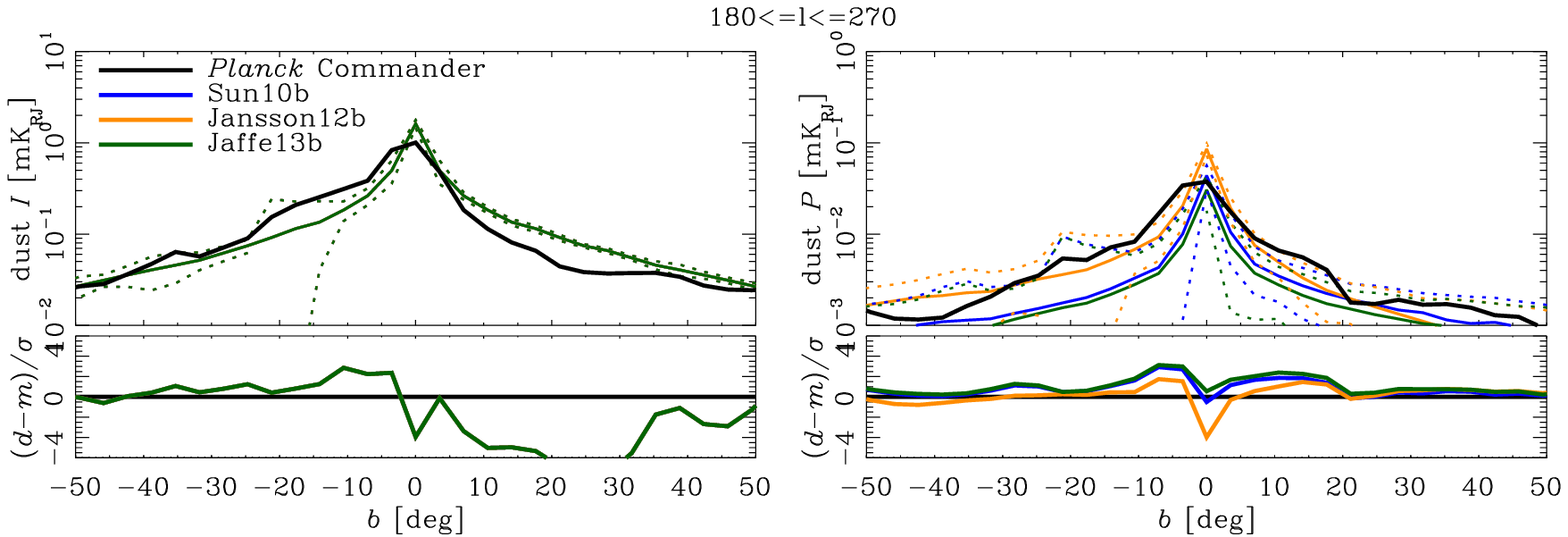}
\caption{ \added{  Dust profiles in latitude for the models as in
  Fig.\,\ref{fig:lat_prof_comp_synch}.   The top shows the inner Galaxy (i.e.,
  $-90\degr<\ell<90\degr$), while the bottom shows the third
  quadrant ($180\degr<\ell<270\degr$, i.e., the outer Galaxy excluding
  the Fan region).  For total intensity on the left, the
  {\commander} dust solution in solid black is 
  compared to the {\planck} 2013 dust model of
  \citet{planck2013-p06b} (black dashed), but the difference is
  not generally visible.  (Since the models shown in colour differ
  only in the magnetic field, which has no impact on the dust total
  intensity, these curves are not distinguishable.) }
\label{fig:lat_prof_comp_dust}
}
\end{figure*}

\begin{figure*}
\includegraphics[]{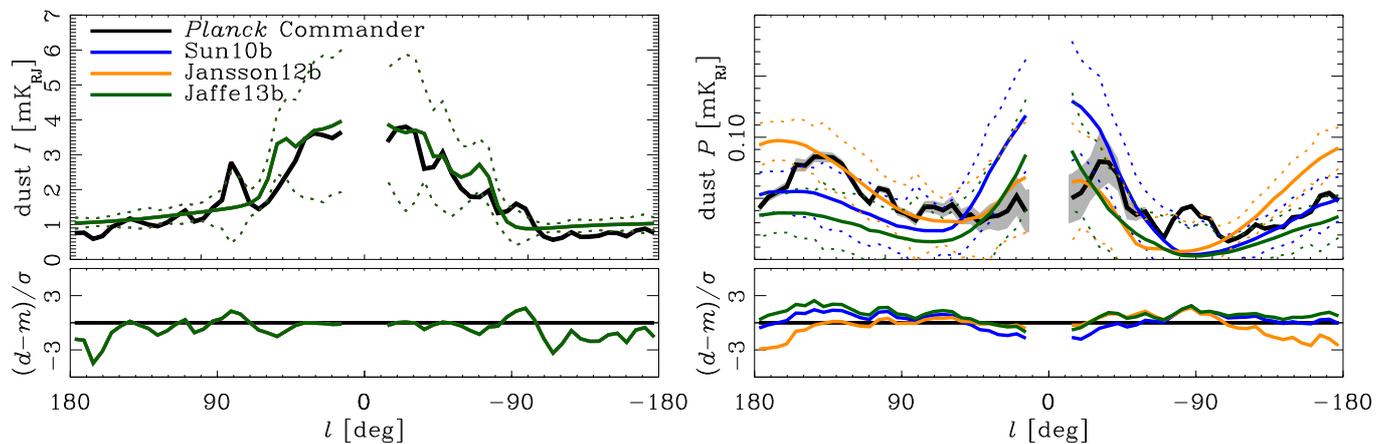}
\caption{ Dust profiles in longitude for the updated models in
  Fig.\,\ref{fig:lat_prof_comp_dust}.  \addedB{The grey band shows the
    uncertainty in the data due to the bandpass leakage discussed in
    \S\,\ref{sec:appendix_systematics_hfi}.}
\label{fig:lon_prof_comp_dust}
}
\end{figure*}

\begin{figure*}
$\begin{array}{cccc}

& { I} & { Q} & { U} \\

& \multicolumn{3}{c}{      \rmn{Data, }d} \\

& 
\includegraphics[trim=0.2cm 0 0.3cm 0, height=0.28\linewidth,angle=90,clip]{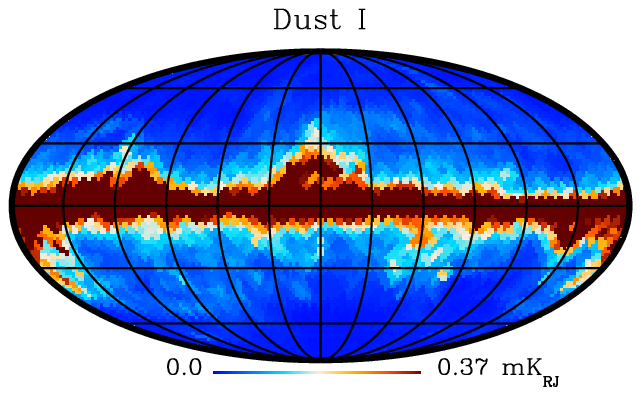} &
\includegraphics[trim=0.2cm 0 0.3cm 0, height=0.28\linewidth,angle=90,clip]{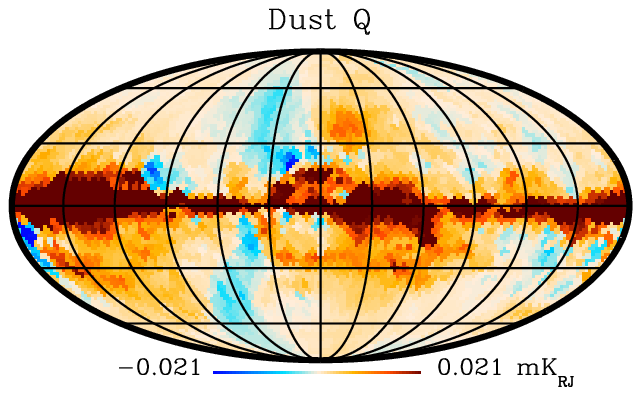} &
\includegraphics[trim=0.2cm 0 0.3cm 0, height=0.28\linewidth,angle=90,clip]{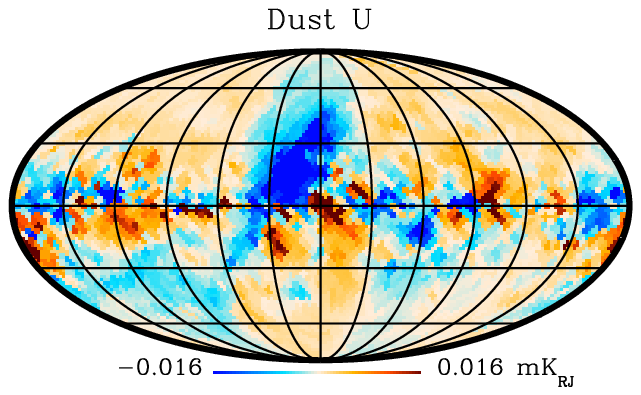} \\

& \multicolumn{3}{c}{      \rmn{Models, }m} \\

\rotatebox{90}{{\hspace{0.5cm} Sun10b}} & 
\includegraphics[trim=0.2cm 0 0.3cm 0, height=0.28\linewidth,angle=90,clip]{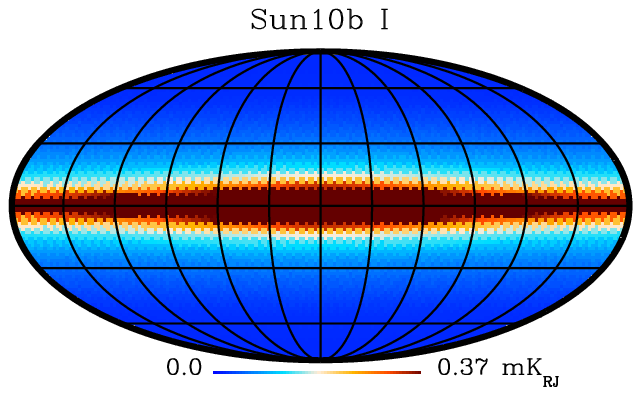} &
\includegraphics[trim=0.2cm 0 0.3cm 0, height=0.28\linewidth,angle=90,clip]{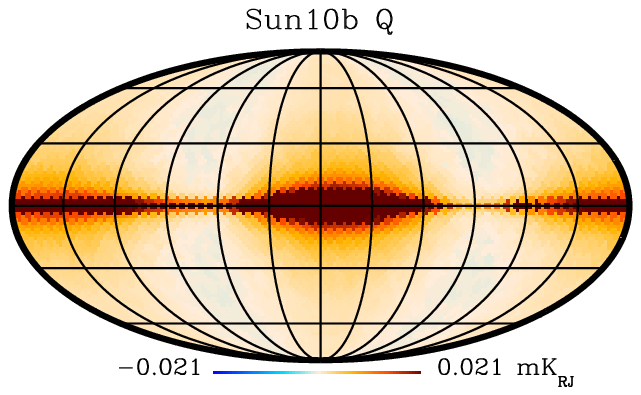} &
\includegraphics[trim=0.2cm 0 0.3cm 0, height=0.28\linewidth,angle=90,clip]{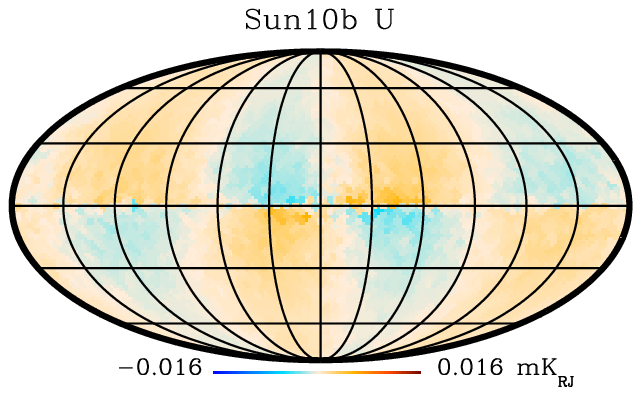} \\

\rotatebox{90}{{\hspace{0.5cm} {\jf}b}} & 
& 
\includegraphics[trim=0.2cm 0 0.3cm 0, height=0.28\linewidth,angle=90,clip]{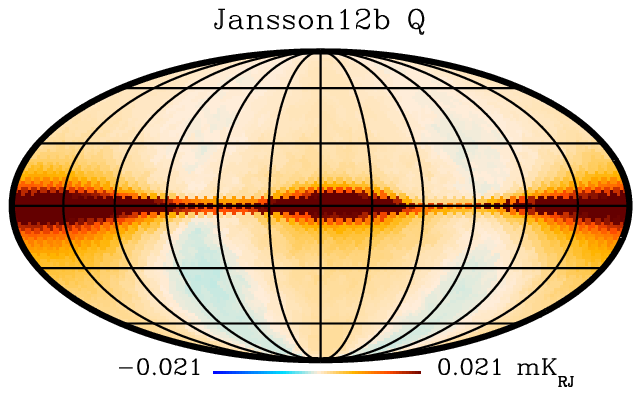} &
\includegraphics[trim=0.2cm 0 0.3cm 0, height=0.28\linewidth,angle=90,clip]{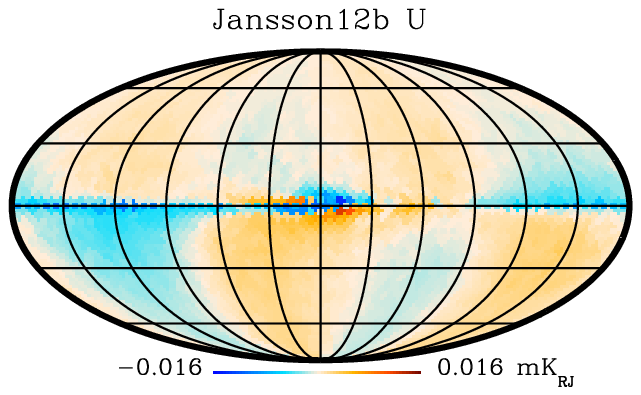} \\

\rotatebox{90}{{\hspace{0.5cm}  Jaffe13b}} & 
& 
\includegraphics[trim=0.2cm 0 0.3cm 0, height=0.28\linewidth,angle=90,clip]{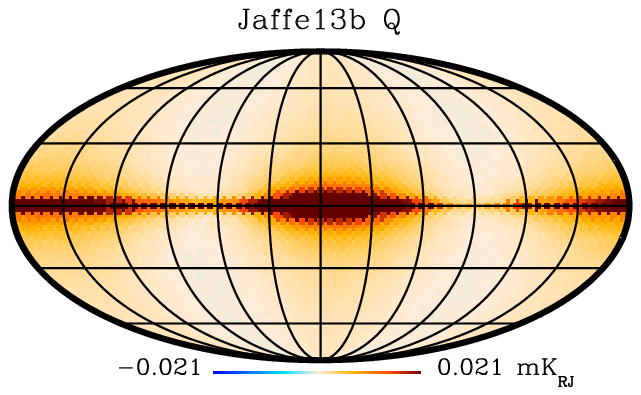} &
\includegraphics[trim=0.2cm 0 0.3cm 0, height=0.28\linewidth,angle=90,clip]{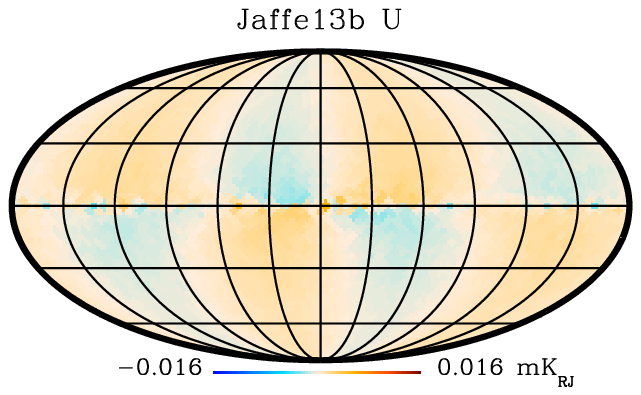} \\

&  \multicolumn{3}{c}{ \rmn{Residuals, } (d-m)/\sigma } \\

\rotatebox{90}{{\hspace{0.5cm} Sun10b}} & 
\includegraphics[trim=0.2cm 0 0.3cm 0, height=0.28\linewidth,angle=90,clip]{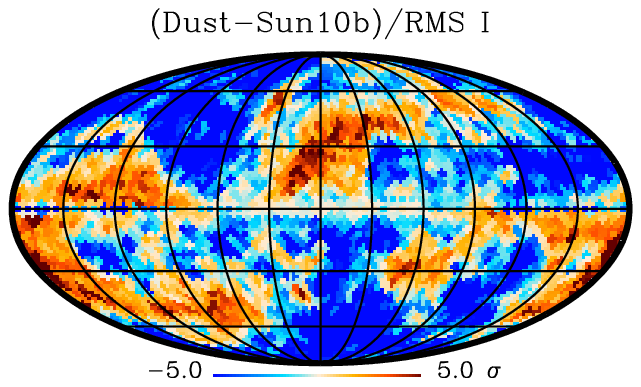} &
\includegraphics[trim=0.2cm 0 0.3cm 0, height=0.28\linewidth,angle=90,clip]{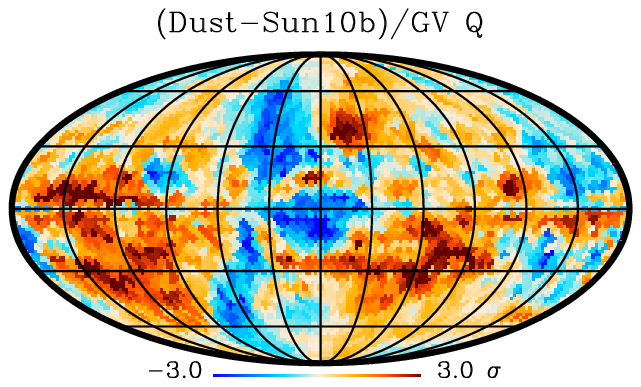} &
\includegraphics[trim=0.2cm 0 0.3cm 0, height=0.28\linewidth,angle=90,clip]{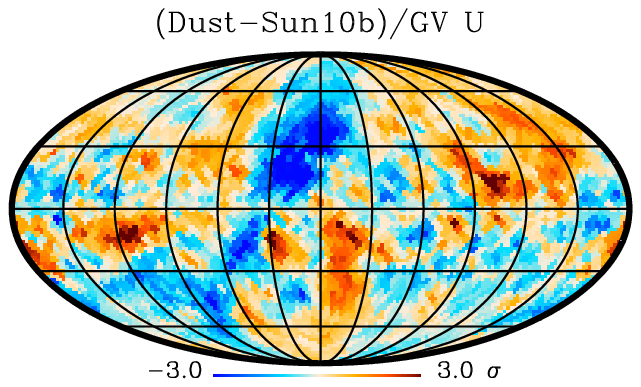} \\

\rotatebox{90}{{\hspace{0.5cm} {\jf}b}} & 
& 
\includegraphics[trim=0.2cm 0 0.3cm 0, height=0.28\linewidth,angle=90,clip]{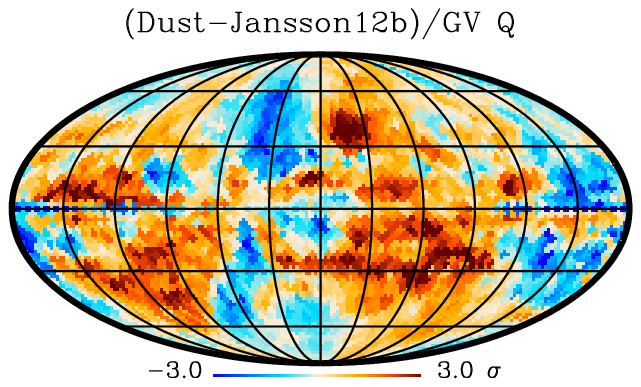} &
\includegraphics[trim=0.2cm 0 0.3cm 0, height=0.28\linewidth,angle=90,clip]{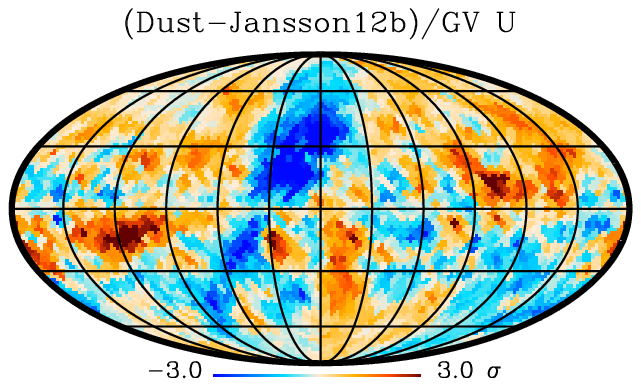} \\

\rotatebox{90}{{\hspace{0.5cm}  Jaffe13b}} & 
& 
\includegraphics[trim=0.2cm 0 0.3cm 0, height=0.28\linewidth,angle=90,clip]{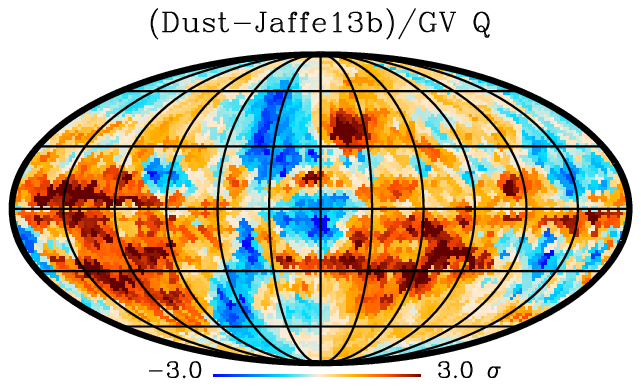} &
\includegraphics[trim=0.2cm 0 0.3cm 0, height=0.28\linewidth,angle=90,clip]{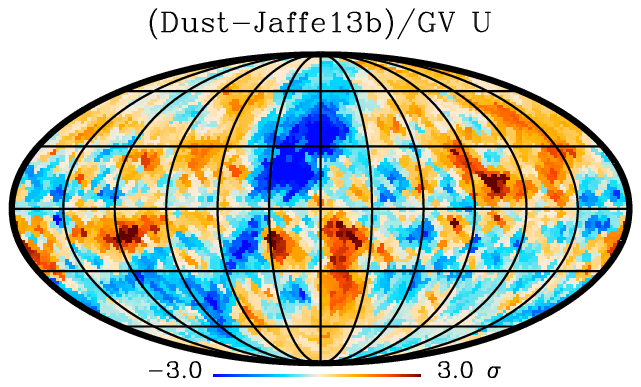} \\

\end{array}$
\caption{ Comparison of the model predictions for dust and the
  {\Planck} data.   The columns from left
  to right are for Stokes $I$, $Q$, and $U$, while the rows are the
  data followed by the prediction for each model, and lastly the
  difference between model and data divided by the uncertainty.  
  \added{ As for synchrotron emission, the polarization uncertainty is
  computed as the ensemble variance predicted by the models.  
  Since dust total intensity is not a function of the magnetic field,
  its uncertainties are computed from the sky rms
  of the dust map.} 
\label{fig:chisq_maps_dust}
}
\end{figure*}

Figures\,\ref{fig:lat_prof_comp_dust} and \ref{fig:lon_prof_comp_dust}
show latitude and longitude profiles, while
Fig.\,\ref{fig:chisq_maps_dust} compares the maps.  

\subsubsection{Profiles and large-scale features}

The latitude profiles in the inner Galaxy shown in
Fig.\,\ref{fig:lat_prof_comp_dust} (top) show that the predicted total
intensity (left) has approximately the right thin-disc thickness in
the inner Galaxy for $|b|\lesssim 10\degr$, but the polarized
intensity (right) is too narrow.  The synchrotron profiles in
Fig.\,\ref{fig:lat_prof_comp_synch} show no such disparity, so this
mismatch in the dust polarization must then be due to a change in the
magnetic fields in the narrow disc where the dust is found but where
the synchrotron is less sensitive.

The dust intensity in the outer Galaxy (Fig.\,\ref{fig:lat_prof_comp_dust} bottom) is \addedC{also}
mis-matched in polarization at low latitudes, but because it is less
well matched in total intensity and in synchrotron emission \addedC{in
  this region}, we cannot
draw any conclusions there.

If this difference is significant (which depends on the expected
GV, a function of the field ordering, in turn a
function of the component separation, etc.) then it points to a
problem in the degree of field ordering in the model as a function of
height in the thin disc.  The model may be correct on the larger scales
probed by the thick synchrotron disc and \addedB{yet} have too \addedB{thinly distributed} an ordered
field component in the disc \addedB{as traced by the dust}.  

At intermediate to high latitudes ($|b|\gtrsim 10\degr$), all models
underpredict significantly the polarized intensity of dust.  

As in Jaffe13, the longitude profiles along the Galactic plane show
that the inner regions of the Galaxy
($-30\degr\lesssim\ell\lesssim30\degr$) are overpredicted in
polarization for some of the models.  Unlike in that work, however, some
updated models can be made to reproduce roughly the right level of
polarization emission in parts of the outer Galaxy.  (The {\jf} model
shows the most extreme case where the fields are completely ordered in much
of the dust-emitting regions, and this then overpredicts the
polarization \addedC{toward the anti-centre}.)  Again, this is due to the changes
in the degree of field ordering that are necessary to fit to the
{\commander} synchrotron solution discussed above.  Since it is not clear
that this change is physically realistic, it is possible that the
fields are more disordered than in our current models tailored to this
synchrotron estimate.  It is therefore not clear that the mismatch discussed by Jaffe13 
between the degree of polarization in the dust and
synchrotron emission in the plane has been resolved.

These issues will be discussed further in \S\,\ref{sec:svd}.  


\subsubsection{Maps and local residuals}  

The left column and fifth row of
Fig.\,\ref{fig:chisq_maps_dust} shows the residuals for the model
dust prediction compared to the 353\,GHz data in total intensity.
Since dust total $I$ is not a function of the magnetic field,
what is seen is only the distribution of the dust emissivity model as
described in \S\,\ref{sec:dust_model}.  For the significance of the
residuals, we divide by the sky rms estimate of the galactic variance
(see \S\,\ref{sec:galactic_variance}) computed on the dust total
intensity map.  

The residual map in total intensity highlights several known
nearby regions by removing the background
Galactic disc.  For example, above the Galactic centre is a strong arc
of emission from the Aquila Rift up through the Ophiuchus region.  
\added{ Since the dust is in the thin disc, the emission at $|b| >
  10\degr$ is very close and included in the Gould Belt system as
  determined using \ion{H}{i} velocities and mapped by \citet{lallement:2014}.
  The Aquila Rift, for instance, is known to be about 80 to 100\,pc
  away (see, e.g., the starlight polarimetry of \citealt{santos:2011}).  This
  is clearly on the ``wall'' of the local cavity, possibly also on the
  swept-up shell of Loop I (even if its centre is at the larger
  distance as per \citealt{planck2014-XXV}).  We also see that the
  intensity minima are not at the Galactic poles but tilted, which is
  consistent with the tilted local ``chimney'' from
  \citet{lallement:2014}.}

The Fan region is also quite distinctly
visible near the plane in the second quadrant.  The model for the dust distribution includes spiral arm
components, so this map shows how the Fan region is bright in dust
emission even on top of the prediction from the Perseus arm ridge.

In polarization, these residuals pick out strongly some of the
features visible in \citet{planck2014-XIX} such as the strong diagonal stretch of
Stokes $U$, implying that the magnetic field is somewhat aligned along
this feature.  This structure is not only associated with the arc above the
Galactic centre but also appears to continue eastwards below the plane.
The Fan region is also quite bright and not accounted for by the
spiral arm model of the dust distribution or the magnetic fields.  

\added{  These structures are well inside the outer
  scale of the magnetic field turbulence and cannot be modelled by the
  methods we use here.}  

\subsection{{\jf}c:  a dust-based magnetic field model}
\label{sec:dust_field_models}

All models optimized for synchrotron underpredict dust
polarization for $|b|\gtrsim 5\degr$.  This is despite the fact that the latitude profile of the
dust total intensity matches observations, and the synchrotron
latitude profiles match in both total and polarized intensities.  As
discussed by Jaffe13, one degeneracy in the synchrotron
modelling is the precise relative distributions of the coherent and
random fields.  The {\cre} distribution is thought to be fairly
smooth, while the dust is thought to be concentrated in
a thin disc with annular and spiral arm modulations.  (This has been modelled in the Milky
Way by \citealt{drimmel:2001} and can be seen directly in {\planck}
observations of M31 in \citealt{planck2014-XXV}.)  Therefore, the dust polarization can be used to
study precisely where the magnetic fields are more or less ordered
relative to these arms and relative to the mid-plane of the Galaxy.

We choose to use the {\jf} model because it has an easy
parametrization for distinct morphological components of the coherent 
and random fields, particularly the disc, halo, and x-shaped components and
the spiral segments.  The high-latitude dust polarization is a function
of what is going on in both the local arm segment (we are situated
near the inner side of segment five) and the next segment inward
(number four, which dominates the high-latitude sky looking towards the
inner Galaxy).  One peculiarity of this model is the presence of jumps
between different \added{spiral} segments in the narrow disc and between the
narrow disc and the thick-disc toroidal component.  The dust latitude
profiles towards the inner and outer Galaxy are each very sensitive to
the details of these transitions because the dust is
so narrowly distributed and therefore all emission above $10\degr$ is
very local.  (This is not the case for synchrotron, which is not as
sensitive to these properties.)  In order to simplify the adjustments
needed to match the data, we shift the arm pattern slightly so that
the region around the observer is located fully within segment five
and segment four does not impact the high-latitude emission.


As shown in Fig.\,\ref{fig:lat_prof_comp_dust}, the
Jansson12b model developed to match synchrotron underpredicts the
polarization at high latitudes, as do all the models.  We therefore decrease the random
component in the local arm segment (number five) and increase its
coherent field amplitude in order to increase the intermediate-latitude polarized intensity.  The synchrotron comes from a much
thicker region, so it is more dependent on the toroidal thick-disc
component and does not change significantly with this adjustment.  

\begin{figure*}
 \includegraphics[clip,trim=0 0 0 0.25cm ,width=\linewidth]{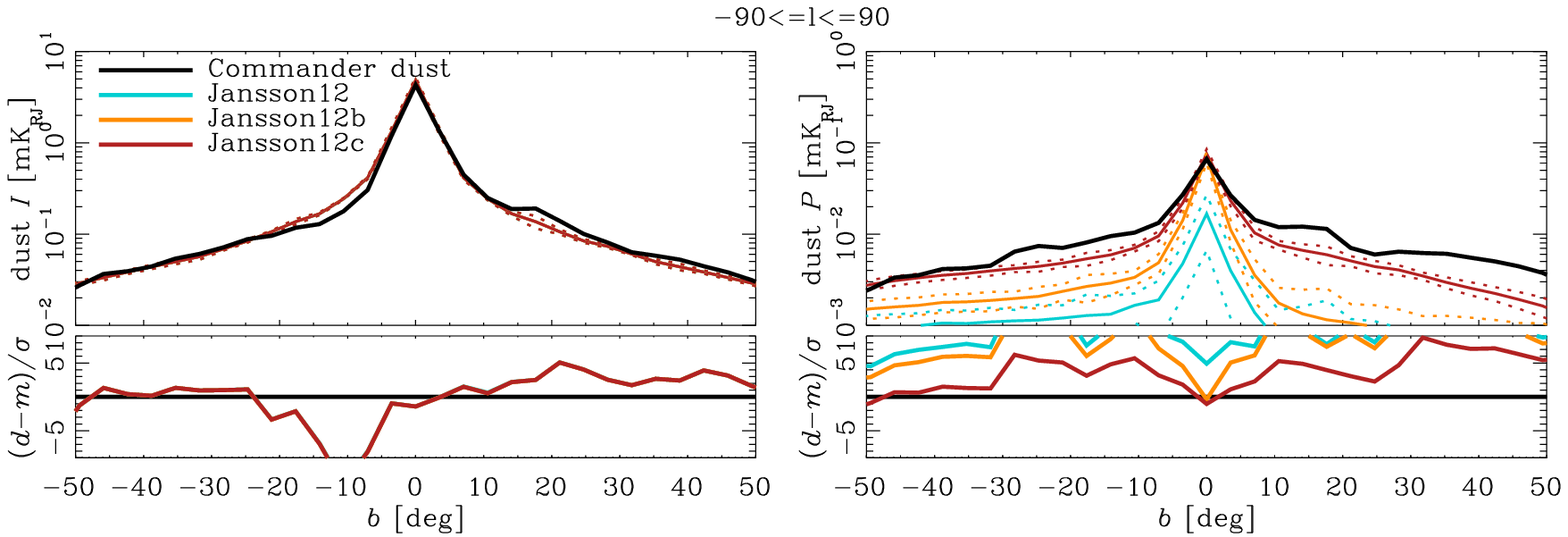}
 \includegraphics[width=\linewidth]{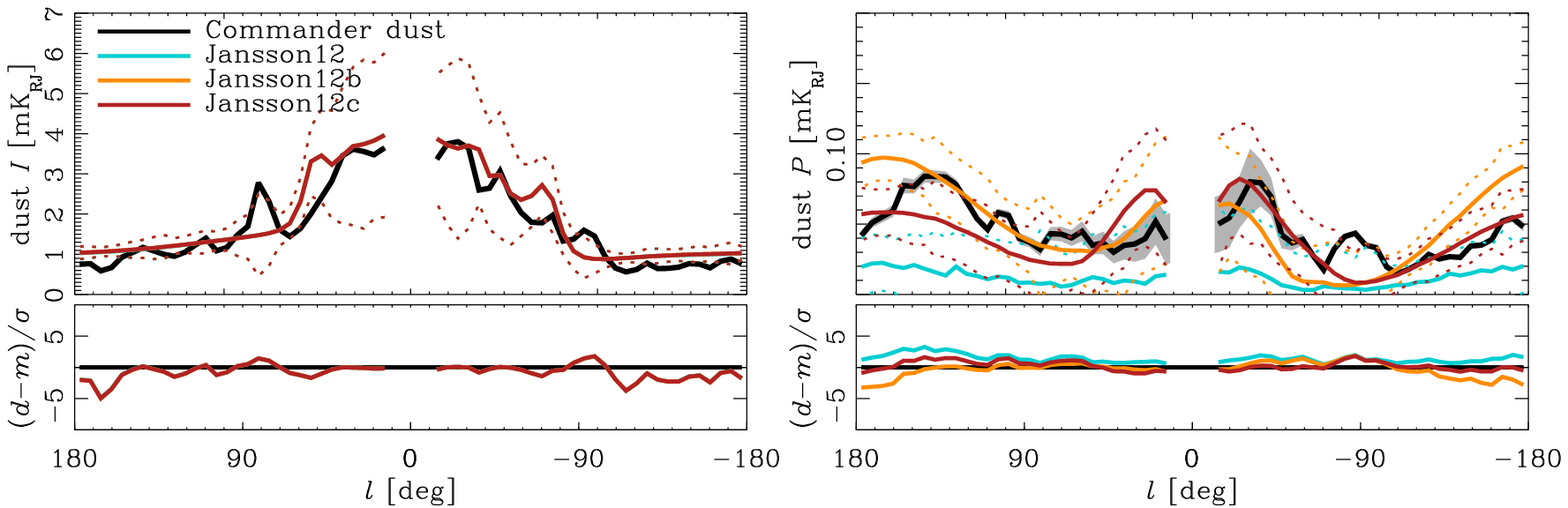} \\
\caption{ Dust latitude profiles for the three {\jf} models.  The top shows the inner Galaxy (i.e.,
  $-90\degr<\ell<90\degr$), while the bottom shows the longitude
  profile along the Galactic plane. 
\label{fig:lat_prof_comp_dust2}
}
\end{figure*}

We refer to these further adjustments as ``{\jf}c'' in
Fig.\,\ref{fig:lat_prof_comp_dust2}.  It is clear that a thorough
exploration of the parameter space would find a better model to fit all
of the available data, but the point of our {\jf}c model is not to
present the definitive solution to the problem; it is to show how the
dust and synchrotron can be used together to reduce, though not
eliminate, degeneracies in the parameter space.

 %

\subsection{Outstanding issues}
\label{sec:svd}

Since neither the {\jf}b model adjusted to match synchrotron nor the
{\jf}c model adjusted to match the dust is a best-fit model, their utility is
primarily as examples of how synchrotron and dust can be used to
probe the fields in the thin and thick discs traced by the two
components with very different scale heights.  These methods, however,
are only as reliable as the observations on which they are based, and
they are of course affected by local structures that the models cannot
take into account.  The differences between the two models are motivated by the fact that
the synchrotron predictions of the first model match the data profiles as a function of
latitude, but the dust profiles do not.  It is not clear, however,
how certain this mismatch is and therefore whether either this
scenario or the ``fix'' in {\jf}c is realistic.

\addedC{The main} limitation of this analysis, as discussed above, is the
uncertainty in the synchrotron spectrum, which is thought to harden in
the Galactic plane in a way not accounted for here.  If the
component-separation procedure included such a spectral hardening near the plane (e.g.,
based on additional information from additional surveys at GHz
frequencies), then this would increase the predicted synchrotron total
intensity in the plane.  (There would then be a corresponding decrease
in the ill-constrained AME component.)  This in turn would require a
{\it decrease} in the degree of magnetic field ordering near the plane
relative to high latitudes.  \addedC{Such a model in the
  plane would resemble that of Jaffe13, which assumed a similar
  synchrotron spectrum.  That work shows that this assumption leads to
an underprediction of the dust polarization in the plane.}  

\addedC{Once the 3D magnetic field models were adjusted to match
such a model of varying synchrotron spectrum, the latitude profile of
dust polarization would be less peaked.  The model would underpredict
for all latitudes, though the precise shape of the profile would
depend on that of the varying synchrotron spectrum. } \added{The entire curve could be
shifted up to match the data by increasing the sub-grid dust
polarization fraction.  That parameter is currently set at $20\,\%$ but
is highly uncertain and, as mentioned above, is a lower limit on the maximum.}

This line of reasoning suggests two things.  Firstly, the 
\addedC{polarized dust emission that is very peaked near the plane} may be an
artefact of the uncertain component separation and synchrotron
spectral index.  \addedB{In other words}, it may not be real.  Secondly, the high degree
of dust polarization predicted by the models in the Galactic plane may also not
be real.  \addedB{It is therefore} not clear whether the problem described by
Jaffe13 \added{(the difficulty reproducing the high level of dust
polarization in the plane) remains or whether the {\jf}c model is
roughly correct.  \addedC{As discussed in Sect.\,\ref{sec:synch_spec},} a variation in the synchrotron spectrum is supported by other observations
of (and plausible reasons for) a \addedC{hardening} of the synchrotron index in
the plane, but it would contribute to the problem \addedC{of
  underpredicted dust polarization} by
decreasing the dust polarization in the plane unless renormalized.
The increase in the intrinsic dust polarization
fraction \addedC{required} to match the data with more disordered fields in the plane
would be quite large, at least a factor of two, which \addedB{is
  unlikely given that even relatively nearby isolated clouds do not
  approach such values;  see, e.g., \citet{planck2014-XIX}.}  In short, these problems 
remain unresolved}, \addedC{and the discrepancy between the
synchrotron and dust polarization degrees is likely to require a more
sophisticated model for the ordering of the magnetic fields in the
thin and thick discs.}
 
Recall also that in order to fit the latitude profile of the dust
total intensity, we implemented a model for the local bubble as a
cavity of radius 150\,pc and height 200\,pc with no dust inside.
Since the dust is confined to a very narrow disc, this removes most of
the dust in the Solar neighbourhood, but there remains high-latitude
dust clearly visible in the logarithmic latitude profile in total
intensity, and yet this emission is not strongly polarized in the
models.  \added{Those models do not include any effect on polarization of
such a bubble.}   One could imagine a scenario wherein the process that
created the local bubble and evacuated much of the dust from the solar
neighbourhood also left a shell of ordered magnetic fields that might
retain enough dust to explain this mismatch.  Such a local phenomenon
would not be reflected in the synchrotron emission that traces a much
thicker disc.  \addedC{To resolve this will require further
  observations that constrain the dust and field distributions in the
  solar neighbourhood, such as using the velocity information from HI
  observations, in addition to more local starlight polarization measurements.} 

Lastly, we see that the inner plane is not well fitted in either
synchrotron or dust emission.  For $10\degr\lesssim|\ell|\lesssim20\degr$, the models
\added{have a roughly similar synchrotron amplitude on average}, but it is
apparent that the polarized synchrotron emission here is climbing
rapidly in a way that the models do not reflect.  By contrast, the
dust polarization is overestimated by most models in this region.  We
do not attempt to model the innermost Galaxy, but clearly the
modelling of the region within the molecular ring is incorrect.
This is a complicated region likely affected by the Galactic bar, 
changes in the star formation, etc.  A study focused on this
region comparing synchrotron and dust would be extremely interesting
for future work.

\section{Conclusions}
\label{sec:discussion}

We have updated three models for the Galactic magnetic fields in the
literature (Sun10, {\jf}, and Jaffe13) in order to match the {\planck}
Commander synchrotron maps.  We use a common {\cre} model from
Orlando13, which has different spatial and spectral
behaviour than the {\cre} models used in the original development of each of
the magnetic field models.  Different {\cre} models result in changes
to both the morphology of the predicted synchrotron emission and the inferred degree of field ordering.  The reference synchrotron
data also have a 
different morphology from, e.g., the {\wmap} component-separation
products used in developing the {\jf} model.  For these reasons, all
of the field models required adjustments to their parameters in order
to match.  Our updates are neither best-fit models nor unique
solutions in a degenerate parameter space, but a full exploration of
these parameters is beyond the scope of this work.  Nevertheless, the updated models
roughly reproduce the basic large-scale morphology of the synchrotron emission
in total and polarized intensity as measured by {\planck}.  

One of the results of this paper is to demonstrate explicitly how the
choice of {\cre} model \addedD{(particularly the spectrum)} and the component separation in the microwave
bands (which is related) affect the results.  Such issues are also 
discussed in \citet{planck2014-a31}, and here we show how they affect in
particular the estimates for the degree of polarization in the synchrotron
emission and therefore the degree of ordering in the magnetic fields.
We use the {\planck} component separation results, but 
these are subject to uncertainties and are unlikely to be reliable
estimates in the Galactic plane, where we are attempting to probe the
magnetic fields through the full Galactic disc.  \added{The resulting
  models, like all of the models in the literature, 
are therefore still subject to significant uncertainties because of these
issues, as will be the case for any such analysis using the {\planck} synchrotron estimate. }  

With the updated magnetic field models, we turn to the predictions for
polarized dust emission and compare \addedD{them} with the {\planck} data at
353\,GHz.  We find that the predictions do not match the dust
emission well, whether using the original magnetic field models from the
literature or our updated models.  In particular, all of the models
predict a narrower distribution of polarized dust emission in the
plane than is observed by {\planck} and underpredict the
polarized emission away from the plane.  
\addedB{Because the synchrotron component separation is uncertain, as
  are the synchrotron spectral index and its latitude variation, the
  vertical variation in the magnetic field ordering is also
  uncertain.  That uncertainty also affects the latitude variation of
  the dust polarization, so this issue is far from understood.}

We then further adjust the {\jf} model parameters in a way that
remains consistent with the synchrotron emission but is also a closer
match to the dust polarization.  This is meant as a proof of concept,
rather than as a physically well-motivated model, and illustrates how
we can, in principle, probe the different fields traced by these two
components.  \addedB{Though this model remains subject to the
  uncertainties discussed extensively in this work, the update
  nevertheless represents the most comprehensive effort to model the
  large-scale Galactic magnetic fields using the combination of
  Faraday RM data, diffuse synchrotron emission, and the new thermal
  dust polarization information brought by the {\planck} data.}

\added{Previous analyses have proceeded from different assumptions
  about component separation and/or about the synchrotron
  spectral index, and this has led to very different models for the
  large-scale fields providing adequate matches to the chosen subsets
  of the available data.
  We have compared these models with each other and updated them for a
  particular set of assumptions, i.e., those made in the {\planck} component
  separation, but we have not overcome these problems.  The
    main result of this paper is \removedB{not an improved model but} an improved
  understanding of the challenges in the analysis and of the
  limitations of the existing models.}

\added{There are, however, several specific points that we have established}:  

\begin{itemize}

\item The original {\jf} model clearly has too large a random
  component, likely due to the {\wmap} MCMC solution being contaminated
  by AME.  This is indicated by the total amount of synchrotron
  emission and by the significantly overpredicted galactic
  variance for synchrotron polarization.  

\item Our updated models may, in contrast, underestimate this random
  component.  This is implied by the observed versus modelled
  variations and may be explained by the fact that the 
  {\planck} {\commander} analysis assumes a very steep synchrotron
  spectrum.  This question remains unanswered, and the original {\jf}
  model and our updates likely represent the extremes that bracket reality.




\item When using the field models adjusted to match the {\commander} 
  synchrotron solution, i.e., with the assumed steep synchrotron spectrum and
  the correspondingly strongly ordered magnetic field model, the
  predicted dust polarization matches roughly the level of
  polarization in the outer regions of the Galactic plane, or can even
  be made to overpredict it.  This is in contrast to the results of
  Jaffe13 but is clearly dependent on the outstanding
  component separation question.

\item We can adjust the {\jf} model to roughly match both synchrotron
  and dust emission.  This model depends strongly on choices
  made in the component-separation process but \addedB{demonstrates how the addition of
  polarized dust emission can improve the detailed modelling in the thin
  Galactic disc.}  

\end{itemize}

\removedB{Though we conclude that the modelling remains challenging,
  t}
\addedB{The prospects for large-scale magnetic field modelling} are quite promising.  Firstly, ongoing
ground-based radio surveys (see Appendix\,\ref{sec:appendix_compsep})
will soon map the sky at several crucial intermediate frequencies and
provide leverage for component-separation algorithms such as {\commander}
\added{via the additional} information about the synchrotron spectrum at low
frequencies.  In the longer term, the Square Kilometre Array
(SKA\footnote{\url{https://www.skatelescope.org/}}) will increase the
sampling of Galactic pulsars by several orders of magnitude, which
will improve our understanding of both the thermal electron
distribution and the magnetic fields in the narrow disc component.
The Gaia mission\footnote{\url{http://sci.esa.int/gaia/}} will provide
millions of extinction measurements in the local quadrant of the Galaxy
that will allow more precise mapping of the dust distribution.  The
combination of SKA and Gaia will therefore greatly advance our ability
to study the fields in the thin disc.  SKA will also improve the
sampling of the extragalactic sources that trace the fields throughout
the Galaxy, including the halo.  The combination of all of
these data will help to study precisely this question of how the
\added{RM data,} synchrotron emission, and dust emission reveal the different regions of the
magnetized ISM.

\begin{acknowledgements}

The Planck Collaboration acknowledges the support of: ESA; CNES, and
CNRS/INSU-IN2P3-INP (France); ASI, CNR, and INAF (Italy); NASA and DoE
(USA); STFC and UKSA (UK); CSIC, MINECO, JA and RES (Spain); Tekes, AoF,
and CSC (Finland); DLR and MPG (Germany); CSA (Canada); DTU Space
(Denmark); SER/SSO (Switzerland); RCN (Norway); SFI (Ireland);
FCT/MCTES (Portugal); ERC and PRACE (EU). A description of the Planck
Collaboration and a list of its members, indicating which technical
or scientific activities they have been involved in, can be found at
\url{http://www.cosmos.esa.int/web/planck/planck-collaboration}.

Some of the results in this paper have been derived using the {\HEALPix}
package.  

We acknowledge the use of the Legacy Archive for Microwave Background
Data Analysis (LAMBDA), part of the High Energy Astrophysics Science
Archive Center (HEASARC). HEASARC/LAMBDA is a service of the
Astrophysics Science Division at the NASA Goddard Space Flight Center.

\end{acknowledgements}

\bibpunct{(}{)}{;}{a}{}{,}  
\bibliographystyle{aat}
\bibliography{manual_bib,Planck_bib,references}

\appendix

\section{Polarization systematics}
\label{sec:appendix_systematics}

For both {\planck} instruments, the dominant polarization systematic in the published maps is the leakage of total intensity signal into polarized intensity due to the bandpass mismatch, i.e., the small differences in the bandpasses of the different detectors used to measure orthogonal polarization orientations.  See  \citet{planck2014-a03} and \citet{planck2014-a08}.  This appendix discusses how we can characterize the effects and, in the case of HFI, compare the different methods used to correct it.  The leakage is largest in the Galactic plane, since it is proportional to the total intensity.  It is also {\em proportionately} worst in the plane due to the lower polarization fraction.   Away from the plane, our analysis is not significantly affected, and here we estimate the effects in the plane. 

\subsection{LFI}
\label{sec:appendix_systematics_lfi}

The LFI leakage is discussed in \citet{planck2014-a03}.  The results along the Galactic plane are shown in Fig.\,\ref{fig:systematics} in comparison with the two low-frequency bands from {\wmap}.  The comparison of the two {\wmap} bands gives an idea of the uncertainty in their correction, and the LFI 30\,GHz data appear to deviate more than this.  This is not unexpected given that the  {\wmap} scan pattern allows the blind separation of the leakage, which is more difficult for LFI.  In the inner Galactic plane, therefore, we should consider an additional systematic error of around $0.06$\,mK.  

Though it is likely that the {\wmap} data are less affected by leakage along the Galactic plane, the {\wmap} solution has a high degree of uncertainty in its largest-scale modes, as discussed in \citet{page:2007}.  As a result, its high-latitude, large-scale morphology, where the signal is low, is unreliable, and it is there that the LFI data are more likely to be accurate.  This issue is discussed in more detail in \citet{planck2014-a12}, which outlines the difficulties in measuring the largest-scale modes of the polarization signal for both {\planck} and {\wmap}.  The cosmological analysis at high latitudes is far more sensitive to these issues than our analysis of the relatively high signal-to-noise Galactic foregrounds near the plane.  Since we do not perform a quantitative fitting in this work, it is sufficient that we compare the data and model morphologies and use these comparisons when judging the significance of the residuals.  

We take the rms variation among the three measures of the synchrotron emission (at 23\,GHz, 30\,GHz, and 33\,GHz) as an estimate of the uncertainty shown by the grey band in the top panel of  Fig.\,\ref{fig:systematics}.  These uncertainties in $Q$ and $U$ are then propagated to polarized intensity and shown as the grey band in Fig.\,\ref{fig:lon_prof_comp_synch}.

\subsection{HFI}
\label{sec:appendix_systematics_hfi}

There are two different leakage corrections included in the HFI data release.  
The default correction for HFI is based on the ground measurements of the bandpasses.  The limitations of this method are mainly the accuracy of those measured bandpasses and the necessary assumption that the dust spectral index and temperature are constant over the sky.  An alternative method is also discussed in \citet{planck2014-a09} that performs a generalized global fit to correct for not only this bandpass leakage but also for calibration and monopole leakage. 

In order to assess the two correction methods, we have looked at the three polarization frequencies of 143, 217, and 353\,GHz (excluding 100\,GHz, which is dominated by CO leakage that is an additional complication at this frequency but \addedC{less important} in the other frequencies).  We extrapolate the different bands to a common frequency in order to compare them with each other.  The variation among the bands may be due to spectral variation in the polarized emission but is more likely due to the bandpass leakage, which is significant on the inner Galactic plane where the total intensity is highest.  We then compare these variations for the two corrections to determine if one is apparently better than the other, and we find that though they perform differently in different regions, they perform similarly overall.  

\begin{figure*}
    \includegraphics[width=180mm]{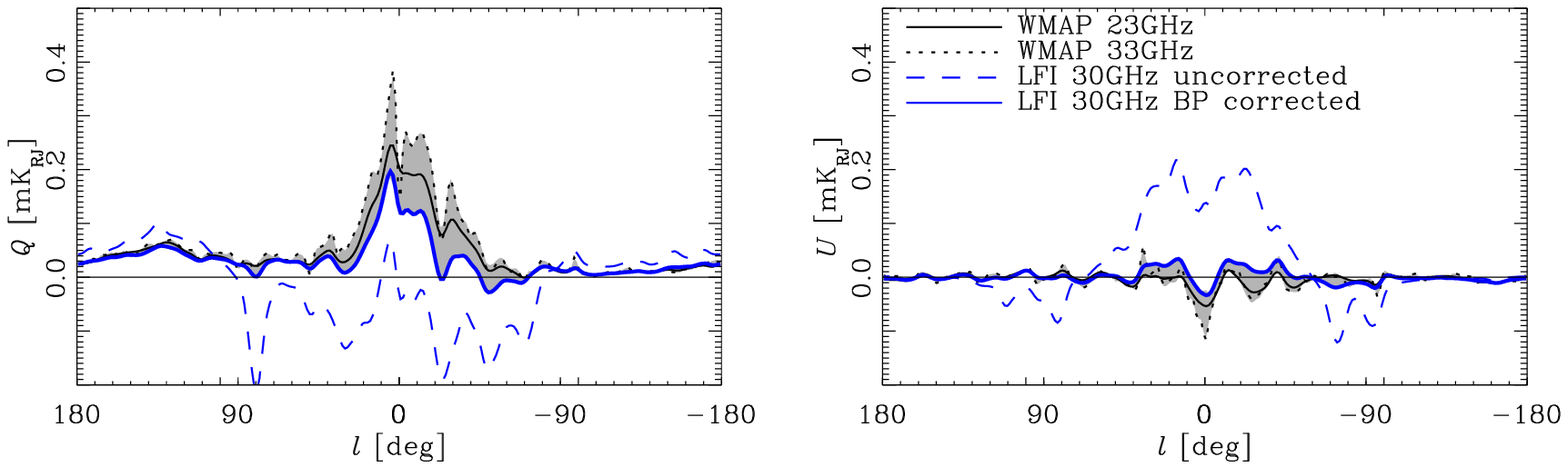}
  \includegraphics[]{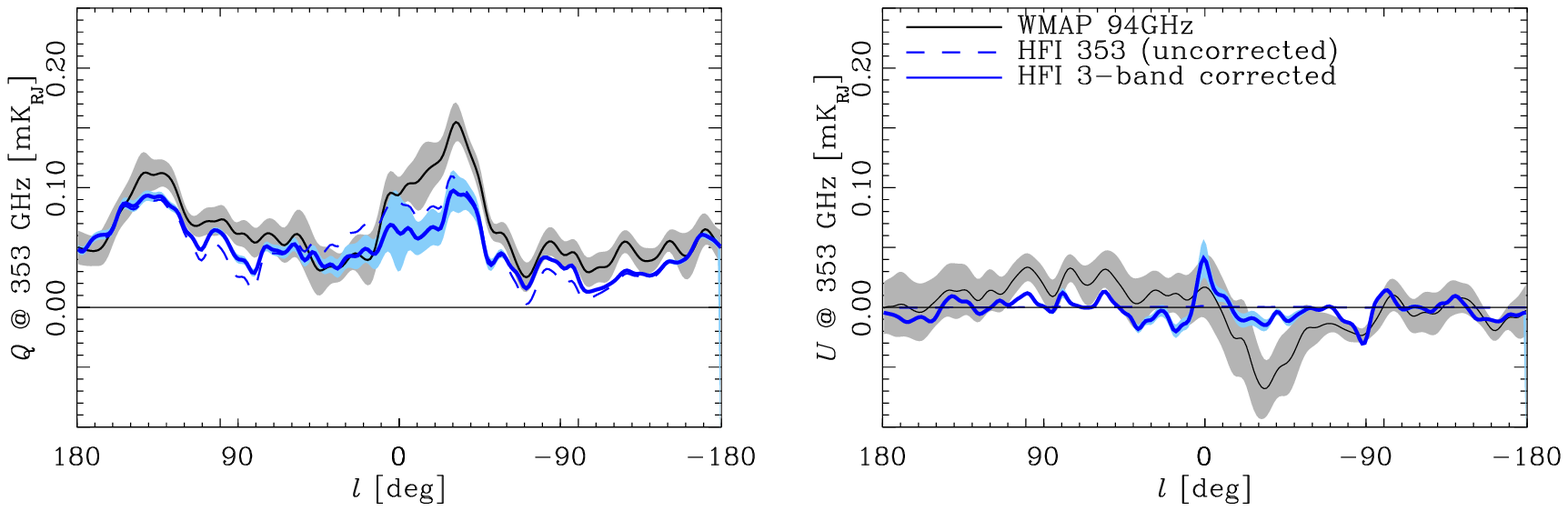}
\caption{ {\em Top}:  comparison of LFI 30\,GHz Stokes $Q$ (left) and $U$ (right) with {\wmap} 23 and 33\,GHz along the Galactic plane smoothed to a full width half maximum (FWHM) of $6\degr$.  \added{All frequencies are over-plotted by extrapolating the {\wmap} data to 30\,GHz using a synchrotron $\beta=-2.95$ \citep{jaffe:2011}.}  The grey band shows the rms variation among them.  \added{ {\em Bottom}:  average profile of three HFI frequencies with ground-based  bandpass leakage correction.  The 143 and 217\,GHz frequencies are extrapolated to 353\,GHz using the {\planck} dust model spectrum of \citet{planck2013-p06b}.  The dark blue curve shows the average of the frequencies with a pale blue band showing the variance.  }
The {\wmap} 94\,GHz data are also extrapolated to 353\,GHz in order to compare its profile in solid black with the grey band showing its variation among different years.  
\label{fig:systematics}
}
\end{figure*}

In Fig.\,\ref{fig:systematics}, we look at the Galactic plane profile at 353\,GHz.  The black curve shows the {\wmap} 94\,GHz data for comparison, with a grey band showing its variation among the different years.  These are extrapolated to 353\,GHz using the {\planck} dust model spectrum of \citet{planck2013-p06b}.  Though {\wmap} can solve for the leakage directly, there is a problem discussed by \cite{page:2007} with the largest-scale modes, which tend to be ill-constrained, an effect that contributes to the grey band indicating the variance among the different years.  
\added{The dark blue curve represents the average of the three HFI frequencies 143, 217, and 353\,GHz, each with the ground-based leakage correction applied.  
The pale blue band then shows the effective uncertainty in the dust polarization profile along the plane computed simply as the variation among the frequencies from their mean.   We examined the same profiles for the alternate correction (not shown) and found that again, while the different corrections are better or worse in different places, neither is significantly better overall. }   

\citet{planck2014-a12} quote an uncertainty of $1\,\mu$K in the polarization at 353\,GHz.  
Figure\,\ref{fig:systematics} shows that this estimate is of the right order on average.  

\added{We therefore use for our dust polarization data set the 353\,GHz frequency band, which maximizes the signal-to-noise, with the ground-based bandpass correction applied.  We estimate the systematic uncertainty using the variation among the frequencies, i.e., the pale blue bands in Fig.\,\ref{fig:systematics}.}  The uncertainties in $Q$ and $U$ are then propagated to polarized intensity and shown as the grey band in Figs.\,\ref{fig:lon_prof_comp_dust} and \ref{fig:lat_prof_comp_dust2}.  

\section{Comparison of synchrotron emission estimates}
\label{sec:appendix_compsep}

\begin{figure*}
\includegraphics[clip,trim=0 0 0 0.2cm ,width=180mm]{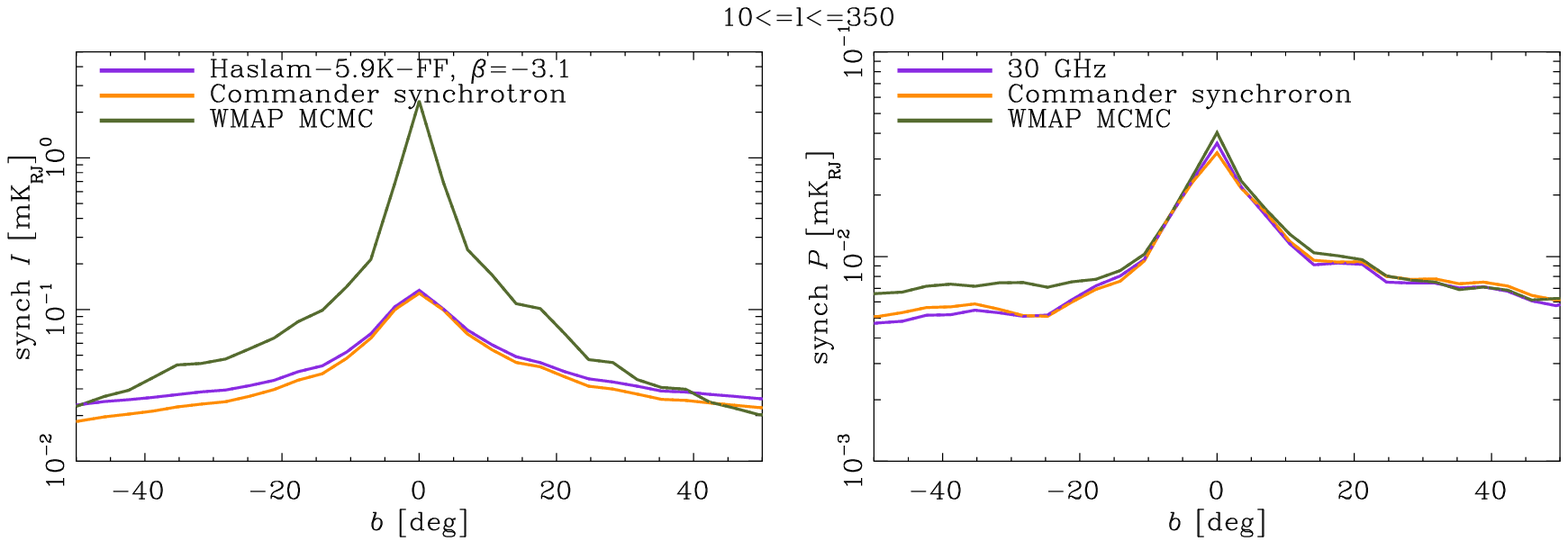}
\includegraphics[width=180mm]{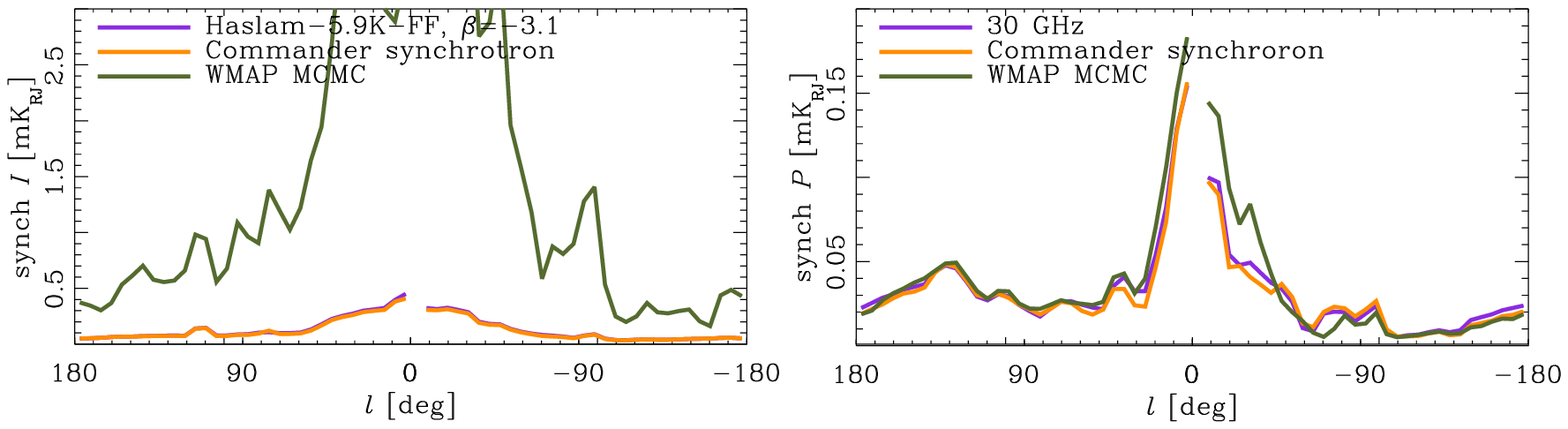}
\caption{ Comparison of synchrotron data sets as latitude (top)
  and longitude (bottom) profiles.
  \added{ For the latitude profiles, the 
  full sky is averaged excluding the inner $10\degr$ ($|b|<10\degr$
  and $|\ell|<10\degr$).  For the longitude profiles, only the pixels
  along the plane are plotted.  In total intensity on the left, the
  {\commander} synchrotron solution,
  which is identical (except for an offset) to the Haslam 408\,MHz
  map,  is compared to the  {\wmap} MCMC synchrotron solution.  In polarization, on the right, \added{the {\commander}
  synchrotron is compared to the LFI 30\,GHz map itself and the
  {\wmap} MCMC synchrotron solution} \addedB{extrapolated to 30\,GHz assuming $\beta=-3$.}  
}
\label{fig:data_comp_lat} 
}
\end{figure*}

We compare the data sets used to develop the magnetic field models
in the literature discussed in \S\,\ref{sec:models} in
Fig.\,\ref{fig:data_comp_lat}.  Recall that
the Sun10 and Jaffe13 models were developed in reference to the
synchrotron total intensity from the 408\,MHz map while the {\jf} model was fitted to the {\wmap} MCMC synchrotron
estimate at 23\,GHz.\footnote{The total intensity synchrotron
estimate used by {\jf} is the {\wmap} MCMC solution
(R. Jansson, private communication).  Specifically, they used the
basic {\wmap} MCMC component separation method described by 
\citet{gold:2011}, i.e., with a synchrotron power law with no
steepening and without fitting any AME component (aka ``spinning
dust'').} 
We plot all data sets at a common frequency of 30\,GHz to match
LFI. For the {\wmap} 23\,GHz maps, the small difference in frequency
makes the plot insensitive to the precise spectrum assumed, and we use
a power law with an index (in brightness temperature) of 
$\beta=-3$. For extrapolating the 408\,MHz map to microwave bands, however, the spectrum
assumed has a large effect on the result, and we do not know its
variations over the sky.  We show \addedB{the result of using
  $\beta=-3.1$}, the effective spectrum of the assumed synchrotron spectral template
used in the {\commander} component separation \citep{planck2014-a12}.  

Another important uncertainty is that of the offset in the 408\,MHz
map.  \cite{haslam:1982} quote 3\,K as the zero-level uncertainty, and
\added{the map contains both CMB and an unresolved extragalactic component.}
\cite{lawson:1987} find an offset of 5.9\,K, including both CMB and
extragalactic components, \added{though the 3\,K uncertainty still applies.}
The purple curve in Fig.\,\ref{fig:data_comp_lat} has this offset
removed.  We compare this curve with the {\planck} {\commander} solution
described in \citet{planck2014-a12}.  As described in that paper, the
synchrotron solution follows the morphology of the 408\,MHz map but
with an independent offset determination from \cite{wehus:2014} of
$9$\,K \added{(consistent with the Lawson offset within the calibration
uncertainty)} and a frequency dependence from a {\galprop} simulated
{\cre} spectrum.  (The 3\,K uncertainty at 408\,MHz is equivalent
to a 5\,$\mu$K uncertainty at 30\,GHz assuming $\beta=-3.1$.) 

The {\planck} {\commander} solution includes the posterior rms
uncertainty at each position, and for the synchrotron total intensity,
this is at the $1\,\%$ level.  
This uncertainty does not, \addedD{however,} take into
account the simplicity of the model and the uncertainties in the
energy spectrum that have a large effect on the microwave intensity
model.  

In polarization, the {\wmap} curve differs
slightly from the {\planck} polarization data.  As described by 
\cite{page:2007}, the {\wmap} polarization processing can solve for
the leakage of intensity, which strongly affects LFI in the Galactic mid-plane
(see \citealt{planck2014-a03}), but {\wmap} also has unconstrained
  large-scale modes that are significant away from the plane.  We
  discuss this in Appendix\,\ref{sec:appendix_systematics}.
  \added{The effect of the slightly different treatment of the
    systematics is visible in the comparison of the {\commander}
    synchrotron versus the LFI 30\,GHz polarization itself.}  

These plots make clear the differences in the data sets that could be
used for modelling the Galactic magnetic fields and indicate that the results
may vary significantly depending on which choices are made.  The
implications for the models in the literature are discussed in
Sects.\,\ref{sec:models} and \ref{sec:discussion}.  

The situation will soon be improved by the completion of surveys such
as: the ongoing C-Band All Sky Survey
(C-BASS\footnote{\url{http://www.astro.caltech.edu/cbass/}},
\citealt{king:2014} and references therein) to map the full sky in
polarization at 5\,GHz; the S-band Parkes All-Sky Survey (S-PASS,
\citealt{2013Natur.493...66C}) at 2.3\,GHz; and the Q-U-I JOint
Tenerife CMB Experiment (QUIJOTE, \citealt{GenovaSantos:2015hc} and
references therein) planned for 10 to 40\,GHz.  These
intermediate-frequency surveys will significantly advance our
understanding of the synchrotron spectral variations and provide
crucial frequencies for parametric component-separation algorithms
like {\commander}.

\section{Model parameters}
\label{sec:appendix_params}

Table \ref{tab:param_updates} lists the changes to the magnetic field
models.  These models are extensively described in the references
given, and we do not reproduce the full list here.  Brief summaries of
the models and methods used are given in \S\,\ref{sec:models}.  Any
parameter not listed here retains its original value from the original
references.  

In Table\,\ref{tab:params_dust}, we specify the dust model we use and list all of its parameters.  This model is described in \S\,\ref{sec:dust}. 

In Table\,\ref{tab:params_cres}, we compare the {\cre} injection and
propagation parameters in two {\galprop} models:  the one used in
  \added{the analysis of \citet{jaffe:2011} on which Jaffe13 is based}, and the one used for the results presented here.

\newcommand{\rdelimN}[4]{
{\multirow{#1}{#2}[#3]{%
   \ensuremath
     {\left\}\vcenter{\hsize=0pt\vrule height 
           \ifnum #1<0 -\fi#1\baselineskip width 0pt}%
       \textrm{#4}\right.}}}
}

\newcommand{\multibraceN}[4]{\rdelimN{#1}{3mm}{#3}{\hspace{1mm}\parbox{#2-4mm}{#4}}}

\begin{table*}[t]                 
\begingroup
\newdimen{\firstcolwidth}
\setlength{\firstcolwidth}{1.8in}
\newlength{\lastcolwidth}
\setlength{\lastcolwidth}{3.3in}
\newdimen\tblskip \tblskip=5pt
\caption{Parameter updates to models in the literature.}                          
\label{tab:param_updates}                            
\nointerlineskip
\vskip -3mm
\footnotesize
\setbox\tablebox=\vbox{
   \newdimen\digitwidth 
   \setbox0=\hbox{\rm 0} 
   \digitwidth=\wd0 
   \catcode`*=\active 
   \def*{\kern\digitwidth}
   \newdimen\signwidth 
   \setbox0=\hbox{+} 
   \signwidth=\wd0 
   \catcode`!=\active 
   \def!{\kern\signwidth}
\halign{\hbox to \firstcolwidth{#\hfil}&
        \hfil#\hfil\tabskip 1em&
        \hfil#\hfil&
        \hfil#\hfil&
        \vtop{\hsize \lastcolwidth\hangafter=1\hangindent=1em\noindent\strut#\strut\par}\tabskip=0pt\cr                          
\noalign{\doubleline}
        Model&Param.&Orig.\ value&New value&Comments\cr                                    
\noalign{\vskip 3pt\hrule\vskip 5pt}
    \multirow{7}{\firstcolwidth}[1.1em]{``Sun10b''\\\citet{sun:2008}\\\citet{Sun:2010}}&
        $\left<B^2_\mathrm{iso}\right>^{1/2}$&
        3\,$\mu$G&
        6.4\,$\mu$G&
        Increased random field component, degenerate with {\cre} normalization.\cr
    &
        $\beta$&
        \dots&
        3&
        Need more ordered fields to reproduce polarized synchrotron, adding ordered {\random} field component using prescription from \citet{jansson:2012b}.\cr
    &
        $r_0^\mathrm{ran}$&
        \dots&
        30\,kpc&
        \multibraceN{5}{\lastcolwidth}{0.8em}{Uniform {\random} field changed to combination of thin and thick exponential discs, i.e.,\\ 
        $\mathbf{B}(z)\propto(1-f^\rmn{ran}_\rmn{disc})\mathrm{sech}^2(z/h_\rmn{halo}^\rmn{ran}) + f^\rmn{ran}_\rmn{disc}\mathrm{sech}^2(z/h_\rmn{disc}^\rmn{ran})$,\\
        $\mathbf{B}(r)\propto \exp \left[-(r-R_{\sun})/r_0^\rmn{ran}\right].$}\cr
    \noalign{\vskip 3pt}
    &
        $f^\rmn{ran}_\rmn{disc}$&
        \dots&
        0.5&
        \cr
    \noalign{\vskip 3pt}
    &
        $h^\mathrm{ran}_\rmn{halo}$&
        \dots&
        3\,kpc&
        \cr
    \noalign{\vskip 3pt}
    &
        $h^\mathrm{ran}_\rmn{disc}$&
        \dots&
        1\,kpc&
        \cr
    \noalign{\vskip 3pt}
    &
        $B_\rmn{c}$&
        2\,$\mu$G&
        0.5\,$\mu$G&
        Dropping the amplitude in the inner Galaxy so as not to overpredict.\cr
\noalign{\vskip 3pt\hrule\vskip 3pt}
    \multirow{10}{\firstcolwidth}[1em]{``Jaffe13b'' \\\citet{jaffe:2010,jaffe:2011,jaffe:2013} 
    }&
        $\left<B^2_\mathrm{iso}\right>^{1/2}$&
        5\,$\mu$G&
        6.5\,$\mu$G&
        \multibraceN{2}{\lastcolwidth}{0.8em}{Reduced global
          normalization of isotropic {\random} component, increasing
          field ordering.  (Here defined at $r=R_\sun$ rather than at $r=0$.)}\cr
    \noalign{\vskip 3pt}
    &
        $f_\rmn{ord}$&
        0.2&
        0.5&
        \cr
    \noalign{\vskip 3pt}
     &
        $a_i$&
        various&
        various&
        Arm amplitudes now relative to global $B_0^\rmn{disc}$
        parameter.  $\left\{1.2,1.3,-3.8,0.3,-0.02\right\}$.\cr
    \noalign{\vskip 2pt}
    &
        $h_\rmn{disc}$&
        6&
        0.1\,kpc&
        \multibraceN{10}{\lastcolwidth}{0.9em}{Jaffe13 vertical profile not previously constrained.  All components now with thin and thick discs.  Random like Sun10 model above. Coherent:\\   
        $B(z)=B_0^\rmn{disc} \rmn{sech}^2(z/h_\rmn{disc}) + B_0^\rmn{halo}\rmn{sech}^2(z/h_\rmn{halo})$.}\cr
    \noalign{\vskip 2pt}
    &
        $h_\rmn{halo}$&
        \dots&
        3\,kpc&
        \cr
    \noalign{\vskip 2pt}
    &
        $h_\rmn{c}$&
        0.5\,kpc&
        0.1\,kpc&
        \cr
    \noalign{\vskip 2pt}
    &
         $B_0^\rmn{halo}$&
         \dots&
         1.38\,$\mu$G&
         \cr
    \noalign{\vskip 3pt}
    &
        $B_0^\rmn{disc}$&
        \dots&
        0.17\,$\mu$G&
        \cr
    \noalign{\vskip 2pt}
    &
        $h^\rmn{ran}_\rmn{halo}$&
        \dots&
        4\,kpc&
        \cr
    \noalign{\vskip 2pt}
    &
        $h^\rmn{ran}_\rmn{disc}$&
        \dots&
        1\,kpc&
        \cr
    \noalign{\vskip 2pt}    
    &
        $r^\rmn{ran}_0$&
        20\,kpc&
        12\,kpc&
        \cr
    \noalign{\vskip 2pt}
    &
        $f_\rmn{disc}^\rmn{ran}$&
        \dots&
        0.1&
        See Sun10 comment above.\cr
\noalign{\vskip 3pt\hrule\vskip 3pt}
    \multirow{4}{\firstcolwidth}[0.9em]{``{\jf}b'' ordered fields\\\cite{jansson:2012b}}&
        $b_6^\mathrm{disc}$&
        $-4.2$&
        $-3.5$&
        Overpredicted outer Galaxy in the plane in total and polarized intensity. (Segment 6 is the Perseus arm.)\cr
    &
        $B_X$&
        4.6&
        1.8&
        Polarization overpredicted at high latitude.\cr
    \noalign{\vskip 2pt}
    &
        $\beta$&
        1.36&
        10&
        Changed strength of ordered random (``striated'') fields, adjusting for different {\cre} normalization.\cr
\noalign{\vskip 3pt\hrule\vskip 3pt}
    \multirow{8}{\firstcolwidth}[1em]{``{\jf}b'' random fields\\\cite{jansson:2012c}}
        &$\left<B^2_\mathrm{iso}\right>^{1/2}$&
        \dots&
        7.8\,$\mu$G&
        Using a GRF simulation, the global normalization of all random components.\cr
    &
        $b^\rmn{disc}_\rmn{even}$&
        various&
        0.8&
        \multibraceN{2}{\lastcolwidth}{0.9em}{Replacing {\random} field dominated by a single arm segment ($b^\rmn{disc}_7=37\,\mu$G) with four roughly equal arms (even-numbered) and four inter-arm regions (odd-numbered). Values now relative to $\left<B^2_\mathrm{iso}\right>^{1/2}$. }\cr
    \noalign{\vskip 2pt}
    &
        $b^\rmn{disc}_\rmn{odd}$&
        various&
        0.4&
        \cr
    \noalign{\vskip 2pt}
    &
        $b^\rmn{disc}_\mathrm{int}$&
        7.63\,$\mu$G&
        0.5&
        \cr
    &
        $B_0$&
        4.68\,$\mu$G&
        0.94&
        When using the GRF scaled by $\left<B^2_\mathrm{iso}\right>^{1/2}$, maintain the same halo amplitude.\cr
\noalign{\vskip 3pt\hrule\vskip 3pt}
    \multirow{5}{\firstcolwidth}[0.8em]{``{\jf}c'' ordered fields\\as {\jf}b
    except:}&
        $B_\rmn{n}$&
        1.4\,$\mu$G&
        1\,$\mu$G&
        \multibraceN{3}{\lastcolwidth}{0.8em}{Reducing the toroidal halo components that partly cancel the disc component; increasing x-shaped halo to compensate high-latitude synchrotron.}\cr
    \noalign{\vskip 2pt}
    &
        $B_\rmn{s}$&
        $-1.1$\,$\mu$G&
        $-0.8$\,$\mu$G&
        \cr
    \noalign{\vskip 2pt}
    &
        $B_\rmn{X}$&
        1.8\,$\mu$G&
        3\,$\mu$G&
        \cr
    &
        $b^\rmn{disc}_2$&
        3\,$\mu$G&
        2\,$\mu$G&
        Reduce inner Galaxy dust polarization.\cr
    &
        $b^\rmn{disc}_4$&
        $-0.8$\,$\mu$G&
        2\,$\mu$G&
        Replacing synchrotron polarization.\cr
    &
        $b^\rmn{disc}_5$&
        $-2$\,$\mu$G&
        $-3$\,$\mu$G&
        Increase high-latitude polarization.\cr
\noalign{\vskip 3pt\hrule\vskip 3pt}
    \multirow{5}{\firstcolwidth}[0.8em]{``{\jf}c'' random fields\\as {\jf}b except:}&
        $b^\rmn{disc}_6$&
        0.8&
        1.6&
        To increase outer-Galaxy synchrotron intensity in the plane.\cr
    &
        $b^\rmn{disc}_\rmn{odd}$&
        0.4&
        0.1&
        To compensate the above change on average. Then to further increase local high-latitude dust polarization, set only $b^\rmn{disc}_5=0$.\cr
    &
        shift&
        \dots&
        0.97&
        Shift the arm pattern by multiplying the $r_{-x}$ parameters from \citet{jansson:2012b} by this factor.\cr
\noalign{\vskip 3pt\hrule\vskip 3pt}}}

\endPlancktablewide                 
\textbf{Notes:}  Where not specified, parameters remain at the values in the
  references.  The notation of the original references is used for
  each model with added sub- or super-scripts as necessary to clarify
  different field components.  (The {\jf} parameter
  that controls the amount of power in ordered random fields relative to that
  of the coherent fields is $\beta$, not to be confused with the temperature spectral index,
  also $\beta$ elsewhere in this work.)\par 
\endgroup
\end{table*}                        

\begin{table*}[t]                 
\begingroup
\setlength{\lastcolwidth}{3in}
\newdimen\tblskip \tblskip=5pt
\caption{Description of the model for the distribution of dust emissivity.}                          
\label{tab:params_dust}                            
\nointerlineskip
\vskip -3mm
\footnotesize
\setbox\tablebox=\vbox{
   \newdimen\digitwidth 
   \setbox0=\hbox{\rm 0} 
   \digitwidth=\wd0 
   \catcode`*=\active 
   \def*{\kern\digitwidth}
   \newdimen\signwidth 
   \setbox0=\hbox{+} 
   \signwidth=\wd0 
   \catcode`!=\active 
   \def!{\kern\signwidth}
\halign{\hfil#\hfil\tabskip 1em&
        \hfil#\hfil&
        \hfil#\hfil&
        \vtop{\hsize \lastcolwidth\hangafter=1\hangindent=1em\noindent\strut#\strut\par}\tabskip=0pt\cr                          
\noalign{\doubleline}
        Param.&Default&Equation&Description\cr                                    
\noalign{\vskip 3pt\hrule\vskip 3pt}
\multispan4\hfil\textsc{Disc component}\hfil\cr
\noalign{\vskip 4pt\hrule\vskip 3pt}
    $r_0$&
    6\,kpc&
    $\rho(r)=\exp(-r/r_0)$&
    Exponential disc scale radius.\cr
    $z_0$&
    0.3\,kpc&
    $\rho(z)=\mathrm{sech}^2(z/z_0)$&
    Exponential disc scale height.\cr
    $R_\rmn{max}$&
    12\,kpc&
    \dots&
    Maximum radius, beyond which $\mathcal{E}_\mathrm{dust}=0$.\cr
\noalign{\vskip 3pt\hrule\vskip 3pt}
\multispan4\hfil\textsc{Spiral arms}\hfil\cr
\noalign{\vskip 4pt\hrule\vskip 3pt}
    $R_\rmn{mol}$&
    5\,kpc&
    \dots&
    Radius of molecular ring.\cr
    $a_i$&
    $\{4, 2, 1.5, 4, 2\}$&
    $\rho_{\mathrm{arm},i} =  a_i \rho_c(d_i)$&
    Amplitude of each of four spiral arms and molecular ring.  Order corresponds to:  Perseus, Sagittarius, Scutum, Norma, molecular ring.  The Sagittarius arm is damped relative to the others, and the amplitudes are relative to the smooth background component.\cr
    &
    &
    $\rho_c(d)= c(r) \rho_\mathrm{c}(z)\rho(r)\exp(-(d/d_0(r))^2)$&
    Amplification factor relative to background.  $d$ is the distance along Galacto-centric $\hat{r}$ to the nearest arm in kpc, computed using $r_i(\phi)$.\cr
    $\phi_{0,i}$&
    $70\degr+90\degr i$&
    $r_i(\phi)=R_s\exp\left[(\phi-\phi_{0,i})/\beta\right]$ and $\beta\equiv1/\tan(\theta_p)$&
    $r(\phi)$ gives the arm radius at a given azimuth, where $\phi_{0,i}$ is the azimuthal orientation of the  spiral around the axis through the Galactic poles.  (Constant $R_\rmn{mol}$ for molecular ring.)\cr
    $\theta_p$&
    $-11\pdeg5$&
    \dots&
    Pitch angle of the spiral arms\cr
    $C_0$&
    5.7&
    $c(r)=\left\{ \begin{array}{ll}C_0&\mbox{if $r\le r_\mathrm{cc}$ }\\C_0 (r/r_\mathrm{cc})^{-3}&\mbox{if $r> r_\mathrm{cc}$}\\\end{array}\right.$&
    Arm amplitude relative to inter-arm, tailing off after $r_\rmn{cc}$.\cr
    $r_\mathrm{cc}$&
    9\,kpc&
    \dots&
    Region of constant arm amplification.\cr
    $d_0$&
    0.1 kpc&
    $d_0(r)=d_0/(c(r)\rho(r))$&
    Defines the base width of arm enhancement, which varies with radius.\cr
    $h_\mathrm{c}$&
    0.04\,kpc&
    $\rho_\mathrm{c}(z) = \mathrm{sech}^2(z/h_{\mathrm c})$&
    Scale height of the spiral arm component.\cr
\noalign{\vskip 3pt\hrule\vskip 3pt}
\multispan4\hfil\textsc{Local bubble}\hfil\cr
\noalign{\vskip 4pt\hrule\vskip 3pt}
    $R_\rmn{LB}$&
    150\,pc&
    \dots&
    Radius of cylindrical region about the observer.\cr
    $h_\rmn{LB}$&
    200\,pc&
    \dots&
    Height of cylindrical region about the observer.\cr
    $a_\rmn{LB}$&
    0&
    $A=\left\{ \begin{array}{ll}1&\mbox{if$r_\odot > R_\rmn{LB}$ and $|z| > h_\rmn{LB}$}\\a_\rmn{LB}& \mbox{if $r_\odot\le R_\rmn{LB}$ and $|z|\le h_\rmn{LB}$}\\\end{array}\right.$&
    Relative amplitude within the local bubble.\cr
\noalign{\vskip 3pt\hrule\vskip 3pt}}}

\endPlancktablewide                 
\textbf{Notes:}  See \S~\ref{sec:dust_model}.  The model is a smooth exponential disc
  plus four logarithmic spiral arms that have a Gaussian
  density profile as a function of radius from the arm ridge: $\mathcal{E}_\mathrm{dust} \propto
 A\left[ \rho(r)\rho(z)  +  \sum_i^{N_\mathrm{arms}}
   \rho_{\mathrm{arm},i}  \right]$. \par
\endgroup
\end{table*}                        

\begin{table}[t]                 
\begingroup
\newdimen\tblskip \tblskip=5pt
\caption{Comparison of the {\cre} injection and diffusion parameters.}                          
\label{tab:params_cres}                            
\nointerlineskip
\vskip -3mm
\footnotesize
\setbox\tablebox=\vbox{
   \newdimen\digitwidth 
   \setbox0=\hbox{\rm 0} 
   \digitwidth=\wd0 
   \catcode`*=\active 
   \def*{\kern\digitwidth}
   \newdimen\signwidth 
   \setbox0=\hbox{+} 
   \signwidth=\wd0 
   \catcode`!=\active 
   \def!{\kern\signwidth}
\halign{\hbox to 1.8in{#\leaderfil}\tabskip 1em&
        \hfil#\hfil&
        \hfil#\hfil\tabskip=0pt\cr
\noalign{\doubleline}
        \omit\hfil Parameter\hfil&z04LMPDS&z10LMPD\cr                                    
\noalign{\vskip 3pt\hrule\vskip 5pt}
    $|z|_\rmn{max}$\,[kpc]&
    4&
    10\cr
    D0\_xx&
    3.4$\times10^{28}$&
    $6\times10^{28}$\cr
    electron\_norm\_Ekin [MeV]&
    3.45$\times10^4$&
    $3.45\times10^4$\cr
    \vtop{\hsize 1.15in\hangafter=1\hangindent=1em\noindent\strut electron\_norm\_flux\\\hspace{0.4cm}[$(\rmn{cm}^{2}\,\rmn{sr}\,\rmn{s}\,\rmn{MeV})^{-1}$]\strut\par}&
    $0.3\times10^{-9}$&
    $0.3\times10^{-9}$\cr
    electron\_g\_0&
    $\left\{1.3,1.6,1.8\right\}^{\rm{a}}$&
    1.6\cr
    electron\_rigid\_br0 [MV]&
    $4\times10^3$&
    $4\times10^3$\cr
    electron\_g\_1&
    2.25&
    2.5\cr
    electron\_rigid\_br [MV]&
    \dots&
    $5\times10^4$\cr
    electron\_g\_2&
    \dots&
    2.2\cr
\noalign{\vskip 3pt\hrule\vskip 3pt}}}

\endPlancktable                    
\tablenote {{\rm a}} The low-energy injection index was 1.8 in \citet{strong:2010},
the preferred value in \citet{strong:2011} was 1.6, and Jaffe13 used
the fitted value of 1.3 from \citet{jaffe:2011}.\par
\textbf{Notes:}  z04LMPDS from \citet{strong:2010} was used in Jaffe13,
  while the newer model z10LMPD from
 Orlando13 was used as the base
  model for the synchrotron spectral template in {\commander} as
  described in \citet{planck2014-a12}, and it is the common model we
  use for all results presented here.   See \S\,\ref{sec:cres} for
  discussion. \par
\endgroup
\end{table}                        

\raggedright
\end{document}